\documentclass[aps,reprint,prb,superscriptaddress]{revtex4-2}
\usepackage{xcolor}
\usepackage{amsmath,amssymb,amsthm,physics,mathtools}
\usepackage{appendix}
\usepackage{braket}
\usepackage{bm}
\usepackage{comment}
\usepackage[normalem]{ulem}
\usepackage{siunitx}
\usepackage{grffile}
\usepackage{algorithm}
\usepackage{algpseudocode}
\newcommand{\rmd}{{\rm d}}

\newcommand{\iu}{{\rm i}}

\newcommand{\kB}{k_{\rm B}}

\newcommand{\vF}{v_{\rm F}}
 \usepackage[
 colorlinks=true,
 citecolor=gray,linkcolor=gray, urlcolor=gray 
 ]{hyperref}

\graphicspath{{./images/}}

\begin{document}
%

\title{Fluctuation-induced antiparallel spin polarization near the boundaries of chiral metals}

\author{Kosuke Yoshimi}
\email{yoshimi-kosuke567@g.ecc.u-tokyo.ac.jp}
\affiliation{Department of Physics, Graduate School of Science, The University of Tokyo, 7-3-1 Hongo, Tokyo 113-0033, Japan}
\author{Yusuke Kato}
\email{yusuke@phys.c.u-tokyo.ac.jp}
\affiliation{Department of Physics, Graduate School of Science, The University of Tokyo, 7-3-1 Hongo, Tokyo 113-0033, Japan}
\affiliation{Department of Basic Science, The University of Tokyo, 3-8-1 Komaba, Tokyo 153-8902, Japan}
\affiliation{Quantum Research Center for Chirality, Institute for Molecular Science, Okazaki, Aichi 444-8585, Japan}

\author{Yuta Suzuki}
\affiliation{Department of Physics, Institute of Science Tokyo, 2-12-1 Ookayama, Tokyo 152-8551, Japan}
\affiliation{RIKEN Center for Emergent Matter Science, Wako, Saitama 351-0198, Japan}

\author{Shuntaro Sumita}
\affiliation{Department of Basic Science, The University of Tokyo, 3-8-1 Komaba, Tokyo 153-8902, Japan}
\affiliation{RIKEN Center for Emergent Matter Science, Wako, Saitama 351-0198, Japan}

\author{Takuro Sato}
\affiliation{Research Center of Integrative Molecular System, Institute for Molecular Science, National Institutes of Natural Sciences, Okazaki, Aichi 444-8585, Japan}
\author{Hiroshi M.~Yamamoto}
\affiliation{Research Center of Integrative Molecular System, Institute for Molecular Science, National Institutes of Natural Sciences, Okazaki, Aichi 444-8585, Japan}
\affiliation{Quantum Research Center for Chirality, Institute for Molecular Science, Okazaki, Aichi 444-8585, Japan}
\author{Yoshihiko Togawa}
\affiliation{
Department of Physics and Electronics, Osaka Metropolitan University, 1-1 Gakuencho, Sakai, Osaka 599-8531, Japan }
\affiliation{Quantum Research Center for Chirality, Institute for Molecular Science, Okazaki, Aichi 444-8585, Japan}


\date{\today}

\begin{abstract}
The spin response of chiral conductors to nonequilibrium electrical fluctuations remains largely unexplored. We develop a low-frequency semiclassical Boltzmann theory coupled to Gauss's law for a chiral metal with spin-orbit coupling of hedgehog type, treating impurity scattering beyond the conventional relaxation-time approximation. We first determine the quadratic response to a local ac current and its frequency dependence, and then show that zero-mean stationary current fluctuation and electric-field fluctuation near boundaries generate finite time-averaged spin polarizations in the two boundary regions of the chiral metal.
The polarizations are normal to the boundaries and antiparallel: for one chirality they point inward at both boundaries, and for the other they point outward. Within this model, the dominant contribution arises from the linear Edelstein effect driven by a quadratic effective electric field localized near each boundary. This picture may provide a qualitative explanation for CISS-related spin polarization reported in the absence of an applied bias. 
More broadly, we expect other externally maintained stochastic drives to induce spin polarization through the same mechanism.
\end{abstract}
\maketitle

\section{Introduction}
\label{sec: introduction}

Chirality-induced spin selectivity (CISS)~\cite{Naaman2012,Naaman2015,Naaman2018,Naaman2019,Naaman2019apl,Naaman2020-jp,Naaman2020-vz,Waldeck2021,Aiello2022,Bloom2024} has attracted considerable attention because it can produce large spin polarization at room temperature, with the polarization direction determined by chirality. CISS also has potential relevance to spintronics and enantiomer separation.
Following the first observations in photoelectron transmission through chiral molecules~\cite{Ray1999,Goehler2011}, CISS-related phenomena were subsequently reported in inorganic chiral metals under dc current~\cite{Inui2020, Nabei2020,Shiota2021,Shishido2021,Shishido2023}.
These studies marked the extension of CISS research from molecular systems to bulk solid-state responses.

Early studies on chiral molecules reported magnetization switching and enantiospecific interactions with magnetic substrates even without an applied transport current~\cite{Ben_Dor2017-am,Banerjee-Ghosh2018}, and interpreted these phenomena in terms of chirality-dependent spin polarization. Related evidence for spin polarization without a bias current was subsequently reported in Refs.~\cite{Miwa_2020,Kondou2022}. References~\cite{Ben_Dor2017-am,Banerjee-Ghosh2018} further proposed antiparallel spin polarizations at the two ends of a chiral molecule to account for the enantiospecific interactions.
These observations motivate us to consider whether nonequilibrium fluctuations may contribute to spin polarization in the absence of an applied bias.

The antiparallel-spin picture was later examined in solid-state superconducting devices driven by an ac current~\cite{nakajima2023,Nakajima_Thesis} to test the proposals of Refs.~\cite{Ben_Dor2017-am,Banerjee-Ghosh2018}.
A subsequent analysis identified the leading dc contribution as quadratic  in the ac-current amplitude~\cite{Yao2024}.
Related chirality-dependent magnetic control has more recently been reported in chiral ionic gating~\cite{Matsuoka2025}.
With these solid-state experiments in mind, we focus here on crystalline systems.

\begin{figure}
    \centering
    \includegraphics[width=0.7\linewidth]{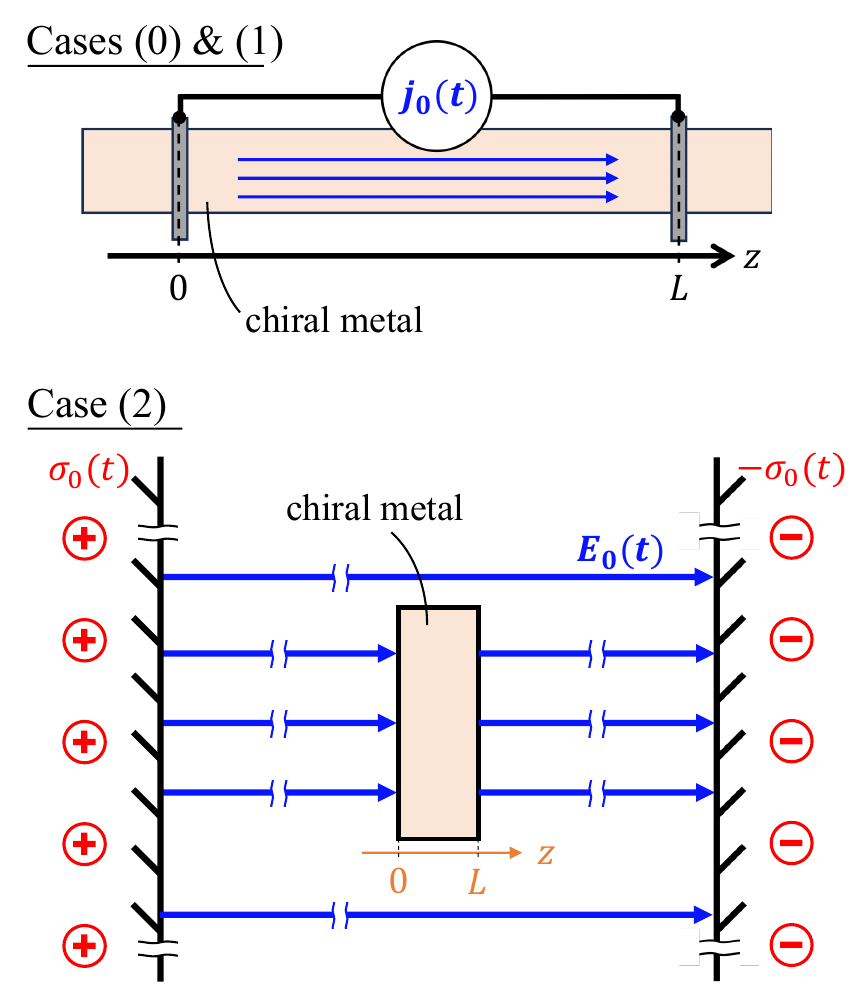}
    \caption{
    Schematics for the setup to measure the response to (0) the local ac current, (1) the local electric current fluctuations and (2) the electric field fluctuations near the surfaces.
    For the cases (0) and (1), the source current in both cases is expressed as $j_0(t)$, which flows between $z=0$ and $z=L$ along $z$-axis.
    For the case (2), we consider the situation where the chiral metal is sufficiently small that the fluctuating electric field generated by the fluctuating surface charge density $\pm\sigma_0(t)$ is along the $z$-axis near its surface.}
    \label{fig_system}
\end{figure}

Previous theoretical studies have addressed spin polarization under externally applied dc and ac drives, using the Boltzmann equation for local current injection \cite{Yoshimi2026} and the nonlinear Kubo formalism for a homogeneous electric field \cite{Inda2026}, respectively. However, spin polarization induced by stochastic electrical fluctuations, rather than by a deterministic drive, remains unexplored.

In this paper, we focus on a nonlinear regime complementary to the linear CISS response under dc transport in chiral metals~\cite{Inui2020,Nabei2020,Shiota2021,Shishido2021,Shishido2023}. Within a band description of a chiral metal, we ask whether zero-mean stochastic electrical fluctuations can generate a finite time-averaged spin polarization through a quadratic response.
We consider two fluctuations setups, (1) and (2), together with a local ac-current setup (0), which serves as a benchmark for the formalism and provides the frequency-dependent response used to analyze case (1).
In case (1), we consider the local injection of nonequilibrium current fluctuation (the upper panel in Fig.~\ref{fig_system}). 
In case (2), we consider nonequilibrium electric-field fluctuations
at the boundaries ($z = 0, L$) of the chiral metal (the lower panel in Fig.~\ref{fig_system}).
The details are explained in Sec.~\ref{sec: set up}.

Our main result is that zero-mean electrical fluctuations generate finite time-averaged spin polarizations localized at the two boundaries. The polarizations are antiparallel and reverse sign under chirality reversal. 
Within the present model, the dominant contribution arises from an effective electric field generated at second order of external drives near the boundaries. This field induces spin polarization through the linear Edelstein effect~\cite{Vasko1979a,Levitov1985,Aronov1989,Edelstein1990,Kato2004a,Silov2004,Ganichev2006,Furukawa2017,Furukawa2021,SuzukiKato2023,Jagoda2023,barts2025}.
This mechanism is common to cases (1) and (2).
These results may provide a qualitative explanation for CISS-related spin polarization reported in the absence of an applied bias. We also discuss how the same mechanism may extend to other zero-mean nonequilibrium stochastic drives, including temperature-gradient fluctuations, in Sec.~\ref{subsec: relation to other sources}.

This paper is organized as follows. Section~\ref{sec: set up} introduces the model of a chiral system and the relevant fluctuations. 
Section~\ref{sec: prerequisites} discusses the boundary conditions and the collision term for solving the Boltzmann equation and outlines the calculation steps.
Section~\ref{sec: spin to ac} establishes the frequency-dependent quadratic spin response to a local ac current, reveals the Edelstein mechanism near the boundaries, and provides the basis for the current-fluctuation response discussed in Sec.~\ref{subsec: local fluctuating electric current injection}.
Section~\ref{sec: spin wo bias input} presents the calculated quadratic spin responses to a local fluctuating electric current and to interfacial fluctuating electric fields. Section~\ref{sec: discussions} discusses the potential applicability of the method to fluctuation-induced spin polarization and outlines possible directions for future work. Section~\ref{sec: summary} summarizes the results. Details of the calculations are given in the Appendices.

Throughout this paper, we set $\hbar=1$ and take $e>0$.
\section{Setup}
\label{sec: set up}
In this section, we describe the details of the cases (0), (1), and (2).
Case (1) is the minimal stochastic extension of the local-current problem studied in Ref.~\cite{Yoshimi2026} and provides a simple setting for understanding how current fluctuations generate a time-averaged spin polarization.
As a distinct form of fluctuation from that considered in case (1), case (2) treats a simple model for the response to a boundary field.
This case may model, for example, low-frequency fluctuations of photoinduced surface charge under illumination.
We model the chiral metal as being placed in a large virtual capacitor with a fluctuating charge density.
This treatment ensures the charge-neutrality condition.
We analytically calculate the quadratic spin responses in these setups by solving the Boltzmann equation together with Gauss's law, thereby explicitly accounting for electric polarization near the boundaries.

For the three setups introduced in Sec.~\ref{sec: introduction}, cases (0) and (1) are driven by the source current $j_0(t)$, whereas case (2) is driven by the boundary electric field $E_0(t)$. These functions are expressed as
\begin{equation}
    j_0(t)=\int_{-\infty}^\infty\frac{\rmd \omega}{\sqrt{2\pi}}j_{0,\omega} e^{-\iu \omega t},\,E_0(t)=\int_{-\infty}^\infty\frac{\rmd \omega}{\sqrt{2\pi}}E_{0,\omega} e^{-\iu \omega t},
\end{equation}
whose Fourier components satisfy, for cases (0), (1), and (2),
\begin{align}
    &\mathrm{(0):}\,j_{0,\omega}=\sqrt{\frac{\pi}{2}}J_0[\delta(\omega-\omega_0)+\delta(\omega+\omega_0)],\label{eq_j0Def_i}\\
    &\mathrm{(1):}\,\langle j_{0,\omega}\rangle=0,\,\langle j_{0,\omega}j_{0,\omega'}\rangle=\sqrt{2\pi} S_{j,\omega}\delta(\omega+\omega'),\label{eq_j0Def_ii}\\
    &\mathrm{(2):}\,\langle E_{0,\omega}\rangle=0,\,\langle E_{0,\omega}E_{0,\omega'}\rangle=\sqrt{2\pi} S_{E,\omega}\delta(\omega+\omega').\label{eq_EDef_iii}
\end{align}
Here $\omega_0$ is the angular frequency of the ac current in case (0), $\langle\cdots\rangle$ denotes an ensemble average, and $S_{j,\omega}$ and $S_{E,\omega}$ are the corresponding fluctuation spectral densities. Throughout this work, these spectra describe externally maintained nonequilibrium fluctuations. 

As a minimal model, we consider a metal with hedgehog-type antisymmetric spin-orbit coupling (SOC)~\cite{Onuki2014,Furukawa2017,Furukawa2021,Shiota2021}, as in our previous work~\cite{Yoshimi2026}.
The Hamiltonian is given by
\begin{equation}
 \mathcal{H}=\frac{k^2}{2m}+\alpha \bm{\sigma}\cdot\bm{k},
 \label{eq: Hamiltonian}
\end{equation}
with $k=|\bm{k}|$ and the Pauli matrices $\bm{\sigma}=(\sigma_x,\sigma_y,\sigma_z)$ in the spin space.
The sign of $\alpha$ distinguishes the two chiralities of the model.
The Hamiltonian is diagonalized as
$\mathcal{H} |\bm{k},\pm\rangle = \varepsilon_{\pm}(k) |\bm{k},\pm\rangle,$
where the dispersions of the spin-split bands are given by $\varepsilon_{\pm}(k)=\frac{k^2}{2m}\pm |\alpha| k$.
For $\alpha>0$ ($\alpha<0$), the spin in the $+$ band is parallel (antiparallel) to the wave vector.

In the following, we assume that $\kB T$ is much smaller than the chemical potential $\mu$, which implies that $\mu$ is almost equal to the Fermi energy $\varepsilon_{\mathrm{F}}$.
The Fermi wave vectors of the spin-split bands for the chemical potential $\mu$ ($>0$) are then given by $k_{\rm F,\pm}=m(\mp |\alpha|+v_{\mathrm{F}})$ with $v_{\mathrm{F}}:=\sqrt{\alpha^2 +2\mu/m}$.
We also assume $|\alpha|\ll v_{\mathrm{F}}$, so that the spin-splitting energy  $\Delta_{\rm SO}\equiv2|\alpha|k_{\rm F}$, $k_{\rm F}\equiv \sqrt{2m\mu }$ to leading order in $|\alpha|/v_{\rm F}$, is much smaller than $\mu$. For all frequencies with appreciable weight in the applied drive or fluctuation spectrum [$\omega=\omega_0$ in case (0)], we require $|\omega|,\tau_0^{-1}\ll \Delta_{\rm SO}$,  where  $\tau_0$ is the impurity scattering time  at the Fermi energy, defined in Sec. \ref{subsec: collision integral}.
This separation of the energy scales allows us to describe the system by the band-diagonal distribution functions $f_{\bm{k}\gamma}$.
Note that the collision integral retains both intraband and interband scattering.
We assume that the size $L$ of a chiral metal is much larger than the mean free path $\ell=v_{\rm F}\tau_0$.

\section{Preliminaries}
\label{sec: prerequisites}
In this section, we present the preliminary considerations for the main calculations.
First, in Sec.~\ref{subsec: Boltzmann eq. and boundary conditions}, we present the Boltzmann equation, the source term consistent with charge conservation and boundary conditions.
Second, in Sec.~\ref{subsec: collision integral}, we derive the collision integral for spin-independent nonmagnetic impurity scattering, retaining both intraband and interband processes.
Third, in Sec.~\ref{subsec: bulk responses to dc}, we apply our method to a bulk chiral system and demonstrate that the Edelstein effect is captured within our framework.
Finally, in Sec.~\ref{subsec: calculation steps}, we outline the calculation procedure for three cases.

Hereafter, we omit the variables $z$ and $t$ in some expressions for brevity.

\subsection{Boltzmann equation, source term, and boundary conditions}
\label{subsec: Boltzmann eq. and boundary conditions}
For cases (0) and (1), we consider spin responses in the chiral metal to an external current density $j_0(t)$ along the $z$ direction. We use the Boltzmann equation in the presence of the nonmagnetic impurities,
\begin{align}
     \frac{\partial f_{\bm{k}\gamma}}{\partial t} + v_{\bm{k}\gamma}^z \frac{\partial f_{\bm{k}\gamma}}{\partial z} +q E(z,t) \frac{\partial f_{\bm{k}\gamma}}{\partial k_z}= \mathrm{St}[f_{\bm{k}\gamma}] + I_{\bm{k}\gamma}(z,t).\label{eq: Boltzmann equation}
\end{align}
The collision and the source terms will be specified below so that Eq.~\eqref{eq: Boltzmann equation} is consistent with charge conservation.
Here, $f_{\bm{k}\gamma}$ denotes the distribution function for the band index $\gamma = \pm$.
The symbols $\bm{v}_{\bm{k}\gamma}$, $q\,(=\pm e)$ and $E(z,t)$ denote the velocity $\partial\varepsilon_{\gamma}/\partial\bm{k}$, charge of carrier and $z$-directed electric field, respectively.
The source term $I_{\bm{k}\gamma}(z,t)$ describes current injection at $z=0$ and extraction at $z=L$, and is given by
\begin{equation}
I_{\bm{k}\gamma}(z,t):=J_{\bm{k}\gamma}(t)[\delta(z)-\delta(z-L)].
\end{equation}
The explicit form of $J_{\bm{k}\gamma}$ depends on the interfacial details, such as the specularity of the interface between the chiral metal and the electrode, but it is normalized so that  
\begin{equation}
     \frac{1}{\Omega}\sum_{\bm{k}\gamma}q J_{\bm{k}\gamma}(t)=j_0(t).
     \label{eq: source-sum-rule}
\end{equation}
Here $\Omega$ is the volume of the chiral metal.
In our previous work~\cite{Yoshimi2026}, we confirmed that the qualitative features and magnitude of the spin and electric-field distributions considered here are insensitive to the specific form of $J_{\bm{k}\gamma}(t)$.
Therefore, we assume the simplest form of $J_{\bm{k}\gamma}(t)$:
\begin{align}
J_{\bm{k}\gamma}(t)=    
-\frac{3j_0(t)}{2q N_\gamma(\mu)} \left(\frac{v_{\bm{k}\gamma}^z}{\vF}\right)^2
        \frac{\partial f^{(0)}(\varepsilon_{\bm{k}\gamma})}{\partial \varepsilon_{\bm{k}\gamma}},
        \label{eq: Jkgamma_1}
\end{align}
in what follows.
Here $N_\gamma(\varepsilon) = \frac{1}{\Omega} \sum_{\bm{k}}\delta(\varepsilon_{\bm{k}\gamma}-\varepsilon)$ is the density of states of the band $\gamma$.

For case (2), we consider spin responses in the chiral metal to the electric field near the surface $E_0(t)$ along the $z$ direction.
Here, we apply Ziman's phenomenological description~\cite{Ziman1960} to the boundary condition at $z=0$ and $z=L$.
In that description, an incident carrier is specularly reflected with probability $p$ as if the surface is perfectly smooth and scattered isotropically with probability $1-p$ as if the surface is perfectly rough.
In our system, this description is expressed as follows:
\begin{subequations}
    \begin{align}
        f_{\bm{k}\gamma|k_z>0}(z=0)&=\frac{1}{\Omega}\sum_{\bm{k}'\gamma'|k'_z<0}f_{\bm{k}'\gamma'}(z=0)\mathcal{R}_{\bm{k}'\gamma'}^{\bm{k}\gamma},\label{eq_Ziman_z=0}\\
        f_{\bm{k}\gamma|k_z<0}(z=L)&=\frac{1}{\Omega}\sum_{\bm{k}'\gamma'|k'_z>0}f_{\bm{k}'\gamma'}(z=L)\mathcal{R}_{\bm{k}'\gamma'}^{\bm{k}\gamma},\label{eq_Ziman_z=L}
    \end{align}
\end{subequations}
where $\bm{k}:=(k_x,k_y,k_z)$, $\overline{\bm{k}}:=(k_x,k_y,-k_z)$, $\bm{\sigma}_{\bm{k}\gamma}:=\langle\bm{k},\gamma|\bm{\sigma}|\bm{k},\gamma\rangle$ and
\begin{align}
    \mathcal{R}_{\bm{k}'\gamma'}^{\bm{k}\gamma}:=&\,p\cdot(2\pi)^3\delta(\bm{k}'-\overline{\bm{k}})\frac{1+\bm{\sigma}_{\bm{k}\gamma}\cdot {\bm{\sigma}}_{\bm{k}'\gamma'}}{2}\notag\\
    &+(1-p)\cdot\frac{2\delta(\varepsilon_{\bm{k}'\gamma'} - \varepsilon_{\bm{k}\gamma})}{N(\varepsilon_{\bm{k}\gamma})}.\label{eq_Ziman_scatteringrate}
\end{align}
Here $N(\varepsilon)$ is the density of one-particle states $N(\varepsilon)=\frac{1}{\Omega}\sum_{\bm{k}\gamma}\delta(\varepsilon-\varepsilon_{\bm{k}\gamma})$.
In this paper, we consider the case where the surface is perfectly rough (i.e. $p=0$) for simplicity.
Note that, in the case (2), $I_{\bm{k}\gamma}(z,t)$ in Eq.~\eqref{eq: Boltzmann equation} is unnecessary.

For all three cases, we solve the Boltzmann equation~\eqref{eq: Boltzmann equation} perturbatively by expanding the distribution function in powers of   $j_0(t)$ or $E_0(t)$,  and treating the resulting equations order by order.
Let $f_{\bm{k} \gamma}^{(i)}(z, t)$ be the $i$th-order correction to the distribution function.
Then, we define the corresponding electric-current and spin densities as
\begin{align}
      j_e^{(i)}(z,t)&=\frac{1}{\Omega}\sum_{\bm{k}\gamma}qv_{\bm{k}\gamma}^z f^{(i)}_{\bm{k}\gamma}(z,t),
      \label{eq: je-def}\\
      S^{(i)}_z(z,t) &= \frac{1}{\Omega}\sum_{\bm{k}\gamma}\frac{\sigma_{z,\bm{k}\gamma}}{2} f^{(i)}_{\bm{k}\gamma}(z,t).
      \label{eq: s-def}
 \end{align}
We can also expand the electric field in the chiral metal $E(z,t)$ as $E(z,t)=E^{(1)}(z,t)+E^{(2)}(z,t)+\cdots$, where $E^{(i)}(z,t)$ denotes the $i$th order ($i=1,2$) in $j_0(t)$ for cases (0) and (1), and in $E_0(t)$ for case (2).
For case (2), the boundary conditions are $E^{(1)}(z=0,t)=E^{(1)}(z=L,t)=E_0(t)$, and $E^{(2)}(z=0,t)=E^{(2)}(z=L,t)=0$.

\subsection{Collision integral for nonmagnetic impurity scattering}
\label{subsec: collision integral}
We model the impurities as randomly distributed $\delta$-function potentials of strength  $v_0$, with impurity density $n_{\rm imp}$, in a three-dimensional system of volume $\Omega$.
The collision integral is then obtained from Fermi's golden rule as
\begin{align}
     {\rm St}[f_{\bm{k}\gamma}]
    =\frac{2\pi v_0^2n_{\text{imp}}}{\Omega}\sum_{\bm{k}', \gamma'}
    |&\braket{\bm{k}', \gamma'|\bm{k}, \gamma}|^2[f_{\bm{k}'\gamma'} - f_{\bm{k}\gamma}]\notag\\
    &\qquad\times \delta(\varepsilon_{\bm{k}'\gamma'} - \varepsilon_{\bm{k}\gamma}).
    \label{eq: Boltzmann col int}
\end{align}
The factor $|\braket{\bm{k}', \gamma'|\bm{k}, \gamma}|^2$, represented as
\begin{equation}
    |\braket{\bm{k}' ,\gamma'|\bm{k}, \gamma}|^2
    = \frac{1+\bm{\sigma}_{\bm{k}\gamma}\cdot {\bm{\sigma}}_{\bm{k}'\gamma'}}{2},
    \label{eq: matrix-element}
\end{equation}
is the spinor-overlap probability between the initial and final helicity eigenstates for scattering by a spin-independent impurity.
This expression is a basic form of the spinor-overlap probability and has been widely used to describe scattering in various systems; for example, it has been applied to two-dimensional graphene~\cite{Hwang2007} and topological insulator~\cite{Culcer2010}.
The collision integral~\eqref{eq: Boltzmann col int} also indicates a typical impurity scattering rate, i.e., an inverse of quasiparticle lifetime 
\begin{equation}
 \frac{1}{\tau(\varepsilon)}\equiv \pi v_0^2n_{\text{imp}}\cdot N(\varepsilon).\label{eq: taup definition}
\end{equation}
Then, we obtain the following expression:
\begin{align}
    \mathrm{St}[f_{\bm{k}\gamma}]=-\frac{f_{\bm{k}\gamma}-f_{\bm{k}\gamma}^{(0)}}{\tau(\varepsilon_{\bm{k}\gamma})}+\frac{n(\varepsilon_{\bm{k}\gamma})}{N_0\tau_0}+\frac{2 \bm{\sigma}_{\bm{k}\gamma}\cdot\bm{S}(\varepsilon_{\bm{k}\gamma})}{N_0\tau_0},\label{eq: St[f]}
\end{align}
in terms of the Fermi distribution function $f^{(0)}(\varepsilon)$ with $f_{\bm{k}\gamma}^{(0)}=f^{(0)}(\varepsilon_{\bm{k}\gamma})$, $\tau_0\equiv\tau(\mu)$, $N_0\equiv  N(\mu)$. The excess carrier spectral density and the nonequilibrium spin spectral density stemming from the one-particle states with energy $\varepsilon$ are given by
\begin{align}
     n(\varepsilon,z,t)&=\frac{1}{\Omega}\sum_{\bm{k}\gamma}[f_{\bm{k}\gamma}(z,t)-f^{(0)}(\varepsilon)]\delta (\varepsilon_{\bm{k}\gamma}-\varepsilon),
     \label{eq: n-epsilon}\\
     \bm{S}(\varepsilon,z,t)&=\frac{1}{\Omega}\sum_{\bm{k}\gamma}\frac{\bm{\sigma}_{\bm{k}\gamma}}{2}[f_{\bm{k}\gamma}(z,t)-f^{(0)}(\varepsilon)]\delta (\varepsilon_{\bm{k}\gamma}-\varepsilon).\label{eq: neq-spin}  
\end{align}
We obtain Eq.~\eqref{eq: St[f]} by evaluating the Fermi-golden-rule collision integral \eqref{eq: Boltzmann col int} for the present model, without replacing it by the conventional relaxation-time approximation.
Compared with the treatment in Ref.~\cite{Yoshimi2026}, the present collision integral additionally retains the third term on the right-hand side of Eq.~\eqref{eq: St[f]}.
The first term on the RHS of Eq.~\eqref{eq: St[f]} has the form of the relaxation-time approximation. 
The second term on the RHS of Eq.~\eqref{eq: St[f]} is essential for satisfying the charge conservation law in the interface problems, even though it is typically excluded in the conventional relaxation-time approximation.
The third term on the RHS of Eq.~\eqref{eq: St[f]} originates from the spinor-overlap factor in Eq.~\eqref{eq: matrix-element}, reflecting spin-momentum locking.
This collision integral~\eqref{eq: St[f]} includes both intraband and interband scattering.

Note that, since the electric current and electric field are parallel to the $z$-axis in our setup, rotational symmetry of the Hamiltonian~\eqref{eq: Hamiltonian} about the $z$-axis leads to $S_x=S_y=0$. 
Thus, it is sufficient to consider only $2\sigma_{\bm{k}\gamma}^zS_z(\varepsilon_{\bm{k}\gamma})/N_0\tau_0$.

At low temperatures, the nonequilibrium corrections are confined to a narrow energy window around the Fermi energy. We therefore use 
\begin{equation}
 n(\varepsilon,z,t)\simeq \tilde{n}(z,t)\delta(\varepsilon-\mu),\quad S_z(\varepsilon,z,t)\simeq \tilde{S}_z(z,t)\delta(\varepsilon-\mu)\label{eq_approximation_n&Sz} 
\end{equation}
where 
\begin{equation}
 \tilde{n}(z,t)\equiv \int\rmd \varepsilon n(\varepsilon,z,t)=\frac{1}{\Omega}\sum_{\bm{k}\gamma}[f_{\bm{k}\gamma}-f_{\bm{k}\gamma}^{(0)}]   
\end{equation}
and similarly for $\tilde{S}_z(z,t)$.
It should be noted that $\tilde{n}(z,t)$ has the dimension of a number density, while $\tilde{S}_z(z,t)$ has the dimension of a spin density.
Within the same approximation, $ \tau(\varepsilon)$ can be replaced by $\tau_0=\tau(\mu)$. Applying these approximations consistently preserves
\begin{equation}
\frac{1}{\Omega}\sum_{\bm{k}\gamma}\mathrm{St}[f_{\bm{k}\gamma}]=0,   
\end{equation}
which ensures charge conservation.

\subsection{Bulk responses to dc current}
\label{subsec: bulk responses to dc}
We first discuss the linear response to a uniform dc electric current density $j_0$ without boundaries or the source term $I_{\bm{k}\gamma}(z,t)$, as a prerequisite to addressing the effect of boundaries or local input discussed below.
The Edelstein coefficient obtained here is used to discuss the results in the following sections.
To leading order in $|\alpha|/v_{\rm F}$, the uniform electric field is given by a constant $\mathcal{E}_0 := j_0 / \sigma_0$, where  \begin{equation}\sigma_0= \frac{q^2\tau_0}{\Omega}\sum_{\bm{k},\gamma}(v^{z}_{\bm{k},\gamma})^2 \delta(\varepsilon_{\bm{k},\gamma}-\mu)\end{equation}  is the corresponding conductivity. 
In the spatially uniform case, the excess charge $q n(\varepsilon)$ is zero everywhere.
Multiplying Eq.~\eqref{eq: Boltzmann equation} by $qv_{\bm{k}\gamma}^z\delta(\varepsilon-\varepsilon_{\bm{k}\gamma})$ and $\frac{\sigma_{\bm{k}\gamma}^z}{2}\delta(\varepsilon-\varepsilon_{\bm{k}\gamma})$, and then taking the sum $\frac{1}{\Omega}\sum_{\bm{k}\gamma}$, we obtain the expressions for $j_e^{(1)}$ and $S_{z}^{(1)}$ respectively as follows:
\begin{align}
    \tilde{j}_e^{(1)}&=\sigma_0\mathcal{E}_0-\frac{\alpha}{v_F}\frac{4qv_F}{3}\tilde{S}_z^{(1)},\label{eq_je1Bulk}\\
    \tilde{S}_z^{(1)}&=-\frac{\alpha}{v_F}\frac{3\sigma_0}{2qv_F}\mathcal{E}_0,\label{eq_sz1Bulk}
\end{align}
where we have used the low-temperature approximation~\eqref{eq_approximation_n&Sz}, in which the response is restricted to the Fermi surface.
Here, $\tilde{j}_e$ is defined in the same way as \eqref{eq_approximation_n&Sz}.
The second term on the RHS of Eq.~\eqref{eq_je1Bulk} originates from the third term on the RHS of Eq.~\eqref{eq: St[f]}, which is not captured by the relaxation-time approximation.
Note that this term is much smaller than the first term due to $|\alpha|\ll v_F$.
In addition, Eq.~\eqref{eq_sz1Bulk} corresponds to the Edelstein effect.

\begin{figure}[t]
  \centering
  \includegraphics[width=\columnwidth]{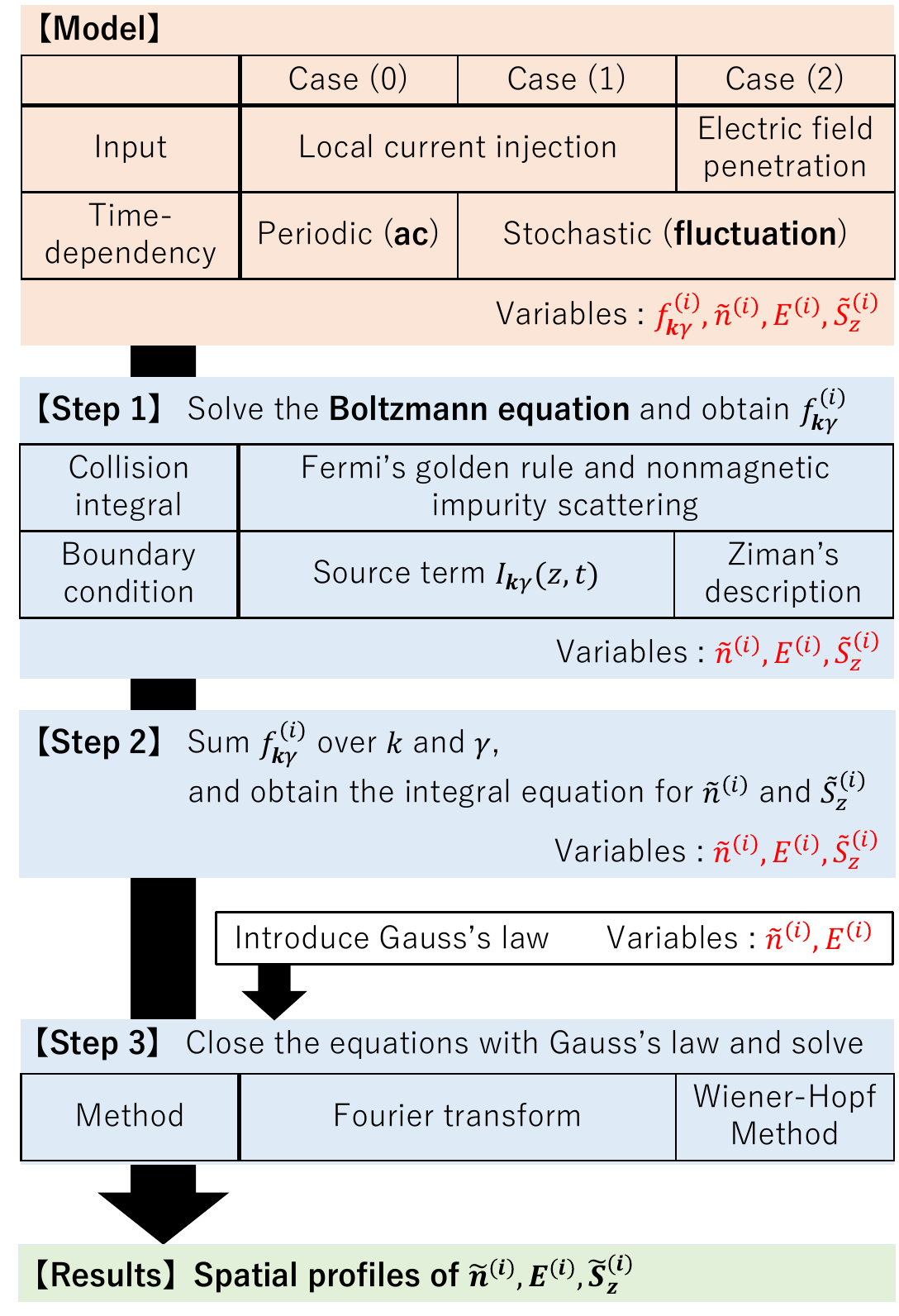}
  \caption{
 Calculation procedure for the problem with boundaries.
First, the Boltzmann equation is solved with $\tilde{n}^{(i)}, E^{(i)}$, and $\tilde{S}_z^{(i)}$ left unspecified, and then summation of $f_{\bm{k}\gamma}^{(i)}$ over $\bm{k}$ and $\gamma$ yield coupled integral equations for $\tilde{n}^{(i)}$, $E^{(i)}$ and $\tilde{S}_z^{(i)}$.
Together with Gauss's law, these equations form a closed set, which is solved by Fourier transformation for cases (0) and (1) and by the Wiener--Hopf method for case (2). }
  \label{fig_procedure}
\end{figure}
\subsection{Calculation steps for systems with boundaries}
\label{subsec: calculation steps}
For systems with boundaries, the Boltzmann equation must be
solved self-consistently with integral equations and Gauss's
law, as summarized in Fig.~\ref{fig_procedure}. We outline the three-step procedure below using simplified expressions; the explicit
derivations are given in the Appendices.
Steps 1--3 correspond to the labels in Fig.~\ref{fig_procedure}.
Because we focus on the time-averaged quadratic response, we consider the zero-frequency ($\omega=0$) component of the quadratic quantities.

\textbf{[Step 1]}
As assumed in Sec.~\ref{sec: set up}, the current path length $L$ is much longer than the mean free path $\ell=v_\mathrm{F}\tau_0$, so that the regions sufficiently far from the boundaries can be regarded as bulk, and hence $f_{\bm{k}\gamma}$ becomes constant there.
Then, the solution to the Boltzmann equation~\eqref{eq: Boltzmann equation} is expressed in a simplified form as 
\begin{align}
    f_{\bm{k}\gamma}^{(i)}(z)&=\int_{z_\mathrm{min}}^{z}\mathrm{d}z'\exp\left(-\frac{z-z'}{v_{\bm{k}\gamma}^z\tau_0}\right)\notag\\
    &\times\left[-qE^{(i)}(z')+\frac{\tilde{n}^{(i)}(z')}{v_{\bm{k}\gamma}^z\tau_0 N_0}+\frac{2 \sigma_{\bm{k}\gamma}^z\tilde{S}^{(i)}_z(z')}{v_{\bm{k}\gamma}^z\tau_0N_0}+\cdots\right].
    \label{eq: f_sol}
\end{align}
For the cases (0) and (1), $z_\mathrm{min}:=-\mathrm{sgn}(v_{\bm{k}\gamma}^z)\cdot\infty$, whereas for the case (2), $z_\mathrm{min}:=0$ when $v_{\bm{k}\gamma}^z>0$ and $z_\mathrm{min}:=\infty$ when $v_{\bm{k}\gamma}^z<0$. 
The ellipsis ($\cdots$) in Eq.~\eqref{eq: f_sol} represents inhomogeneous terms that do not involve the functions $E^{(i)}(z)$, $\tilde{n}^{(i)}(z)$, and $\tilde{S}^{(i)}_z(z)$.
\begin{figure*}
    \includegraphics[pagebox=artbox,width=0.8\textwidth]{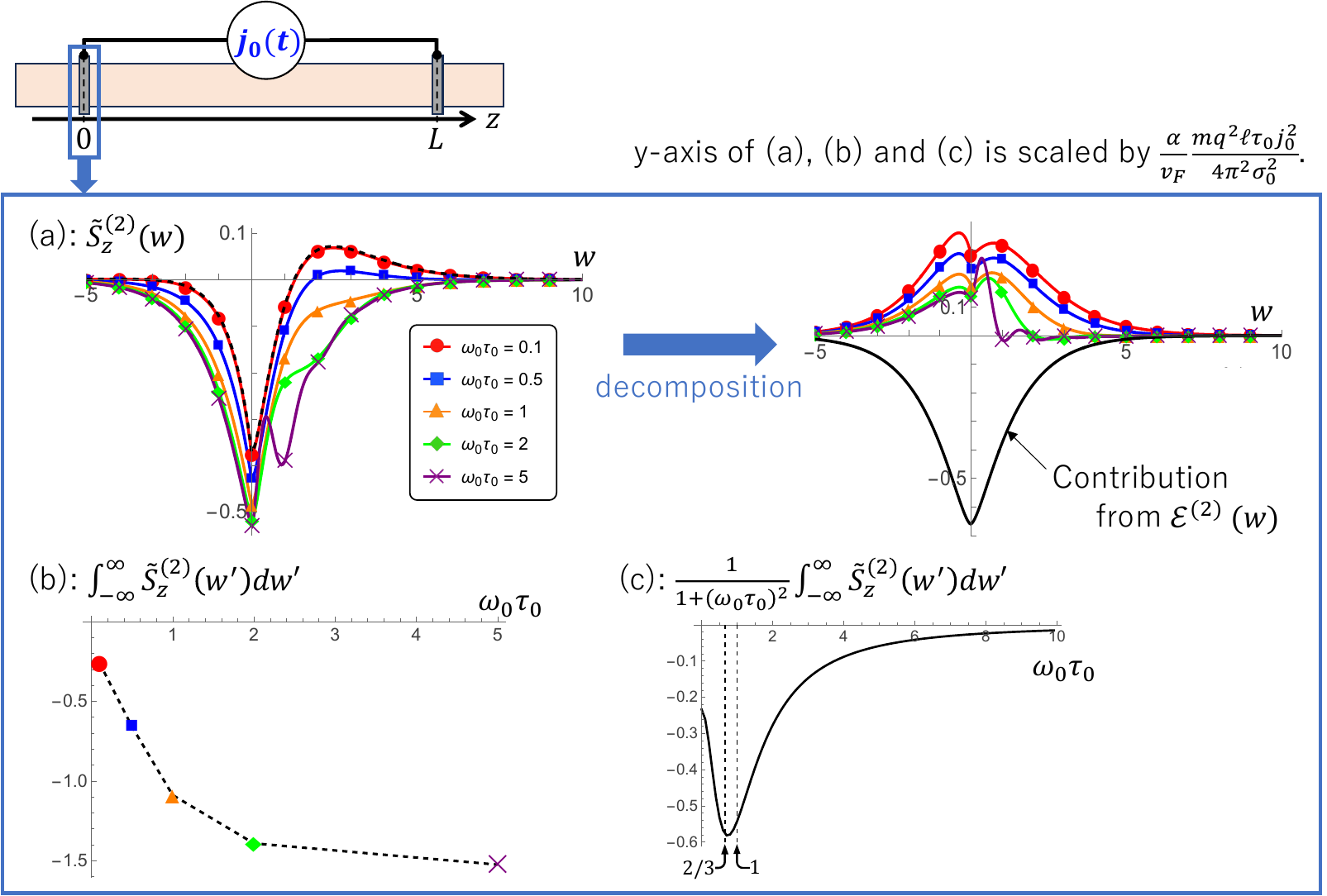}
    \caption{
    Quadratic spin response near the interface $z=0$ and its frequency dependence.
    Panel (a) shows the quadratic spin $\tilde{S}_z^{(2)}(w)$ in case (0) for each $\omega_0\tau_0$.
    The markers on the curves are for visibility only; the actual calculations were performed using denser meshes of $w$.
    The black dashed line corresponds to half of the quadratic spin response due to dc current, $\mathcal{S}_2(w)/2$ defined in Eq.~\eqref{eq_sz2_analytic}.
    The other half is attributed to $2\omega_0$-oscillating term.
    Panel (a) also shows a decomposition into the contribution driven by the effective electric field $\mathcal{E}^{(2)}(w)$ and the remaining contributions.
    The black solid line denotes the contribution from $\mathcal{E}^{(2)}(w)$.
    Panel (b) shows the quadratic spin accumulation $\int_{-\infty}^{\infty}\mathrm{d}w'\tilde{S}_z^{(2)}(w')$.
    The shapes and colors of the points correspond to the legend in (a).
    It converges to a finite value as $\omega_0\tau_0$ increases. Panel (c) shows
    the results obtained by multiplying the results of (b) by $1/[1+(\omega_0\tau_0)^2]$.
    The nonmonotonic variation indicates the presence of the spin relaxation time $3\tau_0/2$, and the vertical dashed lines at $\omega_0\tau_0
=2/3$ and 1 indicate the inverse spin-relaxation and momentum-relaxation time scales, respectively.}
    \label{fig_casei}
\end{figure*}

\textbf{[Step 2]}
Multiplying Eq.~\eqref{eq: f_sol} by $\delta(\varepsilon-\varepsilon_{\bm{k}\gamma})$ and $\frac{\sigma_{\bm{k}\gamma}^z}{2}\delta(\varepsilon-\varepsilon_{\bm{k}\gamma})$, and then taking the sum $\frac{1}{\Omega}\sum_{\bm{k}\gamma}$, we obtain the integral equations for $\tilde{n}^{(i)}(z)$ and $\tilde{S}_{z}^{(i)}(z)$ respectively as follows:
\begin{align}
    &\tilde{n}^{(i)}(z)=\notag\\
    &\int_{z_\mathrm{min}'}^{\infty}\mathrm{d}z'[K_1(z-z')E^{(i)}(z')\notag\\
    &\qquad\quad+K_2(z-z')\tilde{n}^{(i)}(z')+K_3(z-z')\tilde{S}_{z}^{(i)}(z')+\cdots],\label{eq_ni_simplified}
\end{align}
and
\begin{align}
    &\tilde{S}^{(i)}(z)=\notag\\
    &\int_{z'_\mathrm{min}}^{\infty}\mathrm{d}z'[K'_1(z-z')E^{(i)}(z')\notag\\
    &\qquad\quad+K'_2(z-z')\tilde{n}^{(i)}(z')+K'_3(z-z')\tilde{S}_{z}^{(i)}(z')+\cdots].\label{eq_Si_simplified}
\end{align}
The functions $K_{1,2,3}(z),K'_{1,2,3}(z)$ represent integral kernels, with their explicit forms presented in the Appendices~\ref{app: linear spin to time-dependent current}, \ref{app: quadratic spin to time-dependent current}, and \ref{app: quadratic spin to fluctuating electric field}.
The lower limit of integration is given by $z_\mathrm{min}'=-\infty$ for the cases (0) and (1), and $z_\mathrm{min}'=0$ for the case (2).

\textbf{[Step 3]}
The integral equations~\eqref{eq_ni_simplified} and \eqref{eq_Si_simplified} and Gauss's law $\partial E/\partial z=\rho/\epsilon_0=q \int \mathrm{d} \varepsilon n(\varepsilon)/\epsilon_0$ with the electric constant (vacuum permittivity) $\epsilon_0$ form a closed set of equations to determine $E^{(i)}(z)$, $\tilde{n}^{(i)}(z)$, and $\tilde{S}^{(i)}_z(z)$.
Depending on the integration domain $[z_\mathrm{min}',\infty]$, the integral equations are solved using the Fourier transform in cases (0) and (1), and the Wiener-Hopf method~\cite{Milne1921,WienerHopf1931} in case (2).
The Wiener-Hopf method factorizes the kernel into functions analytic in the upper and lower half-planes. The resulting equation is separated into corresponding analytic parts, whose equality on the real axis allows their analytic continuation.
Note that, in addition to the mean free path $\ell$, another characteristic length scale, the Thomas--Fermi screening length $\lambda_{\rm TF}:=\sqrt{\epsilon_0/(q^2 N_0)}$, naturally emerges from Gauss's law.
This screening length is much smaller than the mean free path in typical metals.

\section{Quadratic spin response to a local ac current}
\label{sec: spin to ac}
Before addressing stochastic electrical fluctuations, we first establish the quadratic spin response to a local ac current as case (0). 
This case serves two purposes: demonstrating that a zero-mean periodic drive generates a finite time-averaged interfacial spin polarization whose sign is controlled by chirality, and providing the monochromatic response underlying the fluctuation-induced results in Sec.~\ref{subsec: local fluctuating electric current injection}.
We also clarify that the dominant contribution to the spin polarization originates from the quadratic electric field induced near the boundaries, through the linear Edelstein effect.
This feature is common to cases (1) and (2).

In the limit $\ell\ll L$, the two boundaries can be treated independently, and we therefore focus on the response near $z=0$. Imposing the boundary conditions 
in the bulk limit, we solve the coupled equations to second order in the current amplitude. Introducing the dimensionless coordinate $w=z/\ell$, we obtain
\begin{align}
    \tilde{S}_z^{(2)}(w,t)=\frac{\alpha}{v_F}\frac{mq^2\ell\tau_0 j_0^2}{4\pi^2\sigma_0^2}\mathcal{S}_{\mathrm{d}}(w)+[2\omega_0\textrm{-oscillating term}],\label{eq_sz2_casei}
\end{align}
with a dimensionless function $\mathcal{S}_{\mathrm{d}}(w)$, which depends parametrically on $\omega_0\tau_0$. The detailed derivation is given in Appendices~\ref{app: linear spin to time-dependent current} and \ref{app: quadratic spin to time-dependent current}.  Equation~\eqref{eq_sz2_casei} shows that the zero-frequency spin polarization is quadratic in the current amplitude $j_0$ and odd in the chirality parameter $\alpha$: reversing the chirality reverses the spin polarization, whereas reversing the current direction leaves it unchanged.
The remaining quadratic contribution oscillates at  $2\omega_0$ and vanishes upon time averaging.

Figure~\ref{fig_casei}(a) shows $\mathcal{S}_{\mathrm{d}}(w)$ for $\omega_0\tau_0=0.1-5$.
As $\omega_0\tau_0$ decreases, the quadratic spin $\tilde{S}_z^{(2)}(w)$ in Fig.~\ref{fig_casei}(a) approaches the black dashed line, which corresponds to half of the quadratic spin response due to dc current, $\mathcal{S}_2(w)/2$ defined in Eq.~\eqref{eq_sz2_analytic}.
The other half appears in the $2\omega_0$ harmonic. This agreement provides a direct consistency check on the calculation.

In the following, we first discuss the contribution of the linear Edelstein effect to the spin accumulation in Sec.~\ref{subsec: Edelstein effect from mathcalE} and then examine its frequency dependence in Sec.~\ref{subsec: frequency dependence of spin accumulation}.

\subsection{Spin polarization induced by a quadratic effective electric field via the linear Edelstein effect}
\label{subsec: Edelstein effect from mathcalE}
To understand the results of $S^{(2)}_z (w)$, we note that two electric fields are involved: the Maxwell electric field $E^{(2)}$ and the effective electric field
\begin{equation}
    \mathcal{E}^{(2)}=E^{(2)}-\frac{\partial_z\tilde{n}^{(2)}}{qN_0}.
\end{equation}
The latter includes the contribution from the chemical-potential gradient.
On length scales longer than the Thomas-Fermi screening length, $\mathcal{E}^{(2)}$ and $E^{(2)}$ have the same leading spatial dependence.
As shown by the decomposition in Fig.~\ref{fig_casei}(a), the linear Edelstein response driven by $\mathcal{E}^{(2)}$ dominates the quadratic spin polarization. Since the Edelstein coefficient is proportional to $\alpha$, this mechanism accounts for its chirality-controlled sign, as in the dc local-current problem studied in Ref.~\cite{Yoshimi2026}.

A simple scaling estimate further reproduces the magnitude and parameter dependence of Eq.~\eqref{eq_sz2_casei}.
As a preliminary, we introduce the force-balance relation by referring to our previous work~\cite{Yoshimi2026}.
Hereafter, quantities carrying a subscript $\omega$, such as $A_\omega$, represent the Fourier transform of $A(t)$.
Multiplying the Boltzmann equation by $mv_{\bm{k}\gamma}^z$ and summing over $\bm{k},\gamma$ and integrating by parts, 
we obtain the Fourier-transformed force balance relation 
\begin{align}
    -\frac{j_{m,\omega}^{(2)}(z)}{\tau_{\omega}}+\int_{-\infty}^{\infty}&\frac{\mathrm{d}\omega'}{\sqrt{2\pi}}\rho_{\omega-\omega'}^{(1)}(z)E_{\omega'}^{(1)}(z)\notag\\
    &+\rho^{(0)} E_{\omega}^{(2)}(z)-\frac{\partial\Pi_{zz,\omega}}{\partial z}\sim0.
    \label{eq: j_e^2(z)}
\end{align}
It is useful to refer to Eq.~\eqref{eq_quadraticBE}.
Here $j_{m,\omega}^{(2)}(=m j_{e,\omega}^{(2)}/q)$ denotes the mass current density and 
\begin{equation}
    \Pi_{zz}(z,t) := \frac{1}{\Omega}\sum_{\bm{k}\gamma} m \left(v_{\bm{k}\gamma}^z\right)^2f_{\bm{k}\gamma}^{(2)}(z,t)
    \label{eq: Pizz}
\end{equation}
denotes the second-order momentum flux tensor.
In the bulk region ($z\rightarrow\infty$), this tensor can be expressed as
\begin{align}
    \left.\Pi_{zz,\omega}\right|_{\mathrm{bulk}}\sim2\tau_\omega\int_{-\infty}^{\infty}\frac{\mathrm{d}\omega'}{\sqrt{2\pi}}\mathcal{E}_{0,\omega-\omega'}j_{0,\omega'}
    \label{eq_Pizz_bulk}
\end{align}
using the relation 
\begin{align}
    \left.f_{\bm{k}\gamma,\omega}^{(2)}\right|_{\mathrm{bulk}}=-\tau_\omega\int_{-\infty}^{\infty}\frac{\mathrm{d}\omega'}{\sqrt{2\pi}}q\mathcal{E}_{0,\omega-\omega'}\frac{\partial \left.f_{\bm{k}\gamma,\omega'}^{(1)}\right|_{\mathrm{bulk}}}{\partial k_z}
\end{align}
and $\frac{1}{\Omega}\sum_{\bm{k}\gamma} (v_{\bm{k}\gamma}^z)^2\partial_{k_z}f_{\bm{k}\gamma}^{(1)} \sim-\frac{1}{\Omega}\sum_{\bm{k}\gamma}\partial_{k_z}(v_{\bm{k}\gamma}^z)^2\cdot f_{\bm{k}\gamma}^{(1)}$ follows from integration by parts.
In \eqref{eq: j_e^2(z)}, we also introduce the notations $\tau_\omega:=\tau_0/(1-i\omega\tau_0)$ and $\rho_{\omega}^{(0,1)}=\frac{q}{\Omega}\sum_{\bm{k}\gamma}f^{(0,1)}_{\bm{k}\gamma,\omega}$ for the Fourier-transformed distribution function $f^{(i)}_{\bm{k}\gamma,\omega}$.
This equation represents the balance among the viscous force $-j_{m,\omega}^{(2)}/\tau_\omega$, the Lorentz forces $\int\frac{\mathrm{d}\omega'}{\sqrt{2\pi}}\rho_{\omega-\omega'}^{(1)}(z)E_{\omega'}^{(1)}(z)+\rho^{(0)} E_\omega^{(2)}(z)$, and the hydrodynamic pressure $-\partial\Pi_{zz,\omega}/\partial z$.
For $z\sim O(\ell)$, since the externally applied current density contains no quadratic component $j_{e,\omega}^{(2)}$ and the charge density $\rho_\omega^{(1)}$ decays rapidly to zero within the screening length, Eq.~\eqref{eq: j_e^2(z)} becomes 
\begin{align}
    \frac{\partial\Pi_{zz,\omega}(z)}{\partial z}\sim\rho^{(0)}E_\omega^{(2)}(z).\label{eq_ForceBalanceEq}
\end{align}
Since $\Pi_{zz,\omega}(z)$ is constant with respect to $z$ in the bulk region, Eq.~\eqref{eq_ForceBalanceEq} shows that $E_\omega^{(2)}(z)$, corresponding to its derivative, is localized near the interfaces.

In the following, we discuss the dc component ($\omega=0$) of the electric fields.
Using Eq.~\eqref{eq_Pizz_bulk} and the force-balance relation~\eqref{eq_ForceBalanceEq}, the magnitude of the quadratic electric field $E^{(2)}(z,t)$ can be estimated as
\begin{align}
    E^{(2)}(z,t)\sim\frac{1}{\rho^{(0)}}\partial_z\Pi_{zz}\sim\frac{1}{qk_{\mathrm{F}}^3}\frac{\tau_0\mathcal{E}_0 J_0}{\ell}\sim\frac{q\tau_0}{mv_{\mathrm{F}}}\frac{J_0^2}{\sigma_0^2}.\label{eq_E2_Pizz}
\end{align}
Since $\lambda_{\rm TF}\ll \ell$, this also provides the correct scaling estimate for $\mathcal{E}^{(2)}$ on the length scale relevant to the spin response.
Combining this estimate with the Edelstein coefficient $-(\alpha/v_F)(3\sigma_0/2qv_F)$ given in Eq.~\eqref{eq_sz1Bulk}, we obtain the following estimate for the quadratic spin polarization:
\begin{align}
    \frac{q\tau_0}{mv_{\mathrm{F}}}\frac{J_0^2}{\sigma_0^2}\cdot\frac{-\alpha}{v_\mathrm{F}}\frac{3\sigma_0}{2qv_\mathrm{F}}\sim-\frac{\alpha}{v_\mathrm{F}}mq^2\ell\tau_0\frac{J_0^2}{\sigma_0^2}.
\end{align}
This is indeed consistent with the coefficient in Eq.~\eqref{eq_sz2_casei}.

The basic understanding of the frequency-dependent response established in this section will be used in Sec.~\ref{subsec: local fluctuating electric current injection} to construct the quadratic spin polarization induced by current fluctuation.

\begin{figure*}
 \includegraphics[pagebox=artbox,width=0.9\textwidth]{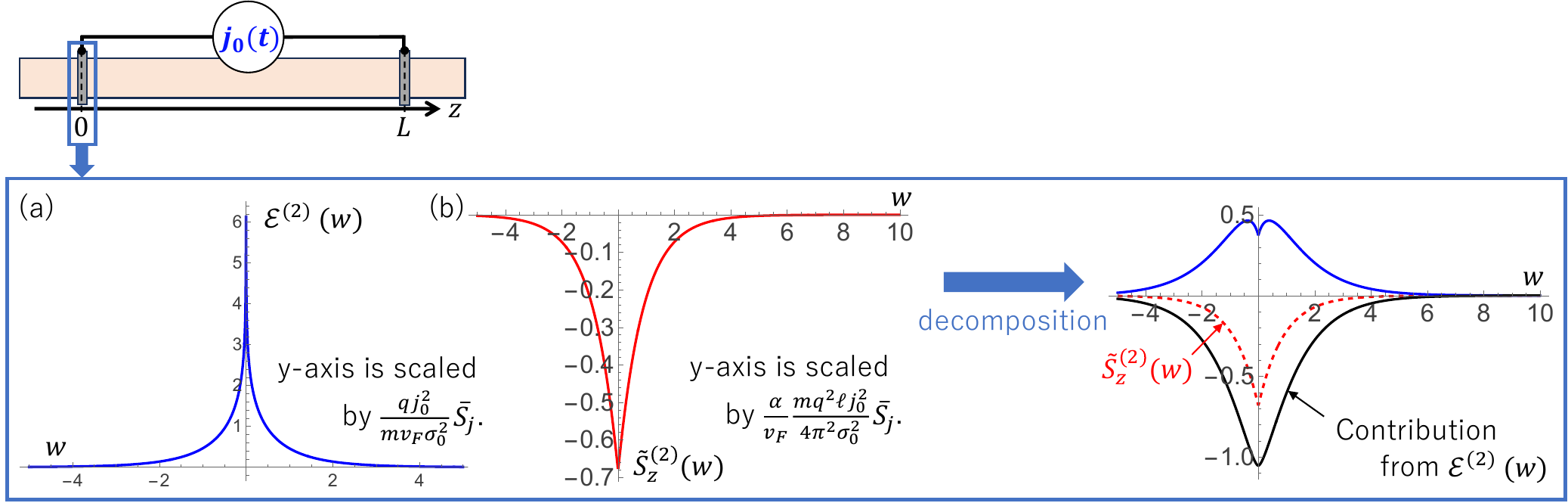}
 \caption{
Enlarged view of the quadratic response to current fluctuations near the interface $z=0$ for $\alpha>0$.
Panels (a) and (b) show the quadratic effective electric field $\mathcal{E}^{(2)}(w)$ and the quadratic spin density $\tilde{S}_z^{(2)}(w)$,
respectively, in case (1) for $\omega_c\tau_0=1$.
In panel (b), the red curve shows the total spin density, while the black and blue curves show the contribution driven by $\mathcal{E}^{(2)}(w)$ and the remaining contributions, respectively. 
}
 \label{fig_caseii}
\end{figure*}

\subsection{Frequency dependence of spin accumulation}
\label{subsec: frequency dependence of spin accumulation}
Figure~\ref{fig_casei}(b) shows the dependence of the quadratic spin accumulation $\int_{-\infty}^{\infty}\mathrm{d}w'\tilde{S}_z^{(2)}(w')$, integrated around the interface at $z=0$, on $\omega_0\tau_0$.
A finite net spin accumulation is generated near the interface over the entire frequency range of the applied ac current.
Moreover, the spin accumulation saturates to a finite value in the high-frequency regime.

At first glance, this saturation may appear counterintuitive, since one might expect the spin accumulation to be unable to follow rapid oscillations of the driving current.
Its origin lies in the frequency independence of the power input.
For a fixed ac current amplitude, the time-averaged power remains constant irrespective of the driving frequency.
Consequently, the dc component of $\Pi_{zz}(z,t)\sim\tau_0E_0j_0$ in the bulk region [Eq.~\eqref{eq_Pizz_bulk}] is independent of $\omega_0\tau_0$.
The spatial distribution of $\Pi_{zz}$ is therefore also unchanged, and so is the dc component of $E^{(2)}(z,t)$, which is determined by the gradient of $\Pi_{zz}$ [Eq.~\eqref{eq_ForceBalanceEq}]. 
As a result, the spin accumulation induced by $E^{(2)}(z,t)$ remains finite even in the high-frequency limit.

For comparison, we next consider the case of a fixed \textit{voltage} amplitude.
Figure \ref{fig_casei}(c) shows the $\omega_0\tau_0$-dependence of the spin accumulation under a fixed applied voltage.
In this case, the time-averaged power input decreases with increasing driving frequency.
For the Drude conductivity $\sigma(\omega)=\sigma_0/(1-i\omega_0\tau_0)$, we have
\begin{equation}
 \left|\frac{\sigma(\omega_0)}{\sigma_0}\right|^2=\frac{1}{1+(\omega_0\tau_0)^2}.
\end{equation}
Since the quadratic spin response is proportional to the square of the current amplitude, the fixed-voltage result is therefore obtained by multiplying the fixed-current result in Fig.~\ref{fig_casei}(b) by this factor.
We can see that the quadratic spin accumulation converges to zero as $\omega_0\tau_0$ increases.
The peak position of this nonmonotonic variation is controlled by the interplay of two characteristic time scales,
namely the momentum relaxation time $\tau_0$ and spin relaxation time $3\tau_0/2$, the latter of which can be obtained by calculating $\sum_{\bm{k}\gamma}\sigma_{\bm{k}\gamma}^z{\rm St}[f_{\bm{k}\gamma}]/2$.
The saturation timescale in Fig.~\ref{fig_casei}(b) is of the order of the spin relaxation time [see Eq.~\eqref{eq_Xi_casei} in Appendix], and multiplying $1/[1+(\omega_0\tau_0)^2]$ effectively introduces a characteristic timescale of order the momentum relaxation time.
Note that the spin relaxation time does not appear in the results within the relaxation-time approximation where the third term in Eq.~\eqref{eq: St[f]} is neglected.
The nonmonotonic frequency dependence therefore reflects the spin-relaxation time $3\tau_0/2$, which is absent in the conventional relaxation-time approximation. Retaining the full collision integral is thus essential for capturing both the spin-relaxation scale and the resulting frequency dependence of the quadratic response.

\section{Quadratic spin responses to stochastic electrical fluctuations}
\label{sec: spin wo bias input}
Section~\ref{subsec: local fluctuating electric current injection} treats locally injected current fluctuation [case (1)], whereas Sec.~\ref{subsec: interfacial fluctuating electric field penetration} treats electric-field fluctuation near the surfaces [case (2)]. 
For the stationary fluctuation defined in Eqs.~\eqref{eq_j0Def_ii} and \eqref{eq_EDef_iii}, we first consider the quadratic response to a pair of frequency components $(\omega,-\omega)$.
We then obtain the total response by integrating these contributions over $\omega$, weighted by the fluctuation spectral density.
Throughout this section, all quadratic quantities denote the zero-frequency components averaged over the fluctuation.

As shown below, in both fluctuation setups, the quadratic carrier density and electric-field responses remain unchanged under chirality reversal, $\alpha\rightarrow -\alpha$, whereas the quadratic spin polarization reverses sign.
Detailed derivations are given in Appendices~\ref{app: quadratic spin to time-dependent current} and \ref{app: quadratic spin to fluctuating electric field}. 
\begin{figure*}
 \includegraphics[pagebox=artbox,width=0.8\textwidth]{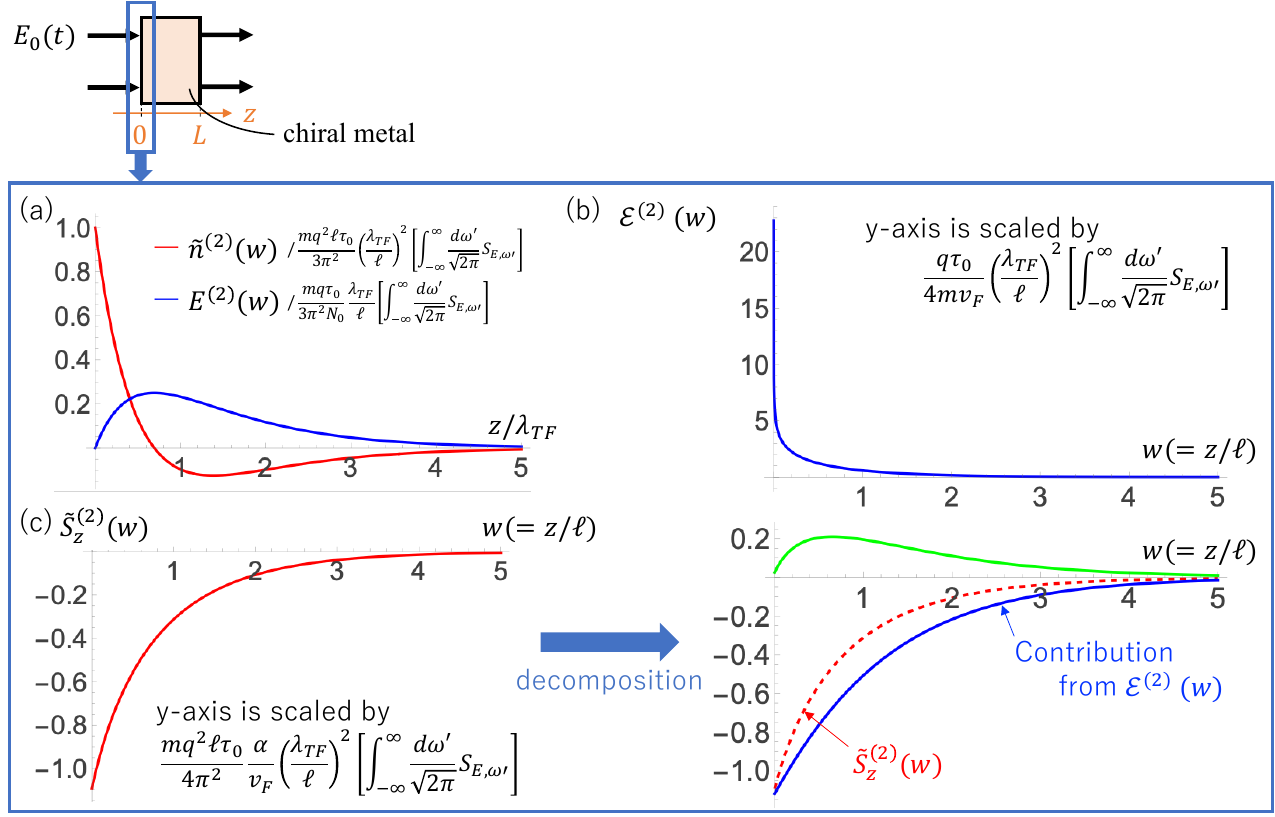}
 \caption{
 Enlarged view near the boundary $z=0$ for case (2). Panel (a) shows the quadratic excess carrier density $\tilde{n}^{(2)}$
 (red) and the Maxwell electric field $E^{(2)}$ (blue) as functions of $z/\lambda_{\rm TF}$
. Panels (b) and (c) show the effective electric field $\mathcal{E}^{(2)}(w)$ and the quadratic spin density 
$\tilde{S}^{(2)}_z(w)$, respectively, as functions of 
$w=z/\ell$. The decomposition shown in panel (c) displays the total spin density (red), the contribution driven by 
$\mathcal{E}^{(2)}(w)$ (blue), and the remaining contributions(green).
 }
 \label{fig_caseiii}
\end{figure*}

\subsection{Response to locally injected current fluctuation}
\label{subsec: local fluctuating electric current injection}
The calculation method is essentially the same as that in Sec.~\ref{sec: spin to ac}.
The resulting quadratic effective electric field $\mathcal{E}^{(2)}(w)$ and the quadratic spin $\tilde{S}_z^{(2)}(w)$ are shown in Fig.~\ref{fig_caseii}.
For the electric current fluctuations in Eq.~\eqref{eq_j0Def_ii}, we take
\begin{equation}
    S_{j, \omega}/J_0^2=\bar{S}_j\cdot\chi_{[-\omega_c,\omega_c]}(\omega),
\end{equation}
where $J_0$ is a reference amplitude of the current density fluctuation and $\chi_A(x)$ denotes the indicator function of the set $A$. This spectrum corresponds to white noise with cutoff $\omega_c$.
We set $\omega_c\tau_0=1$; in Appendix~\ref{app: quadratic spin to time-dependent current}, we show that the results remain qualitatively unchanged for other values of $\omega_c\tau_0$ within the regime where the Boltzmann equation is applicable.
Figure~\ref{fig_caseii}(a) shows that the sign of $\mathcal{E}^{(2)}(w)$ is consistent with the estimate in Eq.~\eqref{eq_E2_Pizz}.
Figure~\ref{fig_caseii}(b) shows that the fluctuating electric current induces the spin density near the boundary.
As in our previous work~\cite{Yoshimi2026} and case (0), the decomposition of the spin $\tilde{S}_z^{(2)}(w)$ indicates that the sign of the spin is primarily determined by the Edelstein effect induced by $\mathcal{E}^{(2)}(w)$ [see the right panel of Fig.~\ref{fig_caseii}(b)].

Here, we make a brief remark concerning the length scale arising from the isotropic spin--orbit coupling considered in this paper.
The spin density decays over the mean-free-path scale. The associated spin-relaxation length is of the same order in the present model and therefore cannot be resolved separately in Fig.~\ref{fig_caseii}.

At the opposite boundary, the result follows from (A) the invariance of the quadratic response under $j_0(t) \rightarrow -j_0(t)$ and (B) the equivalence of the two boundaries under a 180$^\circ$ rotation of the isotropic model and symmetric contact geometry, which maps $z\rightarrow L-z$ and reverses the $z$ component of spin.
This can be expressed as
\begin{equation}
    \tilde{S}_z^{(2)}(z=L,j_0)\overset{({\rm A})}{=}\tilde{S}_z^{(2)}(z=L,-j_0)\overset{({\rm B})}{=}-\tilde{S}_z^{(2)}(z=0,j_0).\label{eq: symmery-caseii}
\end{equation}
This shows that the spin polarizations at the two boundaries are antiparallel.

\subsection{Response to electric-field fluctuation near boundaries}
\label{subsec: interfacial fluctuating electric field penetration}
Imposing Ziman's boundary conditions together with $E^{(1)}(z=0,t)=E_0(t),\quad E^{(2)}(z=0)=0$, we solve the equations analytically to second order in $E_0(t)$.
At the order retained here, the quadratic responses are proportional to the integrated spectral weight $\int \mathrm{d}\omega S_{E,\omega}$, while their normalized spatial profiles are independent of its detailed spectral shape. We therefore do not assume a specific spectral shape for case (2), apart from requiring its appreciable weight to lie within the low-frequency validity regime of the Boltzmann equation.

Figure~\ref{fig_caseiii}(a) shows the quadratic excess carrier density $\tilde{n}^{(2)}$ and the Maxwell electric field $E^{(2)}$ as functions of $z/\lambda_{\rm TF}$. 
The leading responses in Fig.~\ref{fig_caseiii}(a) are independent of chirality and confined to the Thomas-Fermi screening region.
Note that this result is consistent with the boundary condition $E^{(2)}(z=0)=0$.
Because $E^{(2)}(z,t)$ vanishes both at the boundary and in the bulk while $qE^{(2)}(z,t)>0$ between them, Gauss's law requires the carrier-density response to change sign, producing the dipole-like profile shown in Fig.~\ref{fig_caseiii}(a).
The relation $qE^{(2)}(z,t)>0$ can be understood as follows.
We first note the following relation, which holds in the screening region to leading order in $\lambda_{\mathrm{TF}}/\ell$: $\partial_z\Pi_{zz}(z,t)\sim\rho^{(0)}\partial_z[\tilde{n}^{(2)}(z,t)/qN_0]$.
The derivation of this relation is shown in Eq.~\eqref{eq_Pizz_result}--\eqref{eq_Pizz_approximation}.
In the following, we focus on this screening region.
The above equation means that $\rho^{(0)}E^{(2)}-\partial_z\Pi_{zz}\sim\rho^{(0)}\mathcal{E}^{(2)}$ holds.
Then, in contrast to our previous work~\cite{Yoshimi2026} and Sec.~\ref{sec: spin to ac}, the force-balance equation becomes $\rho^{(1)}E^{(1)}+\rho^{(0)}\mathcal{E}^{(2)}\sim0$.
By using Gauss's law, we obtain $\rho^{(1)}E^{(1)}=\partial_z(\varepsilon_0[E^{(1)}]^2/2)$.
Because $[E^{(1)}(z,t)]^2$ decays in the bulk, 
\begin{equation}
\rho^{(1)}E^{(1)}=\partial_z(\varepsilon_0[E^{(1)}]^2/2)<0 
\end{equation}
and the above force-balance condition therefore gives 
$\rho^{(0)}\mathcal{E}^{(2)}>0$.
Therefore, we obtain $q\mathcal{E}^{(2)}>0$.
The equation for $q\mathcal{E}^{(2)}(z,t)$,
\begin{equation}
    q\mathcal{E}^{(2)}=qE^{(2)}-\frac{\epsilon_0}{q^2  N_0 }\frac{\partial^2 [qE^{(2)}]}{\partial z^2},
    \label{eq: def-of-E-tilde}
\end{equation}
is useful in revealing the sign of $qE^{(2)}(z,t)$.
Since $E^{(2)}(z=0)=0$ and $E^{(2)}(z=\infty)=0$ hold as the boundary conditions, $E(z)$ cannot have a negative minimum.
Indeed, if $qE^{(2)}(z_0)\leq0$ were a minimum at some $z_0\in(0,\infty)$, then $qE''(z_0)\geq0$ and Eq.~\eqref{eq: def-of-E-tilde} would give $q\mathcal{E}^{(2)}\leq0$, contradicting $q\mathcal{E}^{(2)}>0$.
Thus, $qE^{(2)}(z,t)>0$ holds for $z\in(0,\infty)$.

When subleading terms in $\lambda_{\rm TF}/\ell$ are retained, the carrier-density and electric-field responses acquire tails extending over the mean-free-path scale.
Their combination gives the effective field $\mathcal{E}^{(2)}(w)$ shown in Fig.~\ref{fig_caseiii}(b) as a function of $w=z/\ell$.
The decomposition in Fig.~\ref{fig_caseiii}(c) shows that the Edelstein response driven by $\mathcal{E}^{(2)}(w)$ dominates the spin density and determines its sign.

Regarding the response magnitude, the response to the electric-field fluctuation contains an additional factor $\left(\frac{\lambda_{\rm TF}}{\ell}\right)^2$, relative to the response to the current fluctuation.
This indicates that the response is expected to be larger in systems with larger $\lambda_{\mathrm{TF}}$, such as low carrier density metals and semimetals.

At the opposite boundary, the same symmetry argument used to obtain Eq.~\eqref{eq: symmery-caseii} with $j_0(t)$ replaced by $E_0(t)$ gives
\begin{equation}
    \tilde{S}_z^{(2)}(z=L,E_0)\overset{({\rm A})}{=}\tilde{S}_z^{(2)}(L,-E_0)\overset{({\rm B})}{=}-\tilde{S}_z^{(2)}(0,E_0).
\end{equation}
Thus, the spin polarizations at the two boundaries are antiparallel.

\section{Discussion}
\label{sec: discussions}
The two fluctuation setups considered in Sec.~\ref{sec: spin wo bias input} share a common mechanism. Although the stochastic input has zero mean, its quadratic response produces an effective electric field localized near a boundary. 
The chirality-dependent linear Edelstein effect converts this field into a time-averaged spin polarization, yielding antiparallel spin polarizations at the two boundaries. 
In the following, we first clarify the scope of our approach in Sec.~\ref{subsec: scope of our approach}, discuss its relation to other sources of spin polarization in Sec.~\ref{subsec: relation to other sources}, and finally consider its possible experimental implications in Sec.~\ref{subsec: implications for experiments}.

\subsection{Scope of our approach}
\label{subsec: scope of our approach}
As discussed in Sec.~\ref{sec: set up}, the band-diagonal Boltzmann description is valid when $|\omega|,\tau_0^{-1}\ll \Delta_{\rm SO}$, so that field-induced interband transitions and interband coherence can be neglected. The cutoff $\omega_c$ introduced in Sec.~\ref{subsec: local fluctuating electric current injection} is chosen to satisfy this condition.
Extending the theory to include these quantum-coherent interband effects remains an important direction for future work.

It is important to distinguish the externally imposed nonequilibrium fluctuations considered here from equilibrium fluctuations. The present formulation applies to fluctuations whose appreciable spectral weight is confined to the low-frequency regime where the band-diagonal Boltzmann equation is valid. A genuine equilibrium fluctuation spectrum of an electronic system is not restricted to this frequency range and therefore lies outside the scope of the present theory.
A consistent treatment would require other formulations, potentially including a Boltzmann-Langevin formulation that enforces detailed balance within the band-diagonal semiclassical regime, or a quantum kinetic treatment beyond this regime~\cite{BixonZwanzig1969,KoganShulman1969,KirkpatrickBelitz2022}.

\subsection{Relation to other sources of spin polarization}
\label{subsec: relation to other sources}
Even for the electrical fluctuations considered here, the magnitude of the spin response may depend on temperature through the fluctuation spectra and transport parameters.
The temperature dependence of CISS remains under debate, with different behaviors reported in molecular and crystalline systems~\cite{Yamamoto2026}.
Our mechanism may also be relevant to the latter systems.
Beyond this implicit temperature dependence, externally maintained nonequilibrium energy-density fluctuations provide a natural extension.
Within a local-equilibrium description, such energy-density fluctuations correspond to temperature fluctuations, suggesting that zero-mean temperature-gradient fluctuations may generate antiparallel spin polarizations near the boundaries through a quadratic response.

This possibility is motivated by the nonlinear thermoelectric effect observed by Arisawa et al.~\cite{Arisawa2024}, in which a voltage proportional to the square of the applied temperature gradient was demonstrated and proposed as a route to utilizing out-of-equilibrium temperature fluctuations.
Establishing the analogous spin response microscopically requires a kinetic theory that consistently treats spin, charge, and energy transport.

Several previous studies have emphasized vibration-assisted mechanisms in CISS, including vibration-induced charge redistribution and exchange-mediated spin polarization \cite{Fransson2021}, vibrational contributions to spin-orbit coupling and dissipation in temperature-dependent CISS~\cite{Das2022}, and variable-range hopping assisted by chiral phonons~\cite{Sano-Kato2024}. These mechanisms concern how molecular vibrations and inelastic transport modify CISS, whereas the present proposal concerns the rectification of nonequilibrium energy-density fluctuations. The two perspectives are complementary.

\begin{figure}
    \centering
    \includegraphics[width=\linewidth]{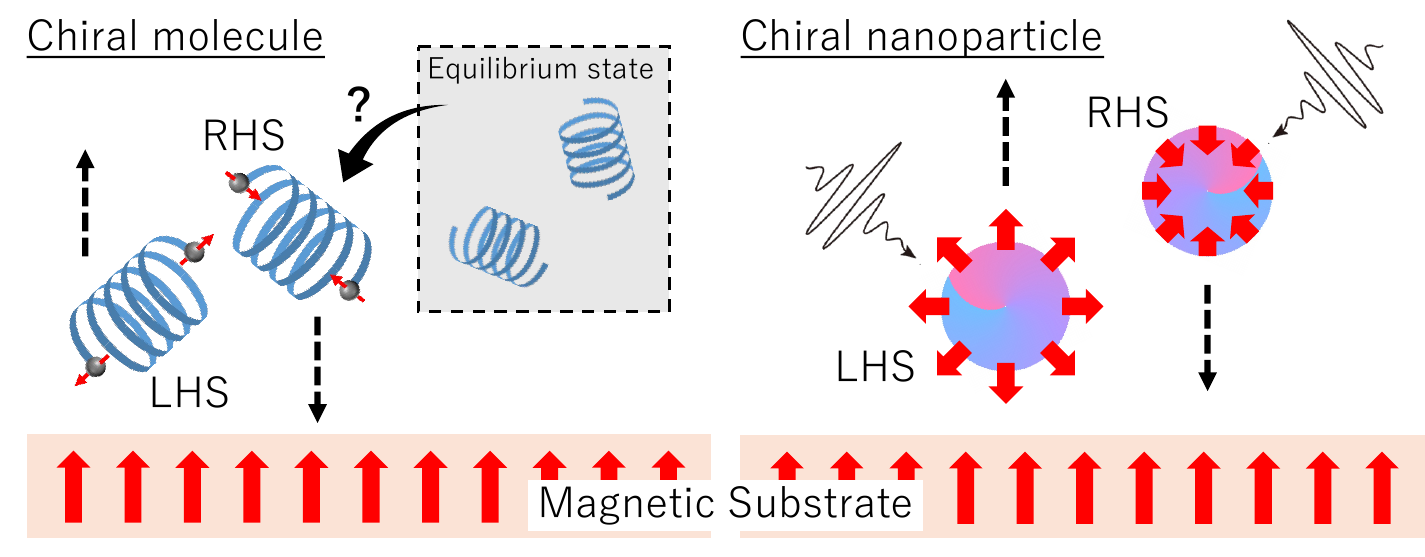}
    \caption{\textit{Left}: 
    Schematic of the enantioseparation mechanism proposed in Ref.~\cite{Banerjee-Ghosh2018}. 
    This indicates that the attraction to the magnetic substrate depends on the orientation of the antiparallel spin polarization, i.e., whether it is directed inward or outward.    
    The mechanism of spin polarization remains under discussion.
    LHS and RHS denote the left-handed and right-handed systems, respectively.
    \textit{Right}: 
    Schematic of the experimental proposal considered here. Analogously to molecular enantioseparation, chiral-metal nanoparticles could be separated using a magnetic substrate through their fluctuation-induced radial spin polarization.}
    \label{fig_discussion}
\end{figure}

\subsection{Implications for experiments}
\label{subsec: implications for experiments}
As a possible experimental implication, nonequilibrium electrical fluctuations—and potentially nonequilibrium energy-density fluctuations—could induce an outward or inward radial spin texture near the surface of a spherical chiral-metal nanoparticle, as illustrated in Fig.~\ref{fig_discussion}.
A quantitative treatment of spherical geometry would require extending the present locally planar analysis.
Such a texture could enable chirality-selective separation by a magnetic substrate, analogous to molecular enantioseparation~\cite{Banerjee-Ghosh2018}. 
The sign of the interaction between the chiral nanoparticle and a substrate magnetized perpendicular to the surface is determined by the spin polarization in the nanoparticle at the point closest to the magnetic substrate and the direction of magnetization of the substrate.
The hedgehog-type radial spin texture is invariant under free rotation of the nanoparticle.
Thus, the interaction between the chiral nanometals and the magnetic substrates remains chirality-dependent.
While chirality-selective transport at the nanoscale is typically achieved using chiral light, such as circularly polarized light or optical vortices, and is often hindered by thermal noise~\cite{Okamoto2022,Lininger2023,Tkachenko2026}, the scheme presented here could enable chirality-selective separation without relying on chiral light and could exploit nonequilibrium fluctuations as a functional resource.
This nanoparticle perspective may also provide a starting point for investigating fluctuation-induced spin polarization in chiral molecules.

\section{Summary}
\label{sec: summary}
We have shown that externally maintained nonequilibrium electrical fluctuations with zero mean generate finite time-averaged spin polarizations in chiral metals through a quadratic response. For both locally injected current fluctuations and electric-field fluctuations near boundaries, a boundary-localized quadratic effective field is converted into spin polarization through the chirality-dependent linear Edelstein effect, producing antiparallel spin polarizations at the two boundaries. 
We have also specified the validity regime of the semiclassical treatment and discussed possible extensions to other nonequilibrium fluctuations, including energy-density fluctuations. 
These results also suggest a possible route to a chirality-selective separation scheme for chiral-metal nanoparticles.

\section*{Acknowledgments}
We thank H. Kusunose and J. Kishine for discussions during the early stages of this work.
We also thank H.~Watanabe for his informative comments on the linear and nonlinear responses of chiral systems.
K.Y. thanks T. Kimura for the helpful comments.
This work was supported by JSPS KAKENHI Grants No.~JP20K03855, No.~JP22J12348, No.~JP23H00291, No. JP23H00091, No.~JP24KJ1036, No. JP24K01331, No. JP25H02113, PRESTO from JST (Grant Number JPMJPR2356), by Joint Research by the Institute for Molecular Science (IMS program No. 23IMS1101), by the grant of OML Project by the National Institutes of Natural Sciences (NINS program No. OML012301), by JST ERATO Grant Number JPMJER2503, by World-leading Innovative Graduate Study Program for Materials Research, Information, and Technology (MERIT-WINGS) of the University of Tokyo, and by JST SPRING, Grant Number JPMJSP2108.

\onecolumngrid
\begin{center}
 {\large\bfseries Appendix}
\end{center}
\appendix
In Appendix, we show the details of the calculations. 
In Appendix~\ref{app: definition}, we define some crucial functions.
In Appendices~\ref{app: linear spin to local dc} and \ref{app: quadratic spin to local dc}, we calculate the linear and quadratic spin response to a local dc current.
In Appendix~\ref{app: approximation}, we propose an approximation which is useful to obtain analytical solutions.
In Appendices~\ref{app: linear spin to time-dependent current} and \ref{app: quadratic spin to time-dependent current}, we calculate the linear and quadratic spin responses to a time-dependent current for the cases (0) and (1).
In Appendix~\ref{app: quadratic spin to fluctuating electric field}, we calculate the quadratic spin response to the fluctuating electric field near the boundaries for the case (2).

In all sections of Appendix, we consider the spin polarization near the boundary $z=0$.
The following calculations are also applicable to the spin polarization near the boundary $z=L$. We introduce a dimensionless variable $w:=z/\ell$.

\section{Definition of some functions}
\label{app: definition}
First, we define the auxiliary functions $F_{n,\omega}(w)$ and $\tilde{F}_{n,\omega}(w),\,n \in \mathbb{Z},\,\omega \in \mathbb{R}$ as 
\begin{align}
 F_{n,\omega}(w) &:= \int_{0}^{1} \mathrm{d}x x^{n-2} e^{-(1-i\omega\tau_0)w/x} \quad (\, w > 0),\\
 \tilde{F}_{n,\omega}(w) &:=
 \begin{cases}
  F_{n,\omega}(|w|) & n = \text{odd}, \\
  \mathrm{sgn}(w) F_{n,\omega}(|w|) & n = \text{even},
 \end{cases}
\end{align}
We also denote the convolution of two functions $A$ and $B$ as
\begin{equation}
 [A * B](w) := \int_{-\infty}^{\infty} \mathrm{d}w' A(w - w') B(w').
\end{equation}
For the case of dc current injection ($\omega=0$), we omit the subscript $\omega$ (e.g., $F_n(w)$) for brevity.

\section{Linear spin response to a local dc current}
\label{app: linear spin to local dc}
\begin{figure}
    \centering
    \includegraphics[width=0.8\linewidth]{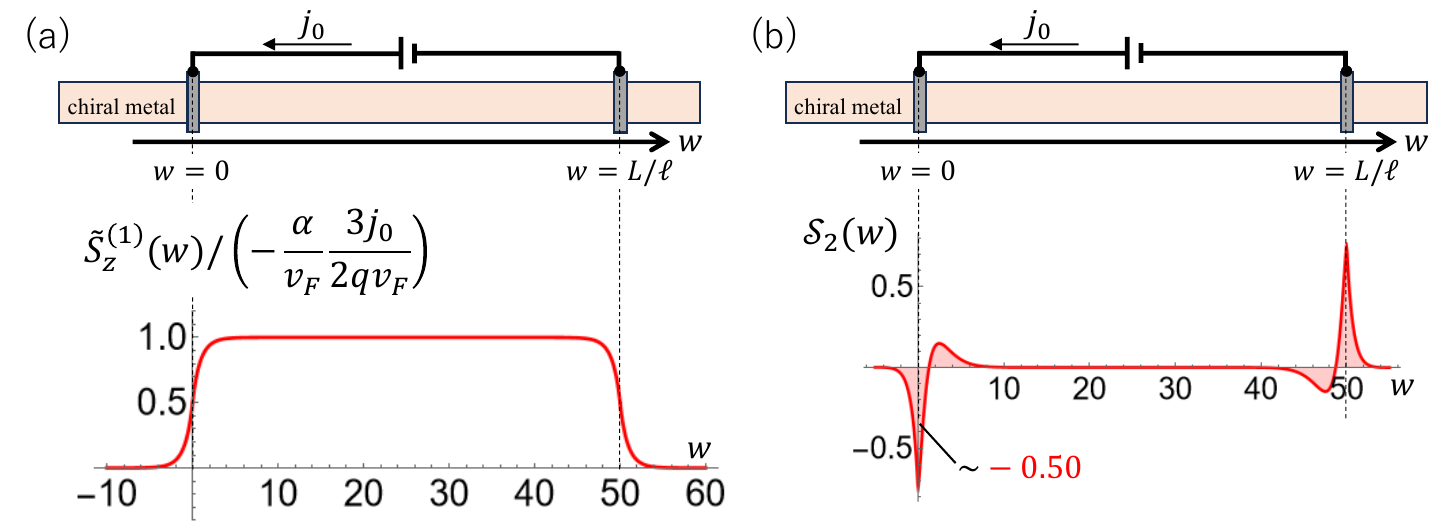}
    \caption{The spatial dependence of (a) $\tilde{S}_z^{(1)}(w)$ and (b) $\tilde{S}_z^{(2)}(w)$ for $\alpha>0$ and $q=-e$. We set $L/\ell=50$ in panels.}
    \label{fig_sz1_sz2_dc}
\end{figure}

Here, as a prerequisite, we discuss the linear responses to a local dc current $j_0$.
The Boltzmann equation for the linear response is expressed as
\begin{equation}
  v_{\bm{k}\gamma}^z\frac{\partial f_{\bm{k}\gamma}^{(1)}}{\partial z}+qE^{(1)}(z)\frac{\partial f_{\bm{k}\gamma}^{(0)}}{\partial k_z}=-\frac{f_{\bm{k}\gamma}^{(1)}}{\tau_0}+\frac{n^{(1)}(\varepsilon_{\bm{k}\gamma},z)}{\tau_0N_0}+\frac{2 \sigma_{\bm{k}\gamma}^z S^{(1)}_z(\varepsilon_{\bm{k}\gamma},z)}{\tau_0N_0}+\frac{3j_0}{2qN_\gamma}\cos^2{\theta}\left(-\frac{\partial f_{\bm{k}\gamma}^{(0)}}{\partial \varepsilon_{\bm{k}\gamma}}\right)\delta(z).
  \label{eq_BE1_dc}
\end{equation}
The solution to Eq.~\eqref{eq_BE1_dc} that satisfies the boundary condition $f_{\bm{k}\gamma}^{(1)} \to 0$ for $z \to -\infty$ is given by
\begin{align}
    &\underline{\mathrm{For}\,z<0:}\notag\\
    &v_{\bm{k}\gamma}^z>0 : f_{\bm{k}\gamma}^{(1)}(z)=\int_{-\infty}^{z}\mathrm{d}z'\exp\left(-\frac{z-z'}{v_{\bm{k}\gamma}^z\tau_0}\right)\left[qE^{(1)}(z')+\frac{\tilde{n}^{(1)}(z')}{v_{\bm{k}\gamma}^z\tau_0N_0}+\frac{2 \sigma_{\bm{k}\gamma}^z S^{(1)}_z(\varepsilon_{\bm{k}\gamma},z)}{v_{\bm{k}\gamma}^z\tau_0N_0}\right]\left(-\frac{\partial f_{\bm{k}\gamma}^{(0)}}{\partial \varepsilon_{\bm{k}\gamma}}\right),\notag\\
    &v_{\bm{k}\gamma}^z<0 : f_{\bm{k}\gamma}^{(1)}(z)=\int_{+\infty}^{z}\mathrm{d}z'\exp\left(-\frac{z-z'}{v_{\bm{k}\gamma}^z\tau_0}\right)\left[qE^{(1)}(z')+\frac{\tilde{n}^{(1)}(z')}{v_{\bm{k}\gamma}^z\tau_0N_0}+\frac{2 \sigma_{\bm{k}\gamma}^z S^{(1)}_z(\varepsilon_{\bm{k}\gamma},z)}{v_{\bm{k}\gamma}^z\tau_0N_0}\right]\left(-\frac{\partial f_{\bm{k}\gamma}^{(0)}}{\partial \varepsilon_{\bm{k}\gamma}}\right)\notag\\
    &\qquad\qquad\qquad\qquad+\frac{3j_0|\cos{\theta}|}{v(\varepsilon_{\bm{k}\gamma})\cdot 2qN_{\gamma}}\exp\left(-\frac{|z|}{|v_{\bm{k}\gamma}^z|\tau_0}\right)\left(-\frac{\partial f_{\bm{k}\gamma}^{(0)}}{\partial \varepsilon_{\bm{k}\gamma}}\right),\notag\\
    &\underline{\mathrm{For}\,z>0:}\notag\\
    &v_{\bm{k}\gamma}^z>0 : f_{\bm{k}\gamma}^{(1)}(z)=\int_{-\infty}^{z}\mathrm{d}z'\exp\left(-\frac{z-z'}{v_{\bm{k}\gamma}^z\tau_0}\right)\left[qE^{(1)}(z')+\frac{\tilde{n}^{(1)}(z')}{v_{\bm{k}\gamma}^z\tau_0N_0}+\frac{2 \sigma_{\bm{k}\gamma}^z S^{(1)}_z(\varepsilon_{\bm{k}\gamma},z)}{v_{\bm{k}\gamma}^z\tau_0N_0}\right]\left(-\frac{\partial f_{\bm{k}\gamma}^{(0)}}{\partial \varepsilon_{\bm{k}\gamma}}\right)\notag\\
    &\qquad\qquad\qquad\qquad+\frac{3j_0\cos{\theta}}{v(\varepsilon_{\bm{k}\gamma})\cdot 2qN_{\gamma}}\exp\left(-\frac{z}{v_{\bm{k}\gamma}^z\tau_0}\right)\left(-\frac{\partial f_{\bm{k}\gamma}^{(0)}}{\partial \varepsilon_{\bm{k}\gamma}}\right),\notag\\
    &v_{\bm{k}\gamma}^z<0 : f_{\bm{k}\gamma}^{(1)}(z)=\int_{+\infty}^{z}\mathrm{d}z'\exp\left(-\frac{z-z'}{v_{\bm{k}\gamma}^z\tau_0}\right)\left[qE^{(1)}(z')+\frac{\tilde{n}^{(1)}(z')}{v_{\bm{k}\gamma}^z\tau_0N_0}+\frac{2 \sigma_{\bm{k}\gamma}^z S^{(1)}_z(\varepsilon_{\bm{k}\gamma},z)}{v_{\bm{k}\gamma}^z\tau_0N_0}\right]\left(-\frac{\partial f_{\bm{k}\gamma}^{(0)}}{\partial \varepsilon_{\bm{k}\gamma}}\right).\label{eq_f1_dc}
\end{align}
Multiplying Eq.~\eqref{eq_f1_dc} by $\delta(\varepsilon-\varepsilon_{\bm{k}\gamma})$, $qv_{\bm{k}\gamma}^z\delta(\varepsilon-\varepsilon_{\bm{k}\gamma})$ and $\frac{\sigma_{\bm{k}\gamma}^z}{2}\delta(\varepsilon-\varepsilon_{\bm{k}\gamma})$, and then taking the sum $\frac{1}{\Omega}\sum_{\bm{k}\gamma}$, we obtain the integral equations 
as
\begin{subequations}
    \begin{align}     0&=\left[\tilde{F}_2*\mathcal{E}^{(1)}\right](w)-\frac{\alpha}{v_F}\cdot \frac{4m}{q\ell N_0} \left[\tilde{F}_2*\tilde{S}_z^{(1)}\right](w)+\frac{j_0}{\sigma_0}\tilde{F}_3(w),\label{eq_dc_convolutionE1}\\        
        \tilde{j}_e^{(1)}(w)&=\frac{3\sigma_0}{2}\left[\tilde{F}_3*\mathcal{E}^{(1)}\right](w)-\frac{\alpha}{v_F}\cdot2qv_F\left[\tilde{F}_3*\tilde{S}_z^{(1)}\right](w)+\frac{3j_0}{2}\tilde{F}_4(w),\label{eq_dc_convolutionje1}\\
        \tilde{S}_z^{(1)}(w)&=-\frac{\alpha}{v_F}\frac{qlN_0}{2}\left[\tilde{F}_3*\mathcal{E}^{(1)}\right](w)+\frac{1}{2}\left[\tilde{F}_3*\tilde{S}_z^{(1)}\right](w)\label{eq_dc_convolutionS1}
    \end{align}
\end{subequations}
where $\mathcal{E}^{(i)}(z):=E^{(i)}(z)-\partial_z\tilde{n}^{(i)}(z)/qN_0$, which corresponds to the effective electric field.
Noting that the leading term in $\tilde{S}_z^{(1)}(w)$ is $O[\alpha/v_{\mathrm{F}}]$, the second term in RHS of Eq.~\eqref{eq_dc_convolutionE1} and Eq.~\eqref{eq_dc_convolutionje1} is found to be $O[(\alpha/v_{\mathrm{F}})^2]$.
Then, we can find the solution of Eq.~\eqref{eq_dc_convolutionE1} as
\begin{align}
    \mathcal{E}^{(1)}(w)=\mathcal{E}_0\Theta(w)+O\left[\left(\frac{\alpha}{v_F}\right)^2\right],\label{eq_dc_E1result}
\end{align}
with the assumption $\alpha\ll v_F$ and the relation
\begin{align}
 \int_{-\infty}^w \tilde{F}_2(w')\mathrm{d}w'=-\tilde{F}_3(w).
\end{align}
Substituting Eq.~\eqref{eq_dc_E1result} into Eq.~\eqref{eq_dc_convolutionje1}, we can also obtain
\begin{align}
    \tilde{j}_e^{(1)}(w)=j_0\Theta(w)+O\left[\left(\frac{\alpha}{v_F}\right)^2\right].
\end{align}
In addition, with a nonzero solution of $x^3+x-\frac{1}{2}\ln\frac{1+x}{1-x}=0$ ($x_0\sim0.946$), we can find the solution to Eq.~\eqref{eq_dc_convolutionS1} as
\begin{align}
    \tilde{S}_z^{(1)}(w)&=-\frac{\alpha}{v_F}\frac{3j_0}{2qv_F}\left\{\int_{1}^{\infty}\mathrm{d}x\frac{x^2\,[2\Theta(w)-\mathrm{sgn}(w)e^{-|w|x}]}{\left(x^3+x+\frac{1}{2}\ln\left|\frac{x-1}{x+1}\right|\right)^2+\frac{\pi^2}{4}}+\frac{2(1-x_0^2)}{3x_0^2-2}[2\Theta(w)-\mathrm{sgn}(w)e^{-|w|x_0}]\right\}\notag\\
    &:=-\frac{\alpha}{v_F}\frac{3j_0}{2qv_F}\mathcal{S}_1(w).
\end{align}
The function $\tilde{S}_z^{(1)}(w)$ converges to $-\frac{\alpha}{v_F}\frac{3j_0}{2qv_F}$ in the limit of $z\rightarrow\infty$ (bulk region).
This result corresponds with Eq.~\eqref{eq_sz1Bulk}.
The result of $\tilde{S}_z^{(1)}(w)$ is plotted in Fig.~\ref{fig_sz1_sz2_dc} (a). 

\section{Quadratic spin response to a local dc current}
\label{app: quadratic spin to local dc}
Here, we discuss the quadratic responses to a local dc current $j_0$.
The Boltzmann equation for the quadratic response is expressed as
\begin{equation}
  v_{\bm{k}\gamma}^z\frac{\partial f_{\bm{k}\gamma}^{(2)}}{\partial z}+qE^{(2)}(z) v_{\bm{k}\gamma}^z\frac{\partial f_{\bm{k}\gamma}^{(0)}}{\partial\varepsilon_{\bm{k}\gamma}}+qE^{(1)}(z)\frac{\partial f_{\bm{k}\gamma}^{(1)}}{\partial k_z}=-\frac{f_{\bm{k}\gamma}^{(2)}}{\tau_0}+\frac{n^{(2)}(\varepsilon_{\bm{k}\gamma},z)}{\tau_0 N_0}+\frac{2 \sigma_{\bm{k}\gamma}^z S^{(2)}_z(\varepsilon_{\bm{k}\gamma},z)}{\tau_0N_0}.
  \label{eq_BE2_dc}
\end{equation}
The solution of Eq.~\eqref{eq_BE2_dc} that satisfies the boundary condition $f_{\bm{k}\gamma}^{(2)} \to 0$ for $z \to -\infty$ is given by
\begin{align}
   f_{\bm{k}\gamma}^{(2)}(z) = \displaystyle\int_{-\infty\times {\rm sgn}(k_z)}^{z}\mathrm{d}z'\exp\left(-\frac{z-z'}{v_{\bm{k}\gamma}^z\tau_0}\right)\left[-qE^{(2)}(z')\frac{\partial f_{\bm{k}\gamma}^{(0)}}{\partial \varepsilon_{\bm{k}\gamma}}+\frac{n^{(2)}(\varepsilon_{\bm{k}\gamma}, z')}{v_{\bm{k}\gamma}^z\tau_0 N_0}+\frac{2 \sigma_{\bm{k}\gamma}^z S^{(2)}_z(\varepsilon_{\bm{k}\gamma},z)}{v_{\bm{k}\gamma}^z\tau_0N_0}-\frac{qE^{(1)}(z')}{v_{\bm{k}\gamma}^z}\frac{\partial f_{\bm{k}\gamma}^{(1)}}{\partial k_z}\right].
\label{eq_f2_dc}
\end{align}

Multiplying Eq.~\eqref{eq_f2_dc} by $\delta(\varepsilon-\varepsilon_{\bm{k}\gamma})$, $qv_{\bm{k}\gamma}^z\delta(\varepsilon-\varepsilon_{\bm{k}\gamma})$, and $\frac{\sigma_{\bm{k}\gamma}^z}{2}\delta(\varepsilon-\varepsilon_{\bm{k}\gamma})$, and then taking the sum $\frac{1}{\Omega}\sum_{\bm{k}\gamma}$, we obtain the integral equations 
\begin{subequations}   
\begin{align}
0=&\left[\tilde{F}_2*\mathcal{E}^{(2)}\right](w)-\frac{\alpha}{v_F}\cdot \frac{4m}{q\ell N_0} \left[\tilde{F}_2*\tilde{S}_z^{(2)}\right](w)+X_n(w),\label{eq_dc_convolutionE2}\\        
    \tilde{j}_e^{(2)}(w)=&\frac{3\sigma_0}{2}\left[\tilde{F}_3*\mathcal{E}^{(2)}\right](w)-\frac{\alpha}{v_F}\cdot2qv_F\left[\tilde{F}_3*\tilde{S}_z^{(2)}\right](w)+X_j(w)\label{eq_dc_convolutionje2}\\
        \tilde{S}_z^{(2)}(w)=&-\frac{\alpha}{v_F}\frac{qlN_0}{2}\left[\tilde{F}_3*\mathcal{E}^{(2)}\right](w)+\frac{1}{2}\left[\tilde{F}_3*\tilde{S}_z^{(2)}\right](w)+X_S(w)\label{eq_dc_convolutionS2}
\end{align}
\end{subequations}
where 
\begin{subequations}   
\begin{align}
&X_n(w)=-\frac{mq\tau_0j_0^2}{\pi^2N_0\sigma_0^2}\cdot w\tilde{F}_1(w)\\
&X_j(w)=-\frac{mq^3\ell^2j_0^2}
{4\pi^2\sigma_0^2}\cdot2\left[\tilde{F}_3(w)+w\tilde{F}_2(w)\right]\\    
&X_S(w)/\left(\frac{\alpha}{v_F}\frac{mq^2\ell\tau_0j_0^2}{4\pi^2\sigma_0^2}\right)
    :=\Theta(w)\left[-2w \tilde{F}_4(w)+w^2 \tilde{F}_1(w)\right]+w \left[\tilde{F}_2(w)+\tilde{F}_4(w)\right]+\tilde{F}_3(w)-\tilde{F}_5(w)\notag\\
    &-\int_{0}^{\infty}\mathrm{d}x\left\{-\frac{3}{2}\left[w\Theta(w)-(w-x)\Theta(w-x)\right]F_3(x)+\left[\frac{w^2}{2}\Theta(w)-\frac{w^2-x^2}{2}\Theta(w-x)\right]F_0(x)\right\}\mathcal{S}_1(w-x)\notag\\
    &-\int_{0}^{\infty}\mathrm{d}x\left\{-\frac{3}{2}\left[w\Theta(w)-(w+x)\Theta(w+x)\right]F_3(x)-\left[\frac{w^2}{2}\Theta(w)-\frac{w^2-x^2}{2}\Theta(w+x)\right]F_0(x)\right\}\mathcal{S}_1(w+x).\label{eq_XS_dc}    
\end{align}
\end{subequations}
Note that in the above equations, only the leading terms of $X(w)$, $X_j(w)$ and $X_S(w)$ under the assumption $\lambda_{\mathrm{TF}}/\ell\ll1$ and $\alpha/v_F\ll1$ are shown.
Noting that the leading term in $\tilde{S}_z^{(2)}(w)$ is $O[\alpha/v_{\mathrm{F}}]$, the second term in RHS of Eq.~\eqref{eq_dc_convolutionE2} and Eq.~\eqref{eq_dc_convolutionje2} is found to be $O[(\alpha/v_{\mathrm{F}})^2]$.
Then, we can obtain
\begin{align}
    \mathcal{E}^{(2)}(w)=\frac{mq\tau_0j_0^2}{2\pi^2N_0\sigma_0^2}\int_1^{\infty}\mathrm{d}x\frac{x^3e^{-|w|x}}{(x^2-1)\left\{\left[x+\frac{1}{2}\ln\left|\frac{x-1}{x+1}\right|\right]^2+\frac{\pi^2}{4}\right\}}+O\left[\left(\frac{\alpha}{v_F}\right)^2\right],
\end{align}
and
\begin{align}
    \tilde{j}_e^{(2)}(w)=0+O\left[\left(\frac{\alpha}{v_F}\right)^2\right],
\end{align}
with the relation
\begin{equation}
\int_{-\infty}^w X(w')\mathrm{d}w'=\frac{2}{3\sigma_0}X_j(w).
\end{equation}

\begin{align}
    \tilde{S}_z^{(2)}(w)&=\int_{-\infty}^{\infty}\frac{ds}{\sqrt{2\pi}}\frac{s^3\left[\frac{\alpha}{qv_F^2}\hat{X}_j(s)+\hat{X}_S(s)\right]}{s^3-s+\arctan(s)}e^{-iws}:=\frac{\alpha}{v_F}\frac{mq^2\ell\tau_0j_0^2}{4\pi^2\sigma_0^2}\mathcal{S}_2(w)\label{eq_sz2_analytic}
\end{align}
where $\hat{f}(s):=\int_{-\infty}^{\infty}\frac{\mathrm{d}w}{\sqrt{2\pi}}f(w)e^{isw}$.
The inverse Fourier transform \eqref{eq_sz2_analytic} is calculated numerically and its result $\mathcal{S}_2(w)$ is shown in Fig.~\ref{fig_sz1_sz2_dc} (b).

\section{Approximation and its validity}
\label{app: approximation}
\begin{figure}
    \centering
    \includegraphics[width=0.9\linewidth]{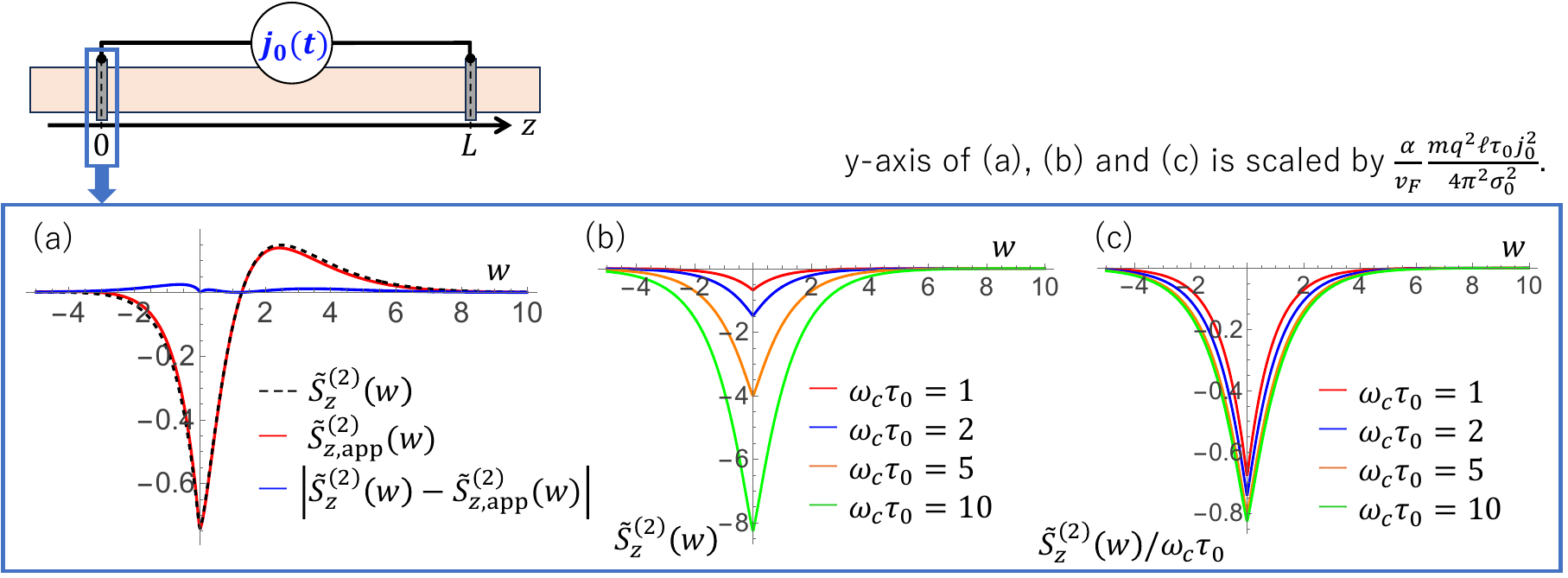}
    \caption{Enlarged view of spatial variations of the quadratic spin polarization near the boundary $z=0$ for $\alpha>0$.
    (a) The black dashed, red solid, and blue solid lines show $\tilde{S}_{z}^{(2)}(w)$, $\tilde{S}_{z,\mathrm{app}}^{(2)}(w)$ and $|\tilde{S}_{z}^{(2)}(w)-\tilde{S}_{z,\mathrm{app}}^{(2)}(w)|$, respectively.
    (b), (c) The spatial dependence of $\tilde{S}_{z}^{(2)}(w)$ and its scaled version by $\omega_c\tau_0$. The red, blue, orange and green solid lines corresponds to the case of $\omega_c\tau_0=1,2,5,10$, respectively.}
    \label{fig_approximation_Caseii}
\end{figure}
Here, as prerequisite, we show the result of $\tilde{S}_z^{(2)}(w)$ with an approximation $\mathcal{S}_1(w)\rightarrow \Theta(w)$ and its validity.
This process is crucial to obtain analytic results in the case of ac current injection, which is much more complex than that of dc current injection due to the time dependence.
The approximation $\mathcal{S}_1(w)\rightarrow \Theta(w)$ corresponds to the neglect of the spatial variations near the boundary that occur on the length scale of the mean free path (e.g. see Fig.~\ref{fig_sz1_sz2_dc} (a)).
By replacing $\mathcal{S}_1(w)$ with $\Theta(w)$ in Eq.~\eqref{eq_XS_dc}, we obtain
\begin{align}
     X_{S,\mathrm{app}}(w)/\left(\frac{\alpha}{v_F}\frac{mq^2\ell\tau_0j_0^2}{4\pi^2\sigma_0^2}\right)=&\Theta(w)\left[-2w \tilde{F}_4(w)+w^2 \tilde{F}_1(w)\right]+w \left[\tilde{F}_2(w)+\tilde{F}_4(w)\right]+\tilde{F}_3(w)-\tilde{F}_5(w)\notag\\
     &+\Theta(w)\left[\frac{w^2}{2}\tilde{F}_1(w)-\frac{3}{2}w\tilde{F}_4(w)\right]-\frac{3}{2}\tilde{F}_5(w)+w\tilde{F}_2(w)+\tilde{F}_3(w).
\end{align}
Figure \ref{fig_approximation_Caseii} (a) compares the exact result $\tilde{S}_z^{(2)}(w)$ with $\tilde{S}_{z,\mathrm{app}}^{(2)}(w)$, which is obtained by substituting this function $X_{S,\mathrm{app}}(w)$ into Eq.~\eqref{eq_sz2_analytic}.
This figure shows that the relative error $|[\tilde{S}_z^{(2)}(w)-\tilde{S}_{z,\mathrm{app}}^{(2)}(w)]/\tilde{S}_z^{(2)}(w)|$, which is expressed as the blue line in Fig.~\ref{fig_approximation_Caseii} (a), is generally of order $10^{-2}$ or smaller, and the relative error of the spin accumulation is $|[\int_{-\infty}^{\infty}\tilde{S}_z^{(2)}(w)\mathrm{d}w-\int_{-\infty}^{\infty}\tilde{S}_{z,\mathrm{app}}^{(2)}(w)\mathrm{d}w]/\int_{-\infty}^{\infty}\tilde{S}_z^{(2)}(w)\mathrm{d}w|\sim 3\times10^{-7}$.
This indicates that the dominant contribution of $\tilde{S}_z^{(1)}(w)$ to $\tilde{S}_z^{(2)}(w)$ originates from the bulk region, while the contribution from the region near the boundary is negligible.

Based on these considerations, $\tilde{S}_{\omega,z}^{(2)}(w)$ is replaced by its value in the bulk region $z=+\infty$ in the calculation of quadratic spin response to ac current injection as well.

\section{Linear spin responses to a time-dependent current}
\label{app: linear spin to time-dependent current}
By using the Fourier transformation in time, the Boltzmann equation for the linear response is expressed as
\begin{equation}
  -i\omega f_{\bm{k}\gamma,\omega}^{(1)}+v_{\bm{k}\gamma}^z\frac{\partial f_{\bm{k}\gamma,\omega}^{(1)}}{\partial z}+qE_{\omega}^{(1)}(z)\frac{\partial f_{\bm{k}\gamma}^{(0)}}{\partial k_z}=-\frac{f_{\bm{k}\gamma,\omega}^{(1)}}{\tau_0}+\frac{n_{\omega}^{(1)}(\varepsilon_{\bm{k}\gamma},z)}{\tau_0N_0}+\frac{2 \sigma_{\bm{k}\gamma}^z S^{(1)}_{z,\omega}(\varepsilon_{\bm{k}\gamma},z)}{\tau_0N_0}+\frac{3j_{0,\omega}}{2qN_\gamma}\cos^2{\theta}\left(-\frac{\partial f_{\bm{k}\gamma}^{(0)}}{\partial \varepsilon_{\bm{k}\gamma}}\right)\delta(z).
\end{equation}
The solution follows from Eq.~\eqref{eq_f1_dc} by replacing $\tau_0$ with $\tau_{\omega}:=\tau_0/(1-i\omega\tau_0)$.
We then obtain the expressions for $j_{\omega,e}^{(1)}$ and $S_{\omega,z}^{(1)}$ respectively as follows:
\begin{subequations}
    \begin{align}
        \frac{-i\omega\tau_0}{1-i\omega\tau_0}\frac{\tilde{n}_{\omega}^{(1)}(w)}{qN_0}&=\left[\tilde{F}_{2,\omega}*\mathcal{E}_{\omega}^{(1)}\right](w)-\frac{\alpha}{v_F}\cdot \frac{4m}{q\ell N_0} \left[\tilde{F}_{2,\omega}*\tilde{S}_{z,\omega}^{(1)}\right](w)+\frac{j_{0,\omega}}{\sigma_0}\tilde{F}_{3,\omega}(w),\label{eq_ac_convolutionE1}\\
        \tilde{j}_{e,\omega}^{(1)}(w)&=\frac{3\sigma_0}{2}\left[\tilde{F}_{3,\omega}*\mathcal{E}_{\omega}^{(1)}\right](w)-\frac{\alpha}{v_F}\cdot2qv_F\left[\tilde{F}_{3,\omega}*\tilde{S}_{z,\omega}^{(1)}\right](w)+\frac{3j_{0,\omega}}{2}\tilde{F}_{4,\omega}(w),\\
        \tilde{S}_{z,\omega}^{(1)}(w)&=-\frac{\alpha}{v_F}\frac{qlN_0}{2}\left[\tilde{F}_{3,\omega}*\mathcal{E}_{\omega}^{(1)}\right](w)+\frac{1}{2}\left[\tilde{F}_{3,\omega}*\tilde{S}_{z,\omega}^{(1)}\right](w),
    \end{align}
\end{subequations}
where $\mathcal{E}_{\omega}^{(i)}(z):=E_{\omega}^{(i)}(z)-\partial_z\tilde{n}_{\omega}^{(i)}(z)/(1-i\omega\tau_0)qN_0$, which corresponds to the effective electric field when $\omega\tau_0=0$.
Solving Eq.~\eqref{eq_ac_convolutionE1} with $\tilde{n}_{\omega}^{(1)}(w)\sim O[\lambda_{\mathrm{TF}}/\ell]$, we obtain the following expressions with the definition of $\sigma_{\omega}:=\sigma_0/(1-i\omega\tau_0)$:
\begin{align}
    \mathcal{E}_{\omega}^{(1)}(w)\sim\frac{j_{0,\omega}}{\sigma_{\omega}}\Theta(w)+O\left[\left(\frac{\alpha}{v_F}\right)^2\right],
\end{align}
due to the assumption $\lambda_{\mathrm{TF}}\ll\ell$ and $\alpha\ll v_F$, and in the bulk region $w=\infty$,
\begin{align}
    \tilde{S}_{z,\omega}^{(1)}(w=\infty)=-\frac{\alpha}{v_F}\frac{3(1-i\omega\tau_0)}{2-3i\omega\tau_0}\frac{j_{0,\omega}}{qv_F}.
\end{align}

\section{Quadratic spin responses to a time-dependent current}
\label{app: quadratic spin to time-dependent current}
By using the Fourier transformation in time, the Boltzmann equation for the linear response is expressed as
\begin{equation}
  -i\omega f_{\bm{k}\gamma,\omega}^{(2)}+v_{\bm{k}\gamma}^z\frac{\partial f_{\bm{k}\gamma,\omega}^{(2)}}{\partial z}+qE_{\omega}^{(2)}(z)\frac{\partial f_{\bm{k}\gamma}^{(0)}}{\partial k_z}+\int_{-\infty}^{\infty}\frac{\mathrm{d}\omega'}{\sqrt{2\pi}}qE_{\omega-\omega'}^{(1)}(z)\frac{\partial f_{\bm{k}\gamma,\omega'}^{(1)}}{\partial k_z}=-\frac{f_{\bm{k}\gamma,\omega}^{(2)}}{\tau_0}+\frac{n_{\omega}^{(2)}(\varepsilon_{\bm{k}\gamma},z)}{\tau_0N_0}+\frac{2 \sigma_{\bm{k}\gamma}^z S^{(2)}_{z,\omega}(\varepsilon_{\bm{k}\gamma},z)}{\tau_0N_0}.
  \label{eq_quadraticBE}
\end{equation}
The solution follows from Eq.~\eqref{eq_f2_dc} by replacing $\tau_0$ with $\tau_{\omega}:=\tau_0/(1-i\omega\tau_0)$.
We then obtain the following equations:
\begin{subequations}
    \begin{align}
        \frac{-i\omega\tau_0}{1-i\omega\tau_0}\frac{\tilde{n}_{\omega}^{(2)}(w)}{qN_0}&=\left[\tilde{F}_{2,\omega}*\mathcal{E}_{\omega}^{(2)}\right](w)-\frac{\alpha}{v_F}\cdot \frac{4m}{q\ell N_0} \left[\tilde{F}_{2,\omega}*\tilde{S}_{z,\omega}^{(2)}\right](w)+X_{n,\omega}(w),\label{eq_ac_convolutionE2}\\
        \tilde{j}_{e,\omega}^{(2)}(w)&=\frac{3\sigma_0}{2}\left[\tilde{F}_{3,\omega}*\mathcal{E}_{\omega}^{(2)}\right](w)-\frac{\alpha}{v_F}\cdot2qv_F\left[\tilde{F}_{3,\omega}*\tilde{S}_{z,\omega}^{(2)}\right](w)+X_{j,\omega}(w),\label{eq_ac_convolutionj2}\\
        \tilde{S}_{z,\omega}^{(2)}(w)&=-\frac{\alpha}{v_F}\frac{qlN_0}{2}\left[\tilde{F}_{3,\omega}*\mathcal{E}_{\omega}^{(2)}\right](w)+\frac{1}{2}\left[\tilde{F}_{3,\omega}*\tilde{S}_{z,\omega}^{(2)}\right](w)+X_{S,\omega}(w),\label{eq_IE_TDsz2}
    \end{align}
\end{subequations}
where
\begin{subequations}
    \begin{align}
        X_{n,\omega}(w)&:=-\frac{q\tau_0j_0^2}{mv_F\sigma_0^2}\left\{\int_{-\infty}^{\infty}\frac{\mathrm{d}\omega'}{\sqrt{2\pi}}\left[1-i(\omega-\omega')\tau_0\right]\frac{j_{0,\omega-\omega'}}{j_0}\frac{j_{0,\omega'}}{j_0}\right\}\cdot w\tilde{F}_{1,\omega}(w),\\
        X_{j,\omega}(w)&:=-\frac{mq^3\ell^2j_0^2}{4\pi^2\sigma_0^2}\cdot2\left\{\int_{-\infty}^{\infty}\frac{\mathrm{d}\omega'}{\sqrt{2\pi}}\left[1-i(\omega-\omega')\tau_0\right]\frac{j_{0,\omega-\omega'}}{j_0}\frac{j_{0,\omega'}}{j_0}\right\}\left[\frac{\tilde{F}_{3,\omega}(w)}{1-i\omega\tau_0}+w\tilde{F}_{2,\omega}(w)\right],\\
        X_{S,\omega}(w)&/\left(\frac{\alpha}{v_F}\frac{mq^2\ell\tau_0j_0^2}{4\pi^2\sigma_0^2}\right)\notag\\
    &:=\Theta(w)\int_{-\infty}^{\infty}\frac{\mathrm{d}\omega'}{\sqrt{2\pi}}\frac{1-i(\omega-\omega')\tau_0}{-i(\omega-\omega')\tau_0}\frac{j_{0,\omega-\omega'}}{j_0}\frac{j_{0,\omega'}}{j_0}\left\{\frac{-\frac{7}{2}+3i\omega'\tau_0}{1-\frac{3}{2}i\omega'\tau_0}\left[-\tilde{F}_{5,\omega}(w)+\tilde{F}_{5,\omega'}(w)\right]\right.\notag\\
    &\qquad\qquad\qquad\qquad\qquad\qquad\left.+\frac{3-3i\omega'\tau_0}{1-\frac{3}{2}i\omega'\tau_0}\left[-(1-i\omega\tau_0)w\tilde{F}_{2,\omega}(w)-\frac{1-i\omega\tau_0}{-i(\omega-\omega')\tau_0}\left(\tilde{F}_{3,\omega}(w)-\tilde{F}_{3,\omega'}(w)\right)\right]\right\}\notag\\
    &\quad+\left\{\int_{-\infty}^{\infty}\frac{\mathrm{d}\omega'}{\sqrt{2\pi}}\left[1-i(\omega-\omega')\tau_0\right]\frac{j_{0,\omega-\omega'}}{j_0}\frac{j_{0,\omega'}}{j_0}\right\}\left[\frac{\tilde{F}_{3,\omega}(w)-\tilde{F}_{5,\omega}(w)}{1-i\omega\tau_0}+w\tilde{F}_{2,\omega}(w)+w\tilde{F}_{4,\omega}(w)\right]\notag\\
    &\quad+\left\{\int_{-\infty}^{\infty}\frac{\mathrm{d}\omega'}{\sqrt{2\pi}}\frac{1-i(\omega-\omega')\tau_0}{1-\frac{3}{2}i\omega'\tau_0}\frac{j_{0,\omega-\omega'}}{j_0}\frac{j_{0,\omega'}}{j_0}\right\}\left[\frac{\tilde{F}_{3,\omega}(w)-\frac{3}{2}\tilde{F}_{5,\omega}(w)}{1-i\omega\tau_0}+w\tilde{F}_{2,\omega}(w)\right].
    \end{align}
\end{subequations}
Note that in the above equations, only the leading terms of $X_{n,\omega}(w)$, $X_{j,\omega}(w)$ and $X_{S,\omega}(w)$ under the assumption $\lambda_{\mathrm{TF}}/\ell\ll1$ and $\alpha/v_F\ll1$ are shown.
With Eq.~\eqref{eq_ac_convolutionE2} and Eq.~\eqref{eq_ac_convolutionj2}, we obtain $\left[\tilde{F}_{2,\omega}*\mathcal{E}_{\omega}^{(2)}\right](w)\sim X_{n,\omega}(w)$ due to $\tilde{n}_{\omega}^{(2)}(w)\sim O[\lambda_{\mathrm{TF}}/\ell]$, and then $\tilde{j}_{e,\omega}^{(2)}(w)=0$ holds.
As a result, Eq.~\eqref{eq_IE_TDsz2} becomes
\begin{align}
     \tilde{S}_{z,\omega}^{(2)}(w)=\frac{1}{2}\left[\tilde{F}_{3,\omega}*\tilde{S}_{z,\omega}^{(2)}\right](w)+X_{\omega}(w)\label{eq_IE_TDsz2_v2}
\end{align}
where
\begin{align}
    X_{\omega}(w)=\frac{\alpha}{v_F}\frac{1}{qv_F}X_{j,\omega}(w)+X_{S,\omega}(w),
\end{align}
which can be written explicitly as
\begin{align}
    X_{\omega}(w)&/\left(\frac{\alpha}{v_F}\frac{mq^2\ell\tau_0j_0^2}{4\pi^2\sigma_0^2}\right)\notag\\
    =&\Theta(w)\int_{-\infty}^{\infty}\frac{\mathrm{d}\omega'}{\sqrt{2\pi}}\frac{1-i(\omega-\omega')\tau_0}{-i(\omega-\omega')\tau_0}\frac{j_{0,\omega-\omega'}}{j_0}\frac{j_{0,\omega'}}{j_0}\left\{\frac{-\frac{7}{2}+3i\omega'\tau_0}{1-\frac{3}{2}i\omega'\tau_0}\left[-\tilde{F}_{5,\omega}(w)+\tilde{F}_{5,\omega'}(w)\right]\right.\notag\\
    &\qquad\qquad\qquad\qquad\qquad\qquad\left.+\frac{3-3i\omega'\tau_0}{1-\frac{3}{2}i\omega'\tau_0}\left[-(1-i\omega\tau_0)w\tilde{F}_{2,\omega}(w)-\frac{1-i\omega\tau_0}{-i(\omega-\omega')\tau_0}\left(\tilde{F}_{3,\omega}(w)-\tilde{F}_{3,\omega'}(w)\right)\right]\right\}\notag\\
    &+\left\{\int_{-\infty}^{\infty}\frac{\mathrm{d}\omega'}{\sqrt{2\pi}}\left[1-i(\omega-\omega')\tau_0\right]\frac{j_{0,\omega-\omega'}}{j_0}\frac{j_{0,\omega'}}{j_0}\right\}\left[\frac{-\tilde{F}_{3,\omega}(w)-\tilde{F}_{5,\omega}(w)}{1-i\omega\tau_0}-w\tilde{F}_{2,\omega}(w)+w\tilde{F}_{4,\omega}(w)\right]\notag\\
    &+\left\{\int_{-\infty}^{\infty}\frac{\mathrm{d}\omega'}{\sqrt{2\pi}}\frac{1-i(\omega-\omega')\tau_0}{1-\frac{3}{2}i\omega'\tau_0}\frac{j_{0,\omega-\omega'}}{j_0}\frac{j_{0,\omega'}}{j_0}\right\}\left[\frac{\tilde{F}_{3,\omega}(w)-\frac{3}{2}\tilde{F}_{5,\omega}(w)}{1-i\omega\tau_0}+w\tilde{F}_{2,\omega}(w)\right].\label{eq_Def_X}
\end{align}
\underline{case (0): $j_{0,\omega}=\sqrt{\frac{\pi}{2}}j_0[\delta(\omega-\omega_0)+\delta(\omega+\omega_0)]$}

In this case, $X_{\omega}(w)$ can be expressed as the sum of terms proportional to $\delta(\omega-2\omega_0)$, $\delta(\omega)$, $\delta(\omega+2\omega_0)$, and so can $\tilde{S}_{z,\omega}^{(2)}(w)$ due to Eq.~\eqref{eq_IE_TDsz2_v2}.
In the following, we calculate the term proportional to $\delta(\omega)$ of $\tilde{S}_{z,\omega}^{(2)}(w)$, $\tilde{S}_{z,\omega}^{(2,\textrm{i})}(w)$, which remains even after time averaging, and denote the term proportional to $\delta(\omega)$ of $X_{\omega}(w)$ as $X_{\omega}^{(\textrm{i})}(w)$.
Then, the equation to be solved is 
\begin{align}
    \tilde{S}_{z}^{(2,\textrm{i})}(t,w)=\frac{1}{2}\left[\tilde{F}_{3,0}*\tilde{S}_{z,t}^{(2,\textrm{i})}\right](w)+X^{(\textrm{i})}(t,w).\label{eq_IE_sz2fin}
\end{align}
As for $X^{(\textrm{i})}(t,w)$, we obtain the time-independent real-valued function,
\begin{align}
    X^{(\textrm{i})}&(t,w)/\left(\frac{\alpha}{v_F}\frac{mq^2\ell\tau_0j_0^2}{4\pi^2\sigma_0^2}\right)\notag\\
    =&\Theta(w)\frac{1+i\omega_0\tau_0}{4i\omega_0\tau_0}\left\{\frac{-\frac{7}{2}+3i\omega_0\tau_0}{1-\frac{3}{2}i\omega_0\tau_0}\left[-\tilde{F}_{5,0}(w)+\tilde{F}_{5,\omega_0}(w)\right]+\frac{3-3i\omega_0\tau_0}{1-\frac{3}{2}i\omega_0\tau_0}\left[-w\tilde{F}_{2,0}(w)-\frac{1}{i\omega_0\tau_0}\left(\tilde{F}_{3,0}(w)-\tilde{F}_{3,\omega_0}(w)\right)\right]\right\}\notag\\
    &+\frac{1+i\omega_0\tau_0}{4}\left[-\tilde{F}_{3,0}(w)-\tilde{F}_{5,0}(w)-w\tilde{F}_{2,0}(w)+w\tilde{F}_{4,0}(w)\right]+\frac{1+i\omega_0\tau_0}{4\left(1-\frac{3}{2}i\omega_0\tau_0\right)}\left[\tilde{F}_{3,0}(w)-\frac{3}{2}\tilde{F}_{5,0}(w)+w\tilde{F}_{2,0}(w)\right]\notag\\
    &+\textrm{c.c.},\label{eq_Xi_casei}
\end{align}
by using Eq.~\eqref{eq_Def_X}.
Equation \eqref{eq_IE_sz2fin} can be solved with the Fourier transformation, and its result is shown in Fig.~\ref{fig_casei}.
\underline{case (1): $\langle j_{0,\omega}\rangle=0,\,\langle j_{0,\omega}j_{0,\omega'}\rangle=\sqrt{2\pi} S_{j,\omega}\delta(\omega+\omega')$}

In this case, $X_{\omega}(w)$ is proportional to $\delta(\omega)$.
Thus, given $S_j(\omega)=S_j(-\omega)$, the inverse Fourier transformation of $X_{\omega}(w)$ becomes time-independent real-valued function and its form is
\begin{align}
    X(t,w)/\left(\frac{\alpha}{v_F}\frac{mq^2\ell\tau_0j_0^2}{4\pi^2\sigma_0^2}\right)=&\Theta(w)\int_{-\infty}^{\infty}\frac{\mathrm{d}\omega'}{\sqrt{2\pi}}\frac{S_j(\omega')}{j_0^2}\mathrm{Re}\left[\frac{1+i\omega'\tau_0}{i\omega'\tau_0}\left\{\frac{-\frac{7}{2}+3i\omega'\tau_0}{1-\frac{3}{2}i\omega'\tau_0}\left[-\tilde{F}_{5,0}(w)+\tilde{F}_{5,\omega'}(w)\right]\right.\right.\notag\\
    &\qquad\qquad\qquad\qquad\qquad\quad\left.\left.+\frac{3-3i\omega'\tau_0}{1-\frac{3}{2}i\omega'\tau_0}\left[-w\tilde{F}_{2,0}(w)-\frac{1}{i\omega'\tau_0}\left(\tilde{F}_{3,0}(w)-\tilde{F}_{3,\omega'}(w)\right)\right]\right\}\right]\notag\\
    &+\left[\int_{-\infty}^{\infty}\frac{\mathrm{d}\omega'}{\sqrt{2\pi}}\frac{S_j(\omega')}{j_0^2}\right]\cdot\left[-\tilde{F}_{3,0}(w)-\tilde{F}_{5,0}(w)-w\tilde{F}_{2,0}(w)+w\tilde{F}_{4,0}(w)\right]\notag\\
    &+\left[\int_{-\infty}^{\infty}\frac{\mathrm{d}\omega'}{\sqrt{2\pi}}\frac{1-\frac{3}{2}(\omega'\tau_0)^2}{1+\frac{9}{4}(\omega'\tau_0)^2}\frac{S_j(\omega')}{j_0^2}\right]\cdot\left[\tilde{F}_{3,0}(w)-\frac{3}{2}\tilde{F}_{5,0}(w)+w\tilde{F}_{2,0}(w)\right].
\end{align}
The equation to be solved is
\begin{align}
    \tilde{S}_{z}^{(2)}(t,w)=\frac{1}{2}\left[\tilde{F}_{3,0}*\tilde{S}_{z,t}^{(2)}\right](w)+X(t,w),
\end{align}
which can be solved by the Fourier transformation.
To illustrate the result, we assume $S_{j}(\omega)/j_0^2\rightarrow \bar{S}_j\cdot\chi_{[-\omega_c,\omega_c]}(\omega)$, where $\chi_A(x)$ denotes the indicator function of the set $A$.
This represents bond limited white noise with cutoff $\omega_c$.
In the main text, we set $\omega_c\tau_0=1$.

\section{Quadratic spin responses to the fluctuating electric field near the boundary}
\label{app: quadratic spin to fluctuating electric field}
\subsection{Introduction of Wiener-Hopf equation for the linear response}
First, we discuss the linear response as prerequisite.
The solution to the Boltmann equation,
\begin{align}
  \frac{f_{\bm{k}\gamma,\omega}^{(1)}(z)}{\tau_{\omega}}+v_{\bm{k}\gamma}^z\pdv{f_{\bm{k}\gamma,\omega}^{(1)}(z)}{z}-qE_{\omega}^{(1)}(z)v_{\bm{k}\gamma}^z\left(-\pdv{f_{\bm{k}\gamma}^{(0)}}{\varepsilon_{\bm{k}\gamma}}\right)=\frac{\tilde{n}_{\omega}^{(1)}(z)}{\tau_0N_0}\left(-\pdv{f_{\bm{k}\gamma}^{(0)}}{\varepsilon_{\bm{k}\gamma}}\right)+\frac{2\sigma_{\bm{k}\gamma}^z\tilde{S}_{z,\omega}^{(1)}(z)}{\tau_0N_0}\left(-\pdv{f_{\bm{k}\gamma}^{(0)}}{\varepsilon_{\bm{k}\gamma}}\right),
\end{align}
is obtained as follows:
\begin{equation}
    \begin{dcases}
        \underline{\mathrm{For}\,v_{\bm{k}\gamma}^z>0:}\\
        f_{\bm{k}\gamma,\omega}^{(1)}(z)=\int_{0}^{z}\mathrm{d}z'\exp{-\frac{z-z'}{v_{\bm{k}\gamma}^z\tau_{\omega}}}\left(qE_{\omega}^{(1)}(z')+\frac{\tilde{n}_{\omega}^{(1)}(z')}{v_{\bm{k}\gamma}^z\tau_0N_0}+\frac{2\sigma_{\bm{k}\gamma}^z\tilde{S}_{z,\omega}^{(1)}(z')}{v_{\bm{k}\gamma}^z\tau_0N_0}\right)\left(-\frac{\partial f_{\bm{k}\gamma}^{(0)}}{\partial \varepsilon_{\bm{k}\gamma}}\right)\\
        \qquad\qquad\,\,+C_{\bm{k}\gamma,\omega}^{(1)}\exp{-\frac{z}{v_{\bm{k}\gamma}^z\tau_{\omega}}},\\
        \underline{\mathrm{For}\,v_{\bm{k}\gamma}^z<0:}\\
        f_{\bm{k}\gamma,\omega}^{(1)}(z)=\int_{+\infty}^{z}\mathrm{d}z'\exp{-\frac{z-z'}{v_{\bm{k}\gamma}^z\tau_{\omega}}}\left(qE_{\omega}^{(1)}(z')+\frac{\tilde{n}_{\omega}^{(1)}(z')}{v_{\bm{k}\gamma}^z\tau_0N_0}+\frac{2\sigma_{\bm{k}\gamma}^z\tilde{S}_{z,\omega}^{(1)}(z')}{v_{\bm{k}\gamma}^z\tau_0N_0}\right)\left(-\frac{\partial f_{\bm{k}\gamma}^{(0)}}{\partial \varepsilon_{\bm{k}\gamma}}\right).
    \end{dcases}
\label{eq_Boltzmann_linear_solution}
\end{equation}
The coefficient $C_{\bm{k}\gamma,\omega}^{(1)}$ is determined by the Ziman's description~\eqref{eq_Ziman_z=0} and \eqref{eq_Ziman_scatteringrate}.
In this paper, we assume $p=0$, i.e.,
\begin{equation}
  f_{\bm{k}\gamma|v_{\bm{k}\gamma}^z>0,\omega}^{(1)}(z=0)=\frac{1}{\Omega}\sum_{\bm{k'}\gamma'|v_{\bm{k'}\gamma'}^z<0}f_{\bm{k'}\gamma',\omega}^{(1)}(z=0)\frac{2\delta(\varepsilon_{\bm{k}\gamma}-\varepsilon_{\bm{k'}\gamma'})}{N(\varepsilon_{\bm{k}\gamma})}.
\end{equation}
Multiplying Eq.~\eqref{eq_Boltzmann_linear_solution} by $\delta(\varepsilon-\varepsilon_{\bm{k}\gamma})$, and then taking the sum $\frac{1}{\Omega}\sum_{\bm{k}\gamma}$, we obtain the following self-consistent equation:
\begin{align}
  \tilde{n}_{\omega}^{(1)}(w)=\frac{\ell N_0}{2}&\left\{\int_{0}^{\infty}\mathrm{d}w'\left[\tilde{F}_{2,\omega}(w-w')qE_{\omega}^{(1)}(w')+\tilde{F}_{1,\omega}(w-w')\frac{\tilde{n}_{\omega}^{(1)}(w')}{\ell N_0}+\tilde{F}_{2,\omega}(w-w')\frac{-4\alpha}{v_F}\frac{\tilde{S}_{\omega}^{(1)}(w')}{\ell N_0}\right]\right.\notag\\
  &\quad\left.-F_{2,\omega}(w)\int_{0}^{\infty}\mathrm{d}w'\left[\tilde{F}_{2,\omega}(w')qE_{\omega}^{(1)}(w')-\tilde{F}_{1,\omega}(w')\frac{\tilde{n}_{\omega}^{(1)}(w')}{\ell N_0}+\tilde{F}_{2,\omega}(w-w')\frac{-4\alpha}{v_F}\frac{\tilde{S}_{\omega}^{(1)}(w')}{\ell N_0}\right]\right\}.\label{eq_n1_original}
\end{align}
By substituting $w=0$ into Eq.~\eqref{eq_n1_original}, we obtain
\begin{align}
    \int_{0}^{\infty}\mathrm{d}w'\left[\tilde{F}_{2,\omega}(w')qE_{\omega}^{(1)}(w')-\tilde{F}_{1,\omega}(w')\frac{\tilde{n}_{\omega}^{(1)}(w')}{\ell N_0}+\tilde{F}_{2,\omega}(w')\frac{-4\alpha}{v_F}\frac{\tilde{S}_{\omega}^{(1)}(w')}{\ell N_0}\right]=-\frac{\tilde{n}_{\omega}^{(1)}(w=0)}{\ell N_0}.\label{eq_n1_w=0}
\end{align}
Using equations \eqref{eq_n1_original}, \eqref{eq_n1_w=0} and the Gauss's law, we obtain 
\begin{align}
  2\frac{\tilde{n}_{\omega}^{(1)}(w)}{N_0}=&F_{2,\omega}(w)\frac{\tilde{n}_{\omega}^{(1)}(0)}{N_0}-F_{3,\omega}(w)\frac{qE_{\omega}^{(1)}(0)\ell}{1-i\omega\tau_0}\notag\\
  &+\int_{0}^{\infty}\mathrm{d}w'\left[-\frac{1}{1-i\omega\tau_0}\left(\frac{\ell}{\lambda_{\mathrm{TF}}}\right)^2\tilde{F}_{3,\omega}(w-w')+\tilde{F}_{1,\omega}(w-w')\right]\frac{\tilde{n}_{\omega}^{(1)}(w')}{N_0},\label{eq_n1_WienerHopf}
\end{align}
for the leading order in $\alpha/v_F$.
This equation is a typical Wiener-Hopf equation, but solving it directly is challenging due to the presence of the special functions $F_{n=1,3,\omega}(w)$ in the kernel.
Note that $F_{1,\omega=0}(w)$ corresponds to incomplete gamma function $\Gamma(0,w)$ .
To address this, first, we approximate $\Gamma(0,w)$ by a sum of exponential functions.
A similar approach using exponential approximations has been considered in previous study~\cite{Pipkin1991}.

\subsection{Approximation of the incomplete gamma function}
We approximate $\Gamma(0,w)$ by the following procedure:
\begin{align}
  &\Gamma(0,w)=\int_{w}^{\infty}dt\frac{e^{-t}}{t}=\int_{1}^{\infty}dt'\frac{e^{-t'w}}{t'}=\int_{\Omega_C}^{\infty}du(1+e^{-u})e^{-\phi(u)w}\sim\int_{\Omega_C}^{U_{\mathrm{max}}}du(1+e^{-u})e^{-\phi(u)w}\notag\\
  &\rightarrow\,\text{Using  Gaussian quadrature on }[\Omega_C,U_{\mathrm{max}}]\text{, we compute }N\text{ pairs }(u_k, v_k)\text{ of nodes and weights}.\notag\\
  &\rightarrow\,\sum_{k=1}^{N}b_ke^{-a_k w}\qquad(a_k:=\phi(u_k),\,b_k:=v_k(1+e^{-u_k}),\,a_1<a_2<\cdots<a_N),\label{eq_IncompleteGammaFunctionApproximation}
\end{align}
where $\phi(u):=\exp[u-\exp(-u)]$ and $\Omega_C$ denotes omega constant.
Here, we introduce the double exponential transformation through $\phi(u)$ to improve numerical accuracy~\cite{TakahasiMori1974}.
Since $\Gamma(0,w)\sim-\gamma-\ln(w)\,[\gamma:\text{Euler's constant}]$ holds as $w\rightarrow0$, this approximation can be regarded as neglecting the logarithmic divergence at $w=0$.
In this paper, we set $U_{\mathrm{max}}=20$ and $N=1000$.
Based on Eq.~\eqref{eq_IncompleteGammaFunctionApproximation}, we define the functions $F_{n,\omega}^{(\text{app})}(w)$ as
\begin{align}
    F_{n,\omega}(w)\sim F_{n,\omega}^{(\text{app})}(w):=\sum_{k=1}^{N}\frac{b_k}{(a_k)^{n-1}}e^{-a_k(1-i\omega\tau_0)w},\label{eq_FnApproximation}
\end{align}
to satisfy the relation
\begin{align}
    F_{n+1,\omega}(w)=(1-i\omega\tau_0)\int_{w}^{\infty}F_{n,\omega}(w')\mathrm{d}w'.\label{eq_Fn+1Fn_relation}
\end{align}

Then, we discuss the validity of this approximation.
In the beginning, we point out that the continuity equation for charge conservation law $-i\omega q\tilde{n}_{\omega}^{(1)}(w)+\partial_wj_e^{(1)}(w)/\ell=0$ is automatically satisfied when Eq.~\eqref{eq_n1_WienerHopf}, \eqref{eq_Fn+1Fn_relation} and $F_{n+1,\omega}(w=0)=1/n$ hold.
This is confirmed by substituting Eq.~\eqref{eq_n1_original} and
\begin{align}
  j_{e,\omega}^{(1)}(w)=qv_F\frac{\ell N_0}{2}&\left\{\int_{0}^{\infty}\mathrm{d}w'\left[\tilde{F}_{3,\omega}(w-w')qE_{\omega}^{(1)}(w')+\tilde{F}_{2,\omega}(w-w')\frac{\tilde{n}_{\omega}^{(1)}(w')}{\ell N_0}\right]\right.\notag\\
  &\quad\left.-F_{3,\omega}(w)\int_{0}^{\infty}\mathrm{d}w'\left[\tilde{F}_{2,\omega}(w')qE_{\omega}^{(1)}(w')-\tilde{F}_{1,\omega}(w')\frac{\tilde{n}_{\omega}^{(1)}(w')}{\ell N_0}\right]\right\},\label{eq_je1}
\end{align}
which is obtained by multiplying Eq.~\eqref{eq_Boltzmann_linear_solution} by $qv_{\bm{k}\gamma}\delta(\varepsilon-\varepsilon_{\bm{k}\gamma})$ and then taking the sum $\frac{1}{\Omega}\sum_{\bm{k}\gamma}$, into the continuity equation.
Under the approximation \eqref{eq_FnApproximation}, however, it is not trivial whether $F_{n+1,\omega}^{(\text{app})}(w=0)=1/n$ holds.
With our exponential approximation, this can be verified with sufficient accuracy as follows:
\begin{equation}
  \left|\sum_{k=1}^{N}\frac{b_k}{a_k}-1\right|\sim2.1\times10^{-9},\quad\left|\sum_{k=1}^{N}\frac{b_k}{(a_k)^2}-\frac{1}{2}\right|<O(10^{-14}),\quad\left|\sum_{k=1}^{N}\frac{b_k}{(a_k)^3}-\frac{1}{3}\right|<O(10^{-14}).
\end{equation}
Therefore, we conclude that the solution to Eq.~\eqref{eq_n1_WienerHopf} satisfies the charge conservation law even within the approximation~\eqref{eq_FnApproximation}.
The question then arises whether the solution of the Wiener-Hopf equation itself is altered by this approximation.
We address this issue in the following.

\begin{figure}
    \centering
    \includegraphics[width=0.9\linewidth]{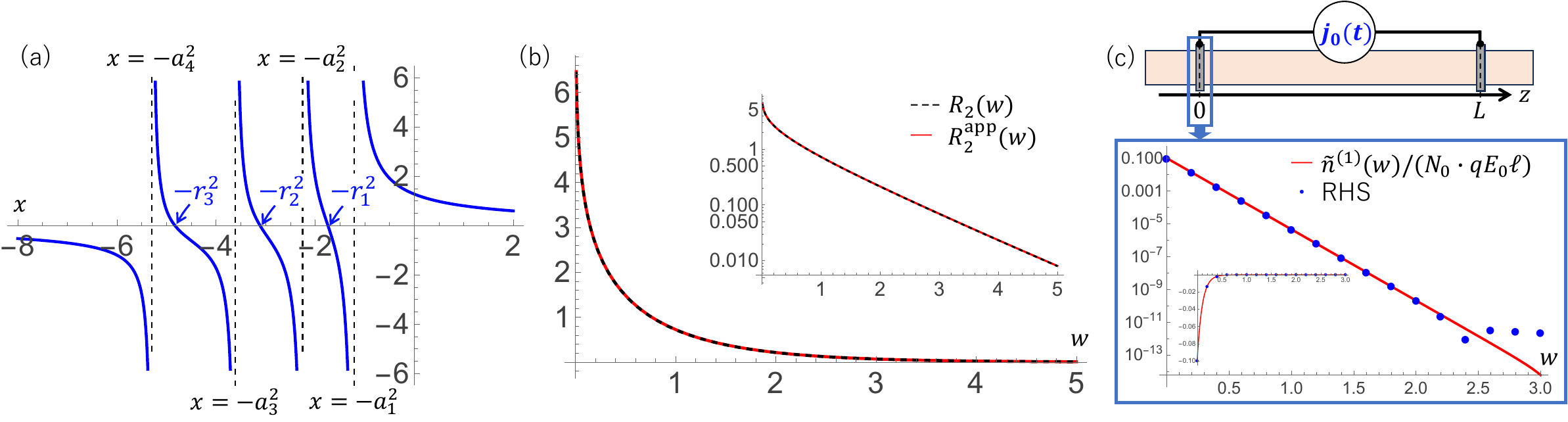}
    \caption{(a) A schematic plot illustrating the qualitative features of $\sum_{k=1}^{N}b_k/a_k(x+a_k^2)$, not a direct plot of the computed data \eqref{eq_IncompleteGammaFunctionApproximation}.
    (b) The red solid and black dashed lines respectively show $R_{2}^{\mathrm{app}}(w)$ in Eq.~\eqref{eq_R2_rkDef} and $R_{2}(w)$ in Eq.~\eqref{eq_R2Def}.
    The inset shows the same graph on a logarithmic vertical scale for clarity.
    (c) The red line shows the spatial dependence of $\tilde{n}^{(1)}(w)$ in Eq.~\eqref{eq_n1_iii} for $\lambda_{\mathrm{TF}}/\ell=10^{-1}$.
    The blue circles show the result of substituting $\tilde{n}^{(1)}(w)$ in Eq.~\eqref{eq_n1_iii} into the RHS of Eq.~\eqref{eq_n1_WienerHopf}.
    The absolute value of $\tilde{n}^{(1)}(w)$ is used in the logarithmic plot.
    The inset shows the same graph on a linear vertical scale for reference.}
    \label{fig_Caseiii_prerequisite}
\end{figure}

In this paragraph, we consider the equations exactly solved in our precious work~\cite{Yoshimi2026},
\begin{subequations}
    \begin{align}
        \tilde{n}^{(1,\mathrm{pre})}(w) =&  \frac{3j_0}{2qv_{\mathrm{F}}} \tilde{F}_3(w) - \frac{1}{2} \left(\frac{\ell}{\lambda_{\mathrm{TF}}}\right)^2[\tilde{F}_3 * \tilde{n}^{(1,\mathrm{pre})}](w) + \frac{1}{2} [\tilde{F}_1 * \tilde{n}^{(1,\mathrm{pre})}](w),\label{eq_n1_previouswork}\\
        \tilde{n}^{(2,\mathrm{pre})}(w) =&-\frac{3m\tau_0j_0\mathcal{E}_0}{2\pi^2N_0}\int_{-\infty}^{\infty}\mathrm{d}w'\left[\Theta(w')-\frac{\text{sgn}(w')}{2}e^{-\frac{\ell}{\lambda_{\mathrm{TF}}}w'}\right][\tilde{F}_1(w-w')-(w-w')\tilde{F}_0(w-w')]\notag\\
        & - \frac{1}{2} \left(\frac{\ell}{\lambda_{\mathrm{TF}}}\right)^2 [\tilde{F}_3 * \tilde{n}^{(2,\mathrm{pre})}](w) + \frac{1}{2} [\tilde{F}_1 * \tilde{n}^{(2,\mathrm{pre})}](w),\label{eq_n2_previouswork}
\end{align}
\end{subequations}
instead of the Wiener-Hopf equation~\eqref{eq_n1_WienerHopf}.
As discussed above, this approximation neglects the logarithmic divergence at the origin.
Since the integration domain includes $w'=w$ in both formulations, it is meaningful to assess the validity of the approximation using Eq.~\eqref{eq_n1_previouswork},~\eqref{eq_n2_previouswork}.
At the first or second order, using Fourier transform, we can obtain the analytic solution, 
\begin{align}
    \tilde{n}_{\mathrm{app}}^{(1,\mathrm{pre})}(w)=&\frac{3j_0\lambda_{\text{TF}}}{2qv_{\mathrm{F}}\ell}e^{-\ell|w|/\lambda_{\text{TF}}},\quad\tilde{n}_{\mathrm{app}}^{(2,\mathrm{pre})}(w)=\frac{3m\tau_0j_0\mathcal{E}_0}{4\pi^3N_0v_F}[R_1*R_2^{\mathrm{app}}](w)
\end{align}
where
\begin{align}
  R_{2}^{\mathrm{app}}(w)=\sqrt{2\pi}\left(\sum_{k=1}^{N}a_k e^{-a_k|w|}-\sum_{k=1}^{N-1}r_ke^{-r_k|w|}\right),\quad\sum_{k=1}^{N}\frac{b_k}{a_k(a_k^2+s^2)}=:\frac{\displaystyle\prod_{k=1}^{N-1}(s^2+r_k^2)}{\displaystyle\prod_{k=1}^{N}(s^2+a_k^2)}.\label{eq_R2_rkDef}
\end{align}
Note that $a_1<r_1<a_2<r_2<\cdots<r_{N-1}<a_N$ holds as shown in Fig.~\ref{fig_Caseiii_prerequisite} (a).
Therefore, $r_k$ can be determined numerically by bisection method.
The exact solutions in our previous work~\cite{Yoshimi2026} are
\begin{align}
    \tilde{n}^{(1,\mathrm{pre})}(w)=&\frac{3j_0\lambda_{\text{TF}}}{2qv_{\mathrm{F}}\ell}e^{-\ell|w|/\lambda_{\text{TF}}},\quad\tilde{n}^{(2,\mathrm{pre})}(w)=\frac{3m\tau_0j_0\mathcal{E}_0}{4\pi^3N_0v_F}[R_1*R_2](w)
\end{align}
where
\begin{align}
    R_2(w)=\sqrt{\frac{\pi}{2}}\int_{1}^{\infty}\mathrm{d}x\frac{x^3 e^{-x|w|}}{(x^2-1)\left[\left\{x+\frac{1}{2}\ln\left(\frac{x-1}{x+1}\right)\right\}^2+\frac{\pi^2}{4}\right]}.\label{eq_R2Def}
\end{align}
Figure \ref{fig_Caseiii_prerequisite} (b) shows $R_2^{\mathrm{app}}(w)$ and $R_2(w)$, which agree well with each other.
Therefore, we conclude that the solutions of the self-consistent equations under the approximation coincide with the exact solutions.

\subsection{Solution of Wiener-Hopf equation for the linear response}
We then apply the approximation~\eqref{eq_FnApproximation} to Eq.~\eqref{eq_n1_WienerHopf}.
The solution can be obtained as
\begin{align}
    \frac{\tilde{n}_{\omega}^{(1)}(w)}{N_0}=-qE_{0,\omega}\ell&\left(\left\{\frac{C_{\omega}}{2}\sum_{m=1}^N\left[\frac{b_mY_m^{11}(\omega)}{a_m}\right]+\frac{1}{2}\sum_{m=1}^N\left[\frac{b_mY_m^{11}(\omega)}{a_m^2}\right]\right\}e^{-\ell w/\lambda_{\text{TF}}}\right.\notag\\
    &\left.+\left\{\frac{C_{\omega}}{2}\sum_{m=1}^N\sum_{n=1}^{N-1}\left[\frac{b_mY_{mn}^{12}(\omega)}{a_m}e^{-r_n(1-i\omega\tau_0)w}\right]+\frac{1}{2}\sum_{m=1}^N\sum_{n=1}^{N-1}\left[\frac{b_mY_{mn}^{12}(\omega)}{a_m^2}e^{-r_n(1-i\omega\tau_0)w}\right]\right\}\right),\label{eq_n1_iii}
\end{align}
where
\begin{align}
    Y_m^{11}(\omega)&:=\frac{\displaystyle\prod_{k=1}^{N}(a_m+a_k)}{\displaystyle\left[a_m+\frac{1}{1-i\omega\tau_0}\frac{\ell}{\lambda_{\mathrm{TF}}}\right]\prod_{k=1}^{N-1}(a_m+r_k)}\cdot\frac{\displaystyle\prod_{\substack{k=1 \\ k\ne m}}^{N}\left[\frac{\ell}{\lambda_{\mathrm{TF}}}-a_k(1-i\omega\tau_0)\right]}{\displaystyle\prod_{k=1}^{N-1}\left[\frac{\ell}{\lambda_{\mathrm{TF}}}-r_k(1-i\omega\tau_0)\right]},\\
     Y_{mn}^{12}(\omega)&:=\frac{\displaystyle\prod_{k=1}^{N}(a_m+a_k)}{\displaystyle\left[a_m+\frac{1}{1-i\omega\tau_0}\frac{\ell}{\lambda_{\mathrm{TF}}}\right]\prod_{k=1}^{N-1}(a_m+r_k)}\cdot\frac{\displaystyle\prod_{\substack{k=1 \\ k\ne m}}^{N}(-r_n+a_k)}{\displaystyle\left[-r_n+\frac{1}{1-i\omega\tau_0}\frac{\ell}{\lambda_{\mathrm{TF}}}\right]\prod_{\substack{k=1 \\ k\ne n}}^{N-1}(-r_n+r_k)}
\end{align}
and
\begin{align}
   C_{\omega}:=\frac{\displaystyle\frac{1}{2}\sum_{m=1}^N\left[\frac{b_mY_m^{11}(\omega)}{a_m^2}\right]+\frac{1}{2}\sum_{m=1}^N\sum_{n=1}^{N-1}\left[\frac{b_mY_{mn}^{12}(\omega)}{a_m^2}\right]}{\displaystyle1-\frac{1}{2}\sum_{m=1}^N\left[\frac{b_mY_m^{11}(\omega)}{a_m}\right]-\frac{1}{2}\sum_{m=1}^N\sum_{n=1}^{N-1}\left[\frac{b_mY_{mn}^{12}(\omega)}{a_m}\right]}.
\end{align}

\begin{figure}
    \centering
    \includegraphics[width=0.9\linewidth]{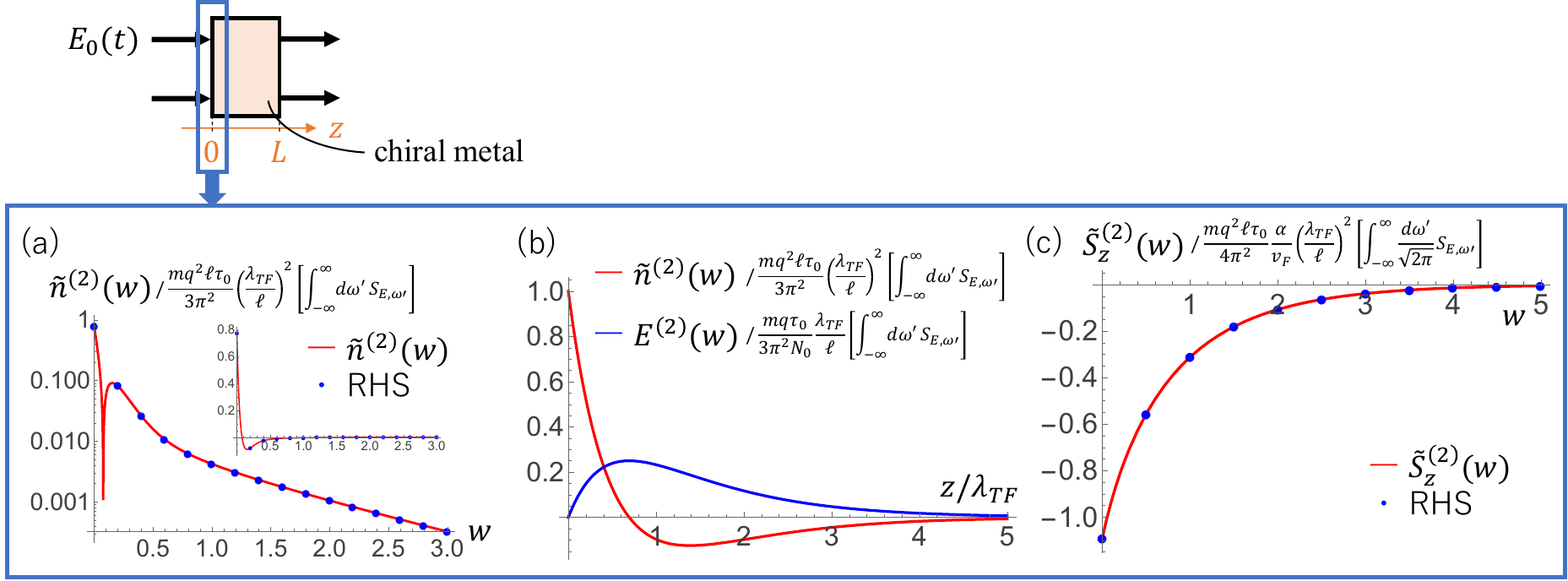}
    \caption{Enlarged view near the boundary $z=0$ for the case (2).
    (a) The red line shows the spatial dependence of $\tilde{n}^{(2)}(w)$ in Eq.~\eqref{eq_n2_iii_solutionfin} for $\lambda_{\mathrm{TF}}/\ell=10^{-1}$.
    The blue circles show the results of substituting $\tilde{n}^{(2)}(w)$ in Eq.~\eqref{eq_n2_iii_solutionfin} into the RHS of Eq.~\eqref{eq_n2_WienerHopf}.
    The absolute value of $\tilde{n}^{(2)}(w)$ is used in the logarithmic plot.
    The inset shows the same graph on a linear vertical scale for reference.
    (b) The red and blue lines show $\tilde{n}^{(2)}(w)$ in Eq.~\eqref{eq_n2_iii_solutionfin_leardingterm} and $E^{(2)}(w)$ in Eq.~\eqref{eq_E2_iii_solutionfin}. Note that the $x$-axis is scaled by the screening length $\lambda_{\mathrm{TF}}$.
    (c) The red line shows the spatial dependence of $\tilde{S}_{z}^{(2)}(w)$ as shown in Eq.~\eqref{eq_sz2_caseiii_fin}. The blue circles show the result of substituting $\tilde{S}_z^{(2)}(w)$ in Eq.~\eqref{eq_sz2_caseiii_fin} into the RHS of Eq.~\eqref{eq_sz2_WienerHopf}.}
    \label{fig_Caseiii}
\end{figure}
Note that we assume $\omega\tau_0\lesssim O[10^0]$ and $\lambda_{\mathrm{TF}}/\ell\ll1$.

Figure~\ref{fig_Caseiii_prerequisite} (c) compares Eq.~\eqref{eq_n1_iii} with the result obtained by substituting it into Eq.~\eqref{eq_n1_WienerHopf} and shows the validity of Eq.~\eqref{eq_n1_iii}.
By calculating $\int_{0}^{w}\mathrm{d}w'\frac{\tilde{n}_{\omega}^{(1)}(w')}{N_0}=q[E_{\omega}^{(1)}(w)-E_{\omega}^{(1)}(0)]l\left(\frac{\lambda_{\mathrm{TF}}}{l}\right)^2$, the electric field $E_{\omega}^{(1)}(w)$ is obtained as
\begin{align}
    E_{\omega}^{(1)}(w)/E_{0,\omega}=&\frac{\ell^2}{\lambda_{\mathrm{TF}}^2}\left(\left\{\frac{C_{\omega}}{2}\sum_{m=1}^N\left[\frac{b_mY_m^{11}(\omega)}{a_m}\right]+\frac{1}{2}\sum_{m=1}^N\left[\frac{b_mY_m^{11}(\omega)}{a_m^2}\right]\right\}\frac{\lambda_{\text{TF}}}{\ell}e^{-\ell w/\lambda_{\text{TF}}}\right.\notag\\
    &\qquad\quad\left.+\frac{C_{\omega}}{2}\sum_{m=1}^N\sum_{n=1}^{N-1}\left[\frac{b_mY_{mn}^{12}(\omega)}{a_mr_n}e^{-r_n(1-i\omega\tau_0)w}\right]+\frac{1}{2}\sum_{m=1}^N\sum_{n=1}^{N-1}\left[\frac{b_mY_{mn}^{12}(\omega)}{a_m^2r_n}e^{-r_n(1-i\omega\tau_0)w}\right]\right).
\end{align}
Here, we use the following numerically verified relations:
\begin{align}
    \frac{C_{\omega}}{2}\sum_{m=1}^N\left[\frac{b_mY_m^{11}(\omega)}{a_m}\right]+\frac{1}{2}\sum_{m=1}^N\left[\frac{b_mY_m^{11}(\omega)}{a_m^2}\right]\sim\frac{\lambda_{\text{TF}}}{\ell},\quad\frac{C_{\omega}}{2}\sum_{m=1}^N\sum_{n=1}^{N-1}\left[\frac{b_mY_{mn}^{12}(\omega)}{a_mr_n}\right]+\frac{1}{2}\sum_{m=1}^N\sum_{n=1}^{N-1}\left[\frac{b_mY_{mn}^{12}(\omega)}{a_m^2r_n}\right]\sim0.
\end{align}
Therefore, we finally obtain
\begin{align}
    \frac{\tilde{n}_{\omega}^{(1)}(w)}{N_0}=-qE_{0,\omega}\ell&\frac{\lambda_{\text{TF}}}{\ell}e^{-\ell w/\lambda_{\text{TF}}},\quad  E_{\omega}^{(1)}(w)=E_{0,\omega}e^{-\ell w/\lambda_{\text{TF}}}.\label{eq_n1E1_iii_fin}
\end{align}
Then, some terms in Eq.~\eqref{eq_Boltzmann_linear_solution} are simpified as
\begin{equation}
    \begin{dcases}
        \underline{\mathrm{For}\,v_{\bm{k}\gamma}^z>0:}\\
        \int_{0}^{z}\mathrm{d}z'\exp{-\frac{z-z'}{v_{\bm{k}\gamma}^z\tau_{\omega}}}\left(qE_{\omega}^{(1)}(z')+\frac{\tilde{n}_{\omega}^{(1)}(z')}{v_{\bm{k}\gamma}^z\tau_0N_0}\right)\left(-\frac{\partial f_{\bm{k}\gamma}^{(0)}}{\partial \varepsilon_{\bm{k}\gamma}}\right)+C_{\bm{k}\gamma,\omega}^{(1)}\exp{-\frac{z}{v_{\bm{k}\gamma}^z\tau_{\omega}}}\\
        \qquad\qquad\qquad\sim-qE_{0,\omega}\ell\left[\frac{\lambda_{\text{TF}}}{\ell}\left(1-\frac{i\omega\tau_0}{\cos\theta}\frac{\lambda_{\text{TF}}}{\ell}\right)e^{-\ell w/\lambda_{\text{TF}}}+\frac{i\omega\tau_0}{\cos\theta}\left(\frac{\lambda_{\text{TF}}}{\ell}\right)^2e^{-(1-i\omega\tau_0)w/\cos\theta}\right]\left(-\frac{\partial f_{\bm{k}\gamma}^{(0)}}{\partial \varepsilon_{\bm{k}\gamma}}\right),\\
        \underline{\mathrm{For}\,v_{\bm{k}\gamma}^z<0:}\\
        \int_{+\infty}^{z}\mathrm{d}z'\exp{-\frac{z-z'}{v_{\bm{k}\gamma}^z\tau_{\omega}}}\left(qE_{\omega}^{(1)}(z')+\frac{\tilde{n}_{\omega}^{(1)}(z')}{v_{\bm{k}\gamma}^z\tau_0N_0}\right)\left(-\frac{\partial f_{\bm{k}\gamma}^{(0)}}{\partial \varepsilon_{\bm{k}\gamma}}\right)\\
        \qquad\qquad\qquad\sim-qE_{0,\omega}\ell\frac{\lambda_{\text{TF}}}{\ell}\left(1-\frac{i\omega\tau_0}{\cos\theta}\frac{\lambda_{\text{TF}}}{\ell}\right)e^{-\ell w/\lambda_{\text{TF}}}\left(-\frac{\partial f_{\bm{k}\gamma}^{(0)}}{\partial \varepsilon_{\bm{k}\gamma}}\right).
    \end{dcases}
\end{equation}
Here, we take the leading terms in $\lambda_{\text{TF}}/\ell$ in each regime characterized by the screening length $\lambda_{\mathrm{TF}}$ and the mean free path $\ell$.
Multiplying Eq.~\eqref{eq_Boltzmann_linear_solution} by $\sigma_{\bm{k}\gamma}^z\delta(\varepsilon-\varepsilon_{\bm{k}\gamma})/2$, and then taking the sum $\frac{1}{\Omega}\sum_{\bm{k}\gamma}$, we obtain the self-consistent equation for the linear spin response:
\begin{align}
    \tilde{S}_{z,\omega}^{(1)}(w)=i\omega\tau_0\frac{m^2\alpha}{2\pi^2}qE_{0,\omega}\ell\left(\frac{\lambda_{\text{TF}}}{\ell}\right)^2\left[F_{2,\omega}(w)+2e^{-\ell w/\lambda_{\text{TF}}}\right]+\frac{1}{2}\int_0^\infty \mathrm{d}w'\tilde{F}_{3,\omega}(w-w')\tilde{S}_{z,\omega}^{(1)}(w').
\end{align}
Then, we can estimate $\tilde{S}_{z,\omega}^{(1)}(w)\sim(\lambda_{\text{TF}}/\ell)^2\cdot O[e^{-\ell w/\lambda_{\text{TF}}}]+(\lambda_{\text{TF}}/\ell)^2\cdot O[e^{-w}]$.
By substituting these results into Eq.~\eqref{eq_Boltzmann_linear_solution}, we obtain
\begin{align}
    f_{\bm{k}\gamma,\omega}^{(1)}(z)=-qE_{0,\omega}\ell\frac{\lambda_{\text{TF}}}{\ell}e^{-\ell w/\lambda_{\text{TF}}}\left(-\frac{\partial f_{\bm{k}\gamma}^{(0)}}{\partial \varepsilon_{\bm{k}\gamma}}\right)+\left(\frac{\lambda_{\text{TF}}}{\ell}\right)^2O[e^{-w}]\label{eq_f1_iii_fin}
\end{align}
for the leading order in $\alpha/v_F$ and $\lambda_{\text{TF}}/\ell$ in each regime characterized by the screening length and the mean free path.

\subsection{Introduction of Wiener-Hopf equation for the quadratic response and its solution}
We next discuss the quadratic response.
The solution to the Boltzmann equation,
\begin{align}
  \frac{f_{\bm{k}\gamma,\omega}^{(2)}(z)}{\tau_{\omega}}+v_{\bm{k}\gamma}^z\pdv{f_{\bm{k}\gamma,\omega}^{(2)}(z)}{z}+\pdv{I_{\bm{k}\gamma,\omega}(z)}{k_z}&-qE_{\omega}^{(2)}(z)v_{\bm{k}\gamma}^z\left(-\pdv{f_{\bm{k}\gamma}^{(0)}}{\varepsilon_{\bm{k}\gamma}}\right)=\frac{\tilde{n}_{\omega}^{(2)}(z)}{\tau_0N_0}\left(-\pdv{f_{\bm{k}\gamma}^{(0)}}{\varepsilon_{\bm{k}\gamma}}\right)+\frac{2\sigma_{\bm{k}\gamma}^z\tilde{S}_{z,\omega}^{(2)}(z)}{\tau_0N_0}\left(-\pdv{f_{\bm{k}\gamma}^{(0)}}{\varepsilon_{\bm{k}\gamma}}\right),
\end{align}
is obtained as
\begin{equation}
    \begin{dcases}
        \underline{\mathrm{For}\,v_{\bm{k}\gamma}^z>0:}\\
        f_{\bm{k}\gamma,\omega}^{(2)}(z)=\int_{0}^{z}\mathrm{d}z'\exp{-\frac{z-z'}{v_{\bm{k}\gamma}^z\tau_{\omega}}}\left[\left(qE_{\omega}^{(2)}(z')+\frac{\tilde{n}_{\omega}^{(2)}(z')}{v_{\bm{k}\gamma}^z\tau_0N_0}+\frac{2\sigma_{\bm{k}\gamma}^z\tilde{S}_{z,\omega}^{(2)}(z')}{v_{\bm{k}\gamma}^z\tau_0N_0}\right)\left(-\frac{\partial f_{\bm{k}\gamma}^{(0)}}{\partial \varepsilon_{\bm{k}\gamma}}\right)-\frac{1}{v^z_{\bm{k}\gamma}}\pdv{I_{\bm{k}\gamma,\omega}(z')}{k_z}\right]\\
        \qquad\qquad\,\,+C_{\bm{k}\gamma,\omega}^{(2)}\exp{-\frac{z}{v_{\bm{k}\gamma}^z\tau_{\omega}}},\\
        \underline{\mathrm{For}\,v_{\bm{k}\gamma}^z<0:}\\
        f_{\bm{k}\gamma,\omega}^{(2)}(z)=\int_{+\infty}^{z}\mathrm{d}z'\exp{-\frac{z-z'}{v_{\bm{k}\gamma}^z\tau_{\omega}}}\left[\left(qE_{\omega}^{(2)}(z')+\frac{\tilde{n}_{\omega}^{(2)}(z')}{v_{\bm{k}\gamma}^z\tau_0N_0}+\frac{2\sigma_{\bm{k}\gamma}^z\tilde{S}_{z,\omega}^{(2)}(z')}{v_{\bm{k}\gamma}^z\tau_0N_0}\right)\left(-\frac{\partial f_{\bm{k}\gamma}^{(0)}}{\partial \varepsilon_{\bm{k}\gamma}}\right)-\frac{1}{v^z_{\bm{k}\gamma}}\pdv{I_{\bm{k}\gamma,\omega}(z')}{k_z}\right]
    \end{dcases}
\label{eq_Boltzmann_quadratic_solution}
\end{equation}
where
\begin{equation}
  I_{\bm{k}\gamma,\omega}(z):=\int_{-\infty}^{+\infty}\frac{\mathrm{d}\omega'}{\sqrt{2\pi}}qE_{\omega-\omega'}^{(1)}(z)f^{(1)}_{\bm{k}\gamma,\omega'}(z).
\end{equation}
The coefficient $C_{\bm{k}\gamma,\omega}^{(2)}$ is determined by the Ziman's description,
\begin{equation}
  f_{\bm{k}\gamma|v_{\bm{k}\gamma}^z>0,\omega}^{(2)}(z=0)=\frac{1}{\Omega}\sum_{\bm{k'}\gamma'|v_{\bm{k'}\gamma'}^z<0}f_{\bm{k'}\gamma',\omega}^{(2)}(z=0)\frac{2\delta(\varepsilon_{\bm{k}\gamma}-\varepsilon_{\bm{k'}\gamma'})}{N(\varepsilon_{\bm{k}\gamma})}.
\end{equation}
In the same way as the case of the linear response, we obtain
\begin{align}
  \tilde{n}_{\omega}^{(2)}(z)=&\frac{N_0}{2}\int_{0}^{\infty}\mathrm{d}z'\left[\tilde{F}_{2,\omega}\left(\frac{z-z'}{\ell}\right)qE_{\omega}^{(2)}(z')+\tilde{F}_{1,\omega}\left(\frac{z-z'}{\ell}\right)\frac{\tilde{n}_{\omega}^{(2)}(z')}{\ell N_0}\right]+P_{\omega}(z)+\frac{\tilde{n}_{\omega}^{(2)}(z=0)}{2}F_{2,\omega}\left(\frac{z}{\ell}\right)\label{eq_n2_WienerHopf}
\end{align}
for the leading order in $\alpha/v_F$ where
\begin{align}
    P_{\omega}(z):=&-\sum_{\gamma}\int_{0}^{1}d(\cos\theta)\int_{-\infty}^{\infty}d\varepsilon\frac{k_{\gamma}^2(\varepsilon)}{4\pi^2v(\varepsilon)}\notag\\
    &\qquad\qquad\times\left\{\int_{0}^{z}\mathrm{d}z'\exp{-\frac{z-z'}{v(\varepsilon)\tau_{\omega}\cos\theta}}\left[\frac{\partial}{\partial\varepsilon}+\frac{1-\cos^2\theta}{v(\varepsilon)k_{\gamma}(\varepsilon)\cos\theta}\frac{\partial}{\partial(\cos\theta)}\right]I_{\gamma,\omega}(\varepsilon,\cos\theta,z')\right.\notag\\
  &\qquad\qquad\qquad\left.-\int_{z}^{\infty}\mathrm{d}z'\exp{-\frac{z'-z}{v(\varepsilon)\tau_{\omega}\cos\theta}}\left[\frac{\partial}{\partial\varepsilon}+\frac{1-\cos^2\theta}{v(\varepsilon)k_{\gamma}(\varepsilon)\cos\theta}\frac{\partial}{\partial(\cos\theta)}\right]I_{\gamma,\omega}(\varepsilon,-\cos\theta,z')\right\}\label{eq_P_Def}
\end{align}
Here, we abbreviate $I_{\gamma,\omega}(\varepsilon_{\bm{k}\gamma},\cos\theta,z)$ as $I_{\bm{k}\gamma,\omega}(z)$.
In the following, we focus on a specific form of $E_{0,\omega}$ as shown in Eq.~\eqref{eq_EDef_iii} to simplify the calculation.\\
\underline{case (2): $\langle E_{0,\omega}\rangle=0,\,\langle E_{0,\omega}E_{0,\omega'}\rangle=\sqrt{2\pi} S_{E,\omega}\delta(\omega+\omega')$}

In this case, using Eq.~\eqref{eq_f1_iii_fin}, $I_{\bm{k}\gamma,\omega}(z)$ becomes
\begin{align}
  I_{\bm{k}\gamma,\omega}(z)&=\int_{-\infty}^{+\infty}\frac{\mathrm{d}\omega'}{\sqrt{2\pi}}qE_{\omega-\omega'}^{(1)}(0)e^{-z/\lambda_{\mathrm{TF}}}\left\{-qE^{(1)}_{\omega'}(0)l\frac{\lambda_{\mathrm{TF}}}{\ell}e^{-z/\lambda_{\mathrm{TF}}}\left(-\frac{\partial f_{\bm{k}\gamma}^{(0)}}{\partial \varepsilon_{\bm{k}\gamma}}\right)+\left(\frac{\lambda_{\text{TF}}}{\ell}\right)^2O[e^{-z/\ell}]\right\}\notag\\
  &\sim\int_{-\infty}^{+\infty}\mathrm{d}\omega'\frac{-q^2\lambda_{\mathrm{TF}}}{\sqrt{2\pi}}\underbracket{E_{\omega-\omega'}^{(1)}(0)E_{\omega'}^{(1)}(0)}_{\rightsquigarrow\sqrt{2\pi}S_{E,\omega'}\delta(\omega)}e^{-2z/\lambda_{\mathrm{TF}}}\left(-\frac{\partial f_{\bm{k}\gamma}^{(0)}}{\partial \varepsilon_{\bm{k}\gamma}}\right)\notag\\
  &=-q^2\lambda_{\mathrm{TF}}\delta(\omega)\left[\int_{-\infty}^{+\infty}\mathrm{d}\omega'S_{E,\omega'}\right]e^{-2z/\lambda_{\mathrm{TF}}}\left(-\frac{\partial f_{\bm{k}\gamma}^{(0)}}{\partial \varepsilon_{\bm{k}\gamma}}\right).\label{eq_Ikgamma}
\end{align}
Here, we use $S_{E,\omega}=S_{E,-\omega}$ and neglect the second term in the first line because the factor $e^{-z/\lambda_{\mathrm{TF}}}$ of the electric field sets the scale of variation to be the screening length, not the mean free path.
Substituting Eq.~\eqref{eq_Ikgamma} into Eq.~\eqref{eq_P_Def}, we obtain
\begin{align}
    P_{\omega}(w)=\frac{mq^2\ell\tau_0}{2\pi^2}\left(\frac{\lambda_{\mathrm{TF}}}{\ell}\right)^2\delta(\omega)\left[\int_{-\infty}^{+\infty}\mathrm{d}\omega'S_{E,\omega'}\right]\left[-\frac{1}{2}\frac{e^{-w}}{1-\frac{1}{2}\frac{\lambda_{\mathrm{TF}}}{\ell}}+\frac{e^{-2\ell w/\lambda_{\mathrm{TF}}}}{1-\frac{1}{4}\frac{\lambda_{\mathrm{TF}}^2}{\ell^2}}\right].
\end{align}
Here, we use $F_2(w)+wF_1(w)=e^{-w}$.
The Wiener-Hopf equation~\eqref{eq_n2_WienerHopf} now can be solved and its solution is given by
\begin{align}
    \frac{\tilde{n}_{\omega}^{(2)}(w)}{N_0}=&\frac{\tilde{n}_{\omega}^{(2)}(0)}{N_0}\left\{\frac{1}{2}\sum_{m=1}^N\left[\frac{b_mY_m^{11}(\omega)}{a_m}\right]e^{-\ell w/\lambda_{\text{TF}}}+\frac{1}{2}\sum_{m=1}^N\sum_{n=1}^{N-1}\left[\frac{b_mY_{mn}^{12}(\omega)}{a_m}e^{-r_n(1-i\omega\tau_0)w}\right]\right\}\notag\\
    &+\frac{mq^2\ell\tau_0}{2\pi^2}\left(\frac{\lambda_{\mathrm{TF}}}{\ell}\right)^2\delta(\omega)\left[\int_{-\infty}^{+\infty}\mathrm{d}\omega'S_{E,\omega'}\right]U(w).\label{eq_n2_iii_solution}
\end{align}
Here, we define $U(w)$ as
\begin{align}
    U(w):=\frac{\displaystyle\prod_{k=1}^{N}\left[\frac{2\ell}{\lambda_{\mathrm{TF}}}+a_k\right]\cdot Q_1(w)}{\displaystyle\left[1-\frac{1}{4}\frac{\lambda_{\mathrm{TF}}^2}{\ell^2}\right]\frac{3\ell}{\lambda_{\mathrm{TF}}}\prod_{k=1}^{N-1}\left[\frac{2\ell}{\lambda_{\mathrm{TF}}}+r_k\right]}-\frac{\displaystyle\prod_{k=1}^{N}\left(a_k+1\right)\cdot Q_2(w)}{\displaystyle\left[2-\frac{\lambda_{\mathrm{TF}}}{\ell}\right]\left[\frac{\ell}{\lambda_{\mathrm{TF}}}+1\right]\prod_{k=1}^{N-1}\left(r_k+1\right)},
\end{align}
where
\begin{align}
    Q_1(w):=-\frac{\displaystyle\prod_{k=1}^{N}\left[\frac{\ell}{\lambda_{\mathrm{TF}}}-a_k\right]\cdot e^{-\frac{\ell w}{\lambda_{\mathrm{TF}}}}}{\displaystyle\frac{\ell}{\lambda_{\mathrm{TF}}}\prod_{k=1}^{N-1}\left[\frac{\ell}{\lambda_{\mathrm{TF}}}-r_k\right]}+\sum_{n=1}^{N-1}\frac{\displaystyle\prod_{k=1}^{N}(-r_n+a_k)\cdot e^{-r_nw}}{\displaystyle\left[-r_n+\frac{\ell}{\lambda_{\mathrm{TF}}}\right]\left[-r_n+\frac{2\ell}{\lambda_{\mathrm{TF}}}\right]\prod_{\substack{k=1 \\ k\ne n}}^{N-1}(-r_n+r_k)}+\frac{\displaystyle\prod_{k=1}^{N}\left[\frac{2\ell}{\lambda_{\mathrm{TF}}}-a_k\right]\cdot e^{-\frac{2\ell w}{\lambda_{\mathrm{TF}}}}}{\displaystyle\frac{\ell}{\lambda_{\mathrm{TF}}}\prod_{k=1}^{N-1}\left[\frac{2\ell}{\lambda_{\mathrm{TF}}}-r_k\right]}
\end{align}
and
\begin{align}
    Q_2(w):=\frac{\displaystyle\prod_{k=1}^{N}\left[\frac{\ell}{\lambda_{\mathrm{TF}}}-a_k\right]\cdot e^{-\frac{\ell w}{\lambda_{\mathrm{TF}}}}}{\displaystyle\left[\frac{\ell}{\lambda_{\mathrm{TF}}}-1\right]\prod_{k=1}^{N-1}\left[\frac{\ell}{\lambda_{\mathrm{TF}}}-r_k\right]}+\sum_{n=1}^{N-1}\frac{\displaystyle\prod_{k=1}^{N}(-r_n+a_k)\cdot e^{-r_nw}}{\displaystyle\left[-r_n+\frac{\ell}{\lambda_{\mathrm{TF}}}\right]\left[-r_n+1\right]\prod_{\substack{k=1 \\ k\ne n}}^{N-1}(-r_n+r_k)}+\frac{\displaystyle\prod_{k=1}^{N}\left(a_k-1\right)\cdot e^{-w}}{\displaystyle\left[\frac{\ell}{\lambda_{\mathrm{TF}}}-1\right]\prod_{k=1}^{N-1}\left(r_k-1\right)}.
\end{align}
Note that, in Eq.~\eqref{eq_f1_iii_fin}, we retain only the leading-order terms in $\lambda_{\mathrm{TF}}/\ell$.
Therefore, it is also necessary to retain only the leading-order terms in $\lambda_{\mathrm{TF}}/\ell$ for the functions in Eq.~\eqref{eq_n2_iii_solution} as follows:
\begin{align}
    &U(w)\sim\frac{2}{3}\left(-e^{-\ell w/\lambda_{\mathrm{TF}}}+2e^{-2\ell w/\lambda_{\mathrm{TF}}}\right),\notag\\
    &Y_m^{11}(\omega=0)\sim\frac{\lambda_{\mathrm{TF}}}{\ell}\frac{\displaystyle\prod_{k=1}^{N}(a_m+a_k)}{\displaystyle\prod_{k=1}^{N-1}(a_m+r_k)},\quad Y_{mn}^{12}(\omega=0)\sim\left(\frac{\lambda_{\mathrm{TF}}}{\ell}\right)^2\frac{\displaystyle\prod_{k=1}^{N}(a_m+a_k)}{\displaystyle\prod_{k=1}^{N-1}(a_m+r_k)}\cdot\frac{\displaystyle\prod_{\substack{k=1 \\ k\ne m}}^{N}(-r_n+a_k)}{\displaystyle\prod_{\substack{k=1 \\ k\ne n}}^{N-1}(-r_n+r_k)}.
\end{align}
Then, we finally obtain
\begin{align}
    \frac{\tilde{n}_{\omega}^{(2)}(w)}{N_0}=&\frac{mq^2\ell\tau_0}{2\pi^2N_0}\left(\frac{\lambda_{\mathrm{TF}}}{\ell}\right)^2\delta(\omega)\left[\int_{-\infty}^{+\infty}\mathrm{d}\omega'S_{E,\omega'}\right]\notag\\
    &\qquad\times\left\{\frac{D}{2}\sum_{m=1}^N\left[\frac{b_mY_m^{11}(0)}{a_m}\right]e^{-\ell w/\lambda_{\text{TF}}}+\frac{D}{2}\sum_{m=1}^N\sum_{n=1}^{N-1}\left[\frac{b_mY_{mn}^{12}(0)}{a_m}e^{-r_nw}\right]+U(w)\right\}\label{eq_n2_iii_solutionfin}\\
    &\sim\frac{mq^2\ell\tau_0}{2\pi^2N_0}\left(\frac{\lambda_{\mathrm{TF}}}{\ell}\right)^2\delta(\omega)\left[\int_{-\infty}^{+\infty}\mathrm{d}\omega'S_{E,\omega'}\right]U(w),\label{eq_n2_iii_solutionfin_leardingterm}
\end{align}
where
\begin{align}
    D:=\frac{U(0)}{\displaystyle1-\frac{1}{2}\sum_{m=1}^N\left[\frac{b_mY_m^{11}(0)}{a_m}\right]-\frac{1}{2}\sum_{m=1}^N\sum_{n=1}^{N-1}\left[\frac{b_mY_{mn}^{12}(0)}{a_m}\right]}\sim U(0).
\end{align}
Figure \ref{fig_Caseiii} (a) compares the unapproximated expression~\eqref{eq_n2_iii_solutionfin} with the result obtained by substituting it into Eq.~\eqref{eq_n2_WienerHopf} and shows the validity of Eq.~\eqref{eq_n2_iii_solutionfin}.
By calculating $\int_{0}^{w}\mathrm{d}w'\frac{\tilde{n}_{\omega}^{(2)}(w')}{N_0}=qE_{\omega}^{(2)}(w)l\left(\frac{\lambda_{\mathrm{TF}}}{l}\right)^2$, the electric field $E_{\omega}^{(2)}(w)$ is obtained as
\begin{align}
    E_{\omega}^{(2)}(w)=\frac{mq\tau_0}{3\pi^2N_0}\frac{\lambda_{\mathrm{TF}}}{\ell}\delta(\omega)\left[\int_{-\infty}^{+\infty}\mathrm{d}\omega'S_{E,\omega'}\right]\left(e^{-\ell w/\lambda_{\mathrm{TF}}}-e^{-2\ell w/\lambda_{\mathrm{TF}}}\right).\label{eq_E2_iii_solutionfin}
\end{align}
The approximated expressions Eq.~\eqref{eq_n2_iii_solutionfin_leardingterm} and Eq.~\eqref{eq_E2_iii_solutionfin} are plotted in Fig.~\ref{fig_Caseiii} (b).
Note that this expression is consistent with the boundary condition $E^{(2)}(w=0,t)=0$.
We also remark on $\mathcal{E}_{\omega}^{(2)}(z)$, which corresponds to the effective electric field when $\omega\tau_0=0$.
Equation~\eqref{eq_n2_WienerHopf} can be rearranged as
\begin{align}
  \tilde{n}_{\omega}^{(2)}(z)=&\frac{qN_0}{2}\int_{0}^{\infty}\mathrm{d}z'\tilde{F}_{2,\omega}\left(\frac{z-z'}{\ell}\right)\mathcal{E}_{\omega}^{(2)}(z')+\frac{\tilde{n}_{\omega}^{(2)}(z)}{1-i\omega\tau_0}-\frac{1}{1-i\omega\tau_0}\frac{\tilde{n}_{\omega}^{(2)}(0)}{2}F_{2,\omega}\left(\frac{z}{\ell}\right)+P_{\omega}(z)+\frac{\tilde{n}_{\omega}^{(2)}(0)}{2}F_{2,\omega}\left(\frac{z}{\ell}\right).
\end{align}
Since $\tilde{n}_{\omega}^{(2)}(z)\propto\delta(\omega)$ holds, we obtain
\begin{align}
    \int_{0}^{\infty}\mathrm{d}z'\tilde{F}_{2,0}\left(\frac{z-z'}{\ell}\right)\mathcal{E}_{\omega}^{(2)}(z')=-\frac{q\tau_0^2}{m}\left(\frac{\lambda_{\mathrm{TF}}}{\ell}\right)^2\delta(\omega)\left[\int_{-\infty}^{+\infty}\mathrm{d}\omega'S_{E,\omega'}\right]\left[-\frac{1}{2}\frac{e^{-w}}{1-\frac{1}{2}\frac{\lambda_{\mathrm{TF}}}{\ell}}+\frac{e^{-2\ell w/\lambda_{\mathrm{TF}}}}{1-\frac{1}{4}\frac{\lambda_{\mathrm{TF}}^2}{\ell^2}}\right].\label{eq_mathcalE2}
\end{align}
For further calculations, we also introduce the following equation:
\begin{align}
    \int_{0}^{\infty}\mathrm{d}z'\tilde{F}_{3,0}\left(\frac{z-z'}{\ell}\right)\mathcal{E}_{\omega}^{(2)}(z')=\frac{q\tau_0^2}{2m}\left(\frac{\lambda_{\mathrm{TF}}}{\ell}\right)^2\delta(\omega)\left[\int_{-\infty}^{+\infty}\mathrm{d}\omega'S_{E,\omega'}\right]e^{-w}+O\left[\left(\frac{\lambda_{\mathrm{TF}}}{\ell}\right)^3\right],\label{eq_mathcalE2_integrated}
\end{align}
which can be obtained by applying the following relation to Eq.~\eqref{eq_mathcalE2}:
\begin{align}
    \int_{0}^{\infty}\mathrm{d}w'\tilde{F}_{3,0}\left(w-w'\right)\mathcal{E}_{\omega}^{(2)}(w')=\int_{w}^{\infty}\mathrm{d}w''\left[\int_{0}^{\infty}\mathrm{d}w'\tilde{F}_{2,0}\left(w''-w'\right)\mathcal{E}_{\omega}^{(2)}(w')\right].
\end{align}
By solving Eq.~\eqref{eq_mathcalE2_integrated} via the Wiener-Hopf method and taking the Fourier transform, $\mathcal{E}^{(2)}(w,t)$ can be obtained as
\begin{align}
    \mathcal{E}^{(2)}(w,t)=\frac{q\tau_0}{4mv_{\mathrm{F}}}\left(\frac{\lambda_{\mathrm{TF}}}{\ell}\right)^2\left[\int_{-\infty}^{+\infty}\frac{\mathrm{d}\omega'}{\sqrt{2\pi}}S_{E,\omega'}\right]\frac{\displaystyle\prod_{k=1}^{N}(a_k+1)}{\displaystyle\prod_{k=1}^{N-1}(r_k+1)}\left\{\frac{\displaystyle\prod_{k=1}^{N}(a_k-1)}{\displaystyle\prod_{k=1}^{N-1}(r_k-1)}e^{-w}-\sum_{m=1}^{N-1}\left[\frac{\displaystyle\prod_{k=1}^{N}(-r_m+a_k)\cdot e^{-r_mw}}{\displaystyle(r_m-1)\prod_{\substack{k=1 \\ k\ne m}}^{N-1}(-r_m+r_k)}\right]\right\}.\label{eq_mathcalE2_caseiii}
\end{align}

Finally, we remark on the momentum flux tensor $\Pi_{zz}(z,t)$ for the discussion in Sec.~\ref{subsec: interfacial fluctuating electric field penetration}.
This tensor can be clearly described using the effective electric field $\mathcal{E}^{(2)}(z,t)$ and given by
\begin{align}
    \Pi_{zz}(z,t)=&\frac{1}{\Omega}\sum_{\bm{k}\gamma}m(v_{\bm{k}\gamma}^z)^2\frac{\tilde{n}^{(2)}(z,t)}{N_0}\left(-\frac{\partial f_{\bm{k}\gamma}^{(0)}}{\partial\varepsilon_{\bm{k}\gamma}}\right)+\frac{qN_0mv_{\mathrm{F}}^2\ell}{2}\int_0^\infty \mathrm{d}w'\tilde{F}_{4}(w-w')\mathcal{E}^{(2)}(w',t)\notag\\
    &-\frac{\lambda_{\mathrm{TF}}}{\ell}\frac{q^2N_0\ell^2}{2}\left[\int_{-\infty}^{+\infty}\frac{\mathrm{d}\omega'}{\sqrt{2\pi}}S_{E,\omega'}\right]\int_0^\infty \mathrm{d}w'[3\tilde{F}_{4}(w-w')+(w-w')\tilde{F}_{3}(w-w')]e^{-2\ell w'/\lambda_{\mathrm{TF}}},
    \label{eq_Pizz_result}
\end{align}
due to
\begin{equation}
    \begin{dcases}
        \underline{\mathrm{For}\,v_{\bm{k}\gamma}^z>0:}\\
        f_{\bm{k}\gamma,\omega}^{(2)}(z)=\frac{\tilde{n}_{\omega}^{(2)}(z)}{N_0}\left(-\frac{\partial f_{\bm{k}\gamma}^{(0)}}{\partial \varepsilon_{\bm{k}\gamma}}\right)+\int_{0}^{z}\mathrm{d}z'\exp{-\frac{z-z'}{v_{\bm{k}\gamma}^z\tau_{\omega}}}\left[q\mathcal{E}_{\omega}^{(2)}(z')\left(-\frac{\partial f_{\bm{k}\gamma}^{(0)}}{\partial \varepsilon_{\bm{k}\gamma}}\right)-\frac{1}{v^z_{\bm{k}\gamma}}\pdv{I_{\bm{k}\gamma,\omega}(z')}{k_z}\right]\\
        \underline{\mathrm{For}\,v_{\bm{k}\gamma}^z<0:}\\
        f_{\bm{k}\gamma,\omega}^{(2)}(z)=\frac{\tilde{n}_{\omega}^{(2)}(z)}{N_0}\left(-\frac{\partial f_{\bm{k}\gamma}^{(0)}}{\partial \varepsilon_{\bm{k}\gamma}}\right)+\int_{+\infty}^{z}\mathrm{d}z'\exp{-\frac{z-z'}{v_{\bm{k}\gamma}^z\tau_{\omega}}}\left[q\mathcal{E}_{\omega}^{(2)}(z')\left(-\frac{\partial f_{\bm{k}\gamma}^{(0)}}{\partial \varepsilon_{\bm{k}\gamma}}\right)-\frac{1}{v^z_{\bm{k}\gamma}}\pdv{I_{\bm{k}\gamma,\omega}(z')}{k_z}\right],
    \end{dcases}
\end{equation}
which is obtained using $\tilde{n}_{\omega}^{(2)}(z)\propto\delta(\omega)$ and holds to leading order in $\alpha/v_{\mathrm{F}}$.
The three terms on RHS of Eq.~\eqref{eq_Pizz_result} are all of order $O[(\lambda_{\mathrm{TF}}/\ell)^2]$.
By considering Eq.~\eqref{eq_n2_iii_solutionfin_leardingterm} and Eq.~\eqref{eq_mathcalE2_caseiii}, the leading term of $\partial_z\Pi_{zz}(z,t)$ within the screening region is found to be
\begin{align}
    \frac{\partial}{\partial z}\Pi_{zz}(z,t)=\frac{1}{\Omega}\sum_{\bm{k}\gamma}m(v_{\bm{k}\gamma}^z)^2\frac{\partial}{\partial z}\left[\frac{\tilde{n}^{(2)}(z,t)}{N_0}\right]\left(-\frac{\partial f_{\bm{k}\gamma}^{(0)}}{\partial\varepsilon_{\bm{k}\gamma}}\right)+O\left[\left(\frac{\lambda_{\mathrm{TF}}}{\ell}\right)^2\right].
\end{align}
We note that the relation $\partial_zU(z)\sim O[\ell/\lambda_{\mathrm{TF}}]$ is used here.
Using integration by parts, we obtain
\begin{align}
    \frac{\partial}{\partial z}\Pi_{zz}(z,t)=\frac{1}{\Omega}\sum_{\bm{k}\gamma}mv_{\bm{k}\gamma}^z\frac{\partial}{\partial z}\left[\frac{\tilde{n}^{(2)}(z,t)}{N_0}\right]\left(-\frac{\partial f_{\bm{k}\gamma}^{(0)}}{\partial k_z}\right)\sim\frac{1}{\Omega}\sum_{\bm{k}\gamma}\frac{\partial mv_{\bm{k}\gamma}^z}{\partial k_z}f_{\bm{k}\gamma}^{(0)}\cdot\frac{\partial}{\partial z}\left[\frac{\tilde{n}^{(2)}(z,t)}{N_0}\right]\sim\rho^{(0)}\frac{\partial}{\partial z}\left[\frac{\tilde{n}^{(2)}(z,t)}{qN_0}\right].
    \label{eq_Pizz_approximation}
\end{align}

\subsection{Solution of Wiener-Hopf equation for the quadratic spin response}
The self-consistent equation for the quadratic spin response $\tilde{S}_{z,\omega}^{(2)}(w)$ is obtained by multiplying Eq.~\eqref{eq_Boltzmann_quadratic_solution} by $\sigma_{\bm{k}\gamma}^z\delta(\varepsilon-\varepsilon_{\bm{k}\gamma})/2$, and then taking the sum $\frac{1}{\Omega}\sum_{\bm{k}\gamma}$.
Using Eq.~\eqref{eq_mathcalE2_integrated}, we obtain the Wiener-Hopf equation for $\tilde{S}_{z,\omega}^{(2)}(w)$ as follows:
\begin{align}
    \tilde{S}_{z,\omega}^{(2)}(w)=\frac{1}{2}\int_0^\infty \mathrm{d}w'\tilde{F}_{3,\omega}(w-w')\tilde{S}_{z,\omega}^{(2)}(w')+\frac{mq^2\ell\tau_0}{4\pi^2}\frac{\alpha}{v_F}\left(\frac{\lambda_{\mathrm{TF}}}{\ell}\right)^2\delta(\omega)\left[\int_{-\infty}^{+\infty}\mathrm{d}\omega'S_{E,\omega'}\right]\left[-e^{-w}+wF_2(w)\right].\label{eq_sz2_WienerHopf}
\end{align}
Here, we retain the only the leading-order terms in $\lambda_{\mathrm{TF}}/\ell$.
The solution to this Wiener-Hopf equation is
\begin{align}
    &\tilde{S}_{z,\omega}^{(2)}(w)\notag\\
    &=\frac{mq^2\ell\tau_0}{4\pi^2}\frac{\alpha}{v_F}\left(\frac{\lambda_{\mathrm{TF}}}{\ell}\right)^2\delta(\omega)\left[\int_{-\infty}^{+\infty}\mathrm{d}\omega'S_{E,\omega'}\right]\notag\\
    &\times\left(-\frac{\displaystyle\prod_{k=1}^{N}(a_k+1)}{\displaystyle\prod_{k=1}^{N}(t_k+1)}\left\{\frac{\displaystyle\prod_{k=1}^{N}(a_k-1)}{\displaystyle\prod_{k=1}^{N}(t_k-1)}e^{-w}+\sum_{m=1}^{N}\left[\frac{\displaystyle\prod_{k=1}^{N}(t_m-a_k)\cdot e^{-t_mw}}{\displaystyle(t_m-1)\prod_{\substack{k=1 \\ k\ne m}}^{N}(t_m-t_k)}\right]\right\}\right.\notag\\
    &\quad\left.+\sum_{n=1}^{N}\frac{b_nR_n}{a_n}\left\{\frac{\displaystyle\prod_{\substack{k=1 \\ k\ne n}}^{N}(-a_n+a_k)}{\displaystyle\prod_{k=1}^{N}(-a_n+t_k)}e^{-a_nw}+\sum_{m=1}^{N}\left[\frac{\displaystyle\prod_{\substack{k=1 \\ k\ne n}}^{N}(-t_m+a_k)\cdot e^{-t_mw}}{\displaystyle(-t_m+a_n)\prod_{\substack{k=1 \\ k\ne m}}^{N}(-t_m+t_k)}\right]-H_n\sum_{m=1}^{N}\left[\frac{\displaystyle\prod_{\substack{k=1 \\ k\ne n}}^{N}(-t_m+a_k)}{\displaystyle\prod_{\substack{k=1 \\ k\ne m}}^{N}(-t_m+t_k)}e^{-t_mw}\right]\right\}\right)\label{eq_sz2_caseiii_fin}
\end{align}
where
\begin{align}
    1-\sum_{k=1}^{N}\frac{b_k}{a_k(a_k^2+s^2)}=:\frac{\displaystyle\prod_{k=1}^{N}(s^2+t_k^2)}{\displaystyle\prod_{k=1}^{N}(s^2+a_k^2)},\quad R_n:=\frac{\displaystyle\prod_{k=1}^{N}(a_n+a_k)}{\displaystyle\prod_{k=1}^{N}(a_n+t_k)},\quad H_n:=\sum_{m=1}^{N}\left[\frac{1}{a_n+a_m}-\frac{1}{a_n+t_m}\right].
\end{align}
Figure \ref{fig_Caseiii} (c) compares Eq.~\eqref{eq_sz2_caseiii_fin} with the result obtained by substituting it into Eq.~\eqref{eq_sz2_WienerHopf} and shows the validity of Eq.~\eqref{eq_sz2_caseiii_fin}.
$t_k$ is determined numerically in a similar way to the case of $r_k$.
\twocolumngrid
\bibliography{reference}

\end{document}